\documentclass[11pt, tightenlines, onecolumn, amsmath, amssymb, notitlepage, a4paper, nofootinbib, superscriptaddress]{revtex4-1}

\usepackage[strict]{revquantum}

\usepackage{graphicx}
\usepackage{algpseudocode} 
\usepackage{algorithm}
\usepackage{subfig}
\usepackage{braket}
\usepackage[margin=1in]{geometry}
\usepackage{bbm}
\usepackage{enumitem}
\usepackage[normalem]{ulem}

\DeclareMathOperator{\tr}{tr}

\algnewcommand\algin{\textbf{in}}

\DeclareMathOperator*{\argmin}{arg\,min}

\newcommand{\nocontentsline}[3]{}
\newcommand{\tocless}[2]{\bgroup\let\addcontentsline=\nocontentsline#1{#2}\egroup}

\begin{document}

\title{Beyond Quantum Noise Spectroscopy: modelling and mitigating noise with quantum feature engineering}

\author{Akram Youssry}
\affiliation{University of Technology Sydney,
		Centre for Quantum Software and Information,
		Ultimo NSW 2007, Australia}
\affiliation{Department of Electronics and Communication Engineering, Faculty of Engineering, Ain Shams University, Cairo, Egypt} 
\author{Gerardo A. Paz-Silva}
\affiliation{Centre for Quantum Dynamics and Centre for Quantum Computation and Communication Technology, Griffith University, Brisbane, Queensland 4111, Australia}

\author{Christopher Ferrie}
\affiliation{University of Technology Sydney,
		Centre for Quantum Software and Information,
		Ultimo NSW 2007, Australia}

\date{\today}

\begin{abstract}

The ability to use quantum technology to achieve useful tasks, be they scientific or industry related, boils down to precise quantum control. In general it is difficult to assess a proposed solution due to the difficulties in characterising the quantum system or device. These arise because of the impossibility to characterise certain components \textit{in situ}, and are exacerbated by noise induced by the environment and active controls. Here we present a general purpose characterisation and control solution making use of a novel deep learning framework composed of quantum features. We provide the framework, sample data sets, trained models, and their performance metrics. In addition, we demonstrate how the trained model can be used to extract conventional indicators, such as noise power spectra.

\end{abstract}


\maketitle
\tableofcontents

%%%%%%%%%%%%%%%%%%%%%%%%%%%%%%%%%%%%%%%%%%%%%%%%%%%%%%%%%
\section{Introduction}

Accurately controlling the dynamics of open quantum systems is a central task in the successful implementation of quantum-enhanced technologies. Doing so to the highest possible level of accuracy involves a two-stage approach: first, quantum noise spectroscopy (QNS)~\cite{qns1,PhysRevLett.107.230501,qns3,qns4, NonGaussianPRL,Ramon,NonGaussianExp,Frey2017, multiqubit, multiaxis, PhysRevA.98.032315, LukazOpt,frey2019, Haas_2019} protocols are used to infer characteristics of the open quantum system  that affect the open quantum system dynamics, and then optimal control routines (OC) exploit this information to minimize the effect of noise and produce high quality gates~\cite{OC1,OC2}. 
    
    In this work, we go beyond the aforementioned approach and pose the problem in a machine learning (ML) context. By doing so, we provide a common language in which the ``learning'' (equivalent to QNS) and ``validation'' (the precursor to OC) cycles are directly related to the objective of controlling an  open quantum system. Notably, we show that doing so considerably extends the real-world applicability of the aforementioned two-stage strategy, as one can forgo some of the non-trivial control and model assumptions necessary for the implementation of sufficiently general QNS protocols. The success of the approach relies on the fact that the ML algorithm learns about the dynamics {\it relative to a given set of control capabilities}, which effectively reduces the complexity of the problem in a way meaningful to experimental constraints.

    The guiding principles of this work were to develop a framework that enables us to have models that are independent on any assumptions. It has to be suitable for estimating physically relevant quantities. And finally, it should have the capacity to do standard tasks such as decoherence suppression and quantum control. 
    
    The proposed method is based on a graybox approach, where all the known relations from quantum mechanics are implemented as custom whitebox layers---quantum features---while the parts that depend on assumptions on noise and control are modeled by standard blackbox machine learning layers. In this paper, we show how to construct such a model. We implement the algorithm and we test it against a set of simulated datasets. The results show high accuracy in terms of prediction error. We also show the possibility of utilizing the trained model to do basic quantum control operations. This paper opens the door for a number of possible novel machine-learning methods in the fields of quantum dynamics and control.

This work complements the existing literature applying classical machine learning to the quantum domain. 
Recently, machine learning and its deep learning framework \cite{deng2014tutorial} have been applied to many areas of quantum information, and physics more generally. Application areas include quantum control \cite{niu2019universal, PhysRevX.8.031086,ostaszewski2019approximation},  characterization of quantum systems \cite{aaronson2018online,ming2018quantum,youssry2019efficient,Youssry_2020}, experiment design \cite{krenn2016automated, melnikov2018active, o2018hybrid}, quantum cryptography \cite{Niemiec2019}, and quantum error correction  \cite{baireuther2018machine,chen2019machine,bausch2020quantum}. A related approach is Bayesian learning which was applied for Hamiltonian learning \cite{wiebe2014quantum, wang2017experimental}, quantum noise spectroscopy \cite{ferrie2018bayesian}, and characterization of devices \cite{lennon2019efficiently}.

The structure of the remainder of the paper is as follows.
The paper starts with overview on the formulation of the problem under consideration in Section \ref{sec:problem}. Section \ref{sec:methods} describes in detail the proposed solution using a graybox ML approach. Next, Section \ref{sec:simulations} discusses the implementation of the proposed method followed by the presenting the numerical results and its significance. Section \ref{sec:conclusions} concludes the paper and gives perspectives on the the possible extensions of this work. 

The Appendix includes additional materials presented for the sake of completeness. First, Appendix \ref{appx:simulator} gives an overview on the implementation details of the quantum simulator used to generate the training and testing datasets. Next, Appendix \ref{appx:Vo} gives a detailed derivation of Equation \ref{equ:Vo} upon which the proposed ML model was built. Appendix \ref{appx:ml} gives a brief introduction to the ML blackbox layers that were utilized in this paper. Finally, Appendix \ref{appx:figures} contains supplementary figures related to the results discussed in Section \ref{sec:simulations}. 

\section{Problem Statement} \label{sec:problem}
In broad terms, our objective is to effectively ``characterize'' and accurately predict the dynamics of a two-level open quantum system, i.e., a qubit interacting with its environment, undergoing user-defined control picked from a fixed set, e.g., consistent with the  control capabilities available to a given experimental platform. In what follows we make this statement precise.

For concreteness we start by choosing a model for our dynamics, although we highlight that the equations we derive below apply generally to classical noise models and can be readily generalized to the scenario of quantum noise models~\cite{multiaxis}. We consider then a qubit evolving under a time-dependent Hamiltonian of the form,
\begin{align}
    H(t) = \frac{1}{2}\left(\Omega + \beta_z(t) + f_z(t) \right)\sigma_z + \frac{1}{2}\left( \beta_x(t) + f_x(t) \right) \sigma_x + \frac{1}{2}\left(\beta_y(t) +  f_y(t)\right) \sigma_y,
\end{align}
where $\Omega$ is the energy gap of the qubit, $\beta_\alpha (t)$ represents the realizations of a classical noise process along the $\alpha-$ direction, and $f_\alpha (t)$ implements the user-defined control pulses along the  $\alpha-$direction, and $\sigma_\alpha$ is the $\alpha$ Pauli matrix. Since we are interested in predicting the dynamics of the qubit in a time interval $[0,T]$, we will be interested in the expectation value $\mathbb{E}\{O(T)\}_\rho$ of observables $O$ at time $T$ given an arbitrary initial state $\rho$ and a choice of $\{f_\alpha(t)\}$. 

While these expectation values contain the necessary information, it will be convenient to further isolate the effect of the noise. To this end we proceed as follows. Our starting point is the usual expression,
$$
\mathbb{E}\{O(T)\}_\rho = 
\langle \Tr[U(T) \rho U(T)^\dagger O]\rangle_c,
$$
where $U(T)= \mathcal{T} e^{- i\int_0^T ds H(s)}$ and $\braket{\cdot}_c$ denotes classical averaging over the noise realizations of the random process $\beta_\alpha(t)$. One can then move to a toggling-frame with respect to the control Hamiltonian, 
\begin{align}
    H_{\text{ctrl}}(t) = \frac{1}{2}\left(\Omega + f_z(t) \right)\sigma_z + \frac{1}{2} f_x(t)  \sigma_x + \frac{1}{2} f_y(t) \sigma_y,
    \label{equ:Hctrl}
\end{align}
inducing a control unitary $U_{\text{ctrl}}(T)$ via, 
\begin{align}
    U_{\text{ctrl}}(T) = \mathcal{T}_+ e^{-i\int_0^T{H_{\text{ctrl}}(s) ds}},
    \label{equ:Uctrl}
\end{align} 
which enables the decomposition,
$$
U(T) = \tilde{U}_I(T) U_{\rm ctrl}(T),
$$
with, 
$$\tilde{U}_I(T) = \mathcal{T}_- e^{-i\int_0^{T}{H_I(s) ds}},$$
the (modified) interaction picture evolution (see Appendix \ref{appx:Vo} for details). In turn, this allows us to rewrite, \begin{align}
    \mathbb{E}\{O(T)\}_\rho = \Tr [V_O(T) U_{\rm ctrl}(T) \rho U_{\rm ctrl}(T)^{\dagger} O],
    \label{equ:Vo}
\end{align}
where  
%
%Assuming the qubit is initialized in state $\rho$ at  $t=0$, and we let it evolve till $t=T$, then we perform a measurement represented by the operator $O$ such that $O^2 = I$, then the expectation of the outcome can be expressed as
%
%
%where $\braket{\cdot}_q$ denotes the usual quantum average of an observable and $\braket{\cdot}_c$ denotes classical averaging over the noise realizations, $U_C$ is the control unitary operator, 
the operator \begin{align}
    V_O(T) = \braket{O^{-1} \tilde{U}_I^{\dagger}(T) O \tilde{U}_I(T)}_c, 
\end{align}
conveniently encodes the influence of the noise. As such, this operator is central to understanding the dynamics of the open quantum system. If our objective is, as is common in optimal control protocols and imperative when quantum-technology applications are considered, to minimize the effect of the noise, e.g., via a dynamical decoupling~\cite{DS1,DS2,DS5} or composite pulses~\cite{DS3,DS4}, then one needs to determine a set of controls for which $V_O \to I$.  Notice that $\tr[ O \braket{U_I^{\dagger} O U_I}_c]$ can be interpreted as the ``overlap' between the observable $O$ and its time evolved version $\braket{{U_I}^{\dagger} O U_I}_c$, which is maximum when the evolution is noiseless. If additionally one wants to implement a quantum gate $G$, then we further require that $U_c \rho U_c ^{\dagger} \to G \rho G^{\dagger}$. Regardless of our objective, it is clear that one needs to be able to predict $V_O(T)$ given (i) the actual noise affecting the qubit and (ii) a choice of control. However, realistically the information available about the noise is limited, and  by the very definition of an open quantum system is something that cannot typically be measured directly.

Fortunately, this limitation can in principle be overcome by quantum noise spectroscopy (QNS) protocols~\cite{qns1,PhysRevLett.107.230501,qns3,qns4, NonGaussianPRL,Ramon,NonGaussianExp,Frey2017, multiqubit, multiaxis, PhysRevA.98.032315, LukazOpt,frey2019}. These protocols exploit the measurable response of the qubit to a known and variable control and the noise affecting it, in order infer information about the noise. The type of accessible information is statistical in nature. That is, without any other information, e.g., about the type of stochastic noise process, the best one can hope to learn are the bath correlation functions $\langle \beta_{\alpha_1}(t_1) \cdots \beta_{\alpha_k} (t_k)\rangle$. If the QNS protocol is sufficiently powerful to characterize the leading correlation functions and matches the model, in principle the inferred information can be plugged into a cumulant expansion or a Dyson series expansion of $V_O(t)$ to successfully obtain an estimate of the operator for any choice of $f_\alpha(t)$, as desired. This has led to a proliferation of increasingly more powerful QNS protocols, including those capable of characterizing the noise model described here~\cite{multiaxis}, some of which have even been experimentally verified~\cite{qns1,PhysRevLett.107.230501,qns3,qns4, NonGaussianExp, frey2019,Frey2017}. More generally, the idea of optimizing control procedures to a known noise spectrum~\cite{OC1} is behind some of the most remarkable coherence times available in the literature~\cite{Wang2017}.   

QNS protocols, however, are not free of complications. The demonstrated success of these protocols relies on the assumptions which support them being satisfied. Different protocols have different assumptions, but they can be roughly grouped into two main flavors:
\begin{itemize}
    \item {\it Assumptions on the noise.---} %TBC
    Existing protocols assume that the only a certain subset of the correlation functions effectively influence the dynamics or, equivalently that a perturbative expansion ov $V_O(t)$ can be effectively truncated to a fixed order. In practice, this is enforced in various ways. For example, demanding that the noise is Gaussian and dephasing or, more generally, that one is working in an appropriately defined ``weak coupling" regime~\cite{multiqubit, NonGaussianPRL,PhysRevA.98.032315}.  
    
    \item {\it Assumptions on the control.---} %TBC
    Many QNS protocols, especially those based around the so-called ``frequency-comb''~\cite{PhysRevLett.107.230501, NonGaussianPRL, multiqubit,multiaxis}, rely on specific control assumptions, such as that pulses are instantaneous. This assumption facilitates the necessary calculations, which ultimately allow the inferring of the noise information. However, it enforces constraints on the control that translate into limitations on the QNS protocol, e.g., a maximum frequency sampling range~\cite{PhysRevLett.107.230501, NonGaussianPRL}. Moreover, experimentally one cannot realize instantaneous pulses, so comb-based QNS protocols are necessarily an approximation with an error that depends on how far the experiment is from satisfying the instantaneous pulse assumption.
\end{itemize}

%% GAPS: Needs a bit more concrete motivation.

This work overcomes these limitations by bypassing the step of inferring the bath correlation functions. We maintain the philosophy of QNS regarding characterizing the open quantum system dynamics, but pose it in a machine learning context. Thus, we address the question: \\

{\noindent \it Can an appropriately designed machine learning algorithm ``learn'' enough about the open quantum system dynamics (relative to a given set of control capabilities), so as to be able to accurately predict its dynamics under an arbitrary element of the aforementioned set of available controls?  } 
\\

We answer positively to this question by  implementing such ML-based approach. Concretely our ML algorithm (i) learns about the open quantum system dynamics and (ii) is capable of accurately estimating -- without  assuming a perturbative expansion -- the operator $V_O(T)$,  and consequently measurement outcomes,  resulting from a control sequence picked from the family  control pulses $\{f_\alpha(t)\}$ specified by an assumed (but in principle arbitrary) set of control capabilities.

%%%%%%%%%%%%%%%%%%%%%%%%%%%%%%%%%%%%%%%%%%%%%%%%%%%%%%%%%%%%%%%%%%%
\section{Methods}\label{sec:methods}
In this section we present in detail the proposed method to solve the problem under consideration. We start by giving an overall summary of our proposed solution. Next, in Section \ref{sec:Vo} we discuss some of the mathematical properties of the $V_O$ operator. This will allow us to find a suitable parameterization that will be useful to build the architecture of the ML model. Next, we present exactly the architecture of the ML model in Section \ref{sec:architecture}. After that, we give an overview on how to train the model in Section \ref{sec:training}. Finally, in Section \ref{sec:performance} we conclude with the metrics used to assess the performance of the proposed model.
%%%%%%%%%%%%%%%%%%%%%%%%%%%%%%%%%%%%%%%%%%%
\subsection{Overview} 
The ML approach naturally matches our control problem, which becomes clear from the following observation. For most optimal control applications, e.g. achieving a target fidelity for a gate acting on an open quantum system, one does not need to have full knowledge of the noise. To see this  
consider a hypothetical scenario where the available control is band limited~\cite{Slepian, Frey2017,PhysRevA.98.032315}, i.e., whose frequency domain representation $F(\omega)$ is compactly supported in a fixed frequency range $|\omega| \leq \Omega_0$. If the response of the open quantum system to the noise~\cite{PazFF} is captured by a convolution of the form $I = \int_{-\infty}^\infty d\omega F(\omega) S(\omega)$, where $S(\omega)$ represents the noise power spectrum, then it is clear that one only needs to know $S(\omega)$ for $|\omega| \leq \Omega_0$. While this statement can be formalized and made more general, we do so Ref.~\cite{Behnam20}, the above example captures a key point: only the ``components'' of the noise that are relevant to the available control need to be characterized. Conversely, this means that a fixed set of resources, e.g., a set of control capabilities, can only provide information about the ``components'' of the noise relevant to them. The above observations make the ML approach particularly well suited for the problem: it is natural to draw the connection between the control problem of ``characterizing a system with respect to a restricted set of control capabilities in order to predict the dynamics under any control such capabilities can generate'' and the fact that the training and testing datasets typical in ML make sense when the datasets are generated in the same way, i.e., by the same ``control capabilities". Of course, the details of the ML approach which can seamless integrate with the quantum control equations are important, and we now provide them. 

In order to address the question presented at the end of section \ref{sec:problem}, we are going to use an ML graybox based approach similar to the one presented in \cite{Youssry_2020}. The basic idea of a graybox is to divide the ML model into two parts, a blackbox part and a whitebox part. The blackbox part is a collection of standard ML layers, such as neural networks (see Appendix \ref{appx:ml} for an overview), that allows us to learn maps between variables without any assumptions on the actual relation. The whitebox part is a collection of customized layers that essentially implement mathematical relations that we are certain of. This approach is better than a full blackbox, because it allows us to estimate physically relevant quantities, and thus enables us to understand more about the physics of the system. In other words, the blackboxes are enforced to learn some abstract representations, but when combined with the whiteboxes we get physically significant quantities. In the parlance of machine learning, these whitebox layers are ``quantum features'', which extract the expect patterns in the data fed to the network.  

%{\bf \color{blue} GAPS: I will: Add emphasis of 'control-capability' related learning..}

In the case of the problem under consideration, we are going to use the blackbox part to estimate some parameters for reconstructing the $V_O$ operators. The reason behind the use of a blackbox for this task is because the calculation of the $V_O$ operators depends on assumptions on the noise and control signals. So, by using a blackbox we get rid of such assumptions. Whereas the whitebox parts would be used for the other standard quantum calculations that we are certain of, such as the time-ordered evolution, and quantum expectations. Thus, we end up with an overall graybox that essentially implements Equation \ref{equ:Vo}, with input representing the control pulses, output corresponding to the classical expectation of quantum observable over the noise, and internal parameters modeling abstractly the noise and its interaction with the control. With this construction, we would be able to estimate important quantities such as $V_O$, and $U_{\text{ctrl}}$. Now, since we are using machine learning, then we will need to perform a training step to learn the parameters of the blackboxes. Thus, the general protocol would be as follows:

\begin{enumerate}
    \item Prepare a training set consisting of pairs of random input pulse sequences  applied to the qubit { (chosen from a fixed and potentially infinite set of allowed sequences)}, and the measured outcomes after evolution.
    \item Initialize the internal parameters of the graybox model.
    \item Train the model for some number of iterations until convergence.
    \item Fix the trained model and use it to predict measurement outcomes for new pulse sequences as well as the $V_O$ operators.
\end{enumerate}

%%%%%%%%%%%%%%%%%%%%%%%%%%%%%%%%%%%%%%%%%%%%
\subsection{Mathematical properties of the $V_O$ operator}\label{sec:Vo}
If we look back into the definition of the $V_O$ operator, we find that it can be expressed as
    \begin{align}
        V_O = O^{-1} \braket{U_I^{\dagger} O U_I}_c,
    \end{align}
    because the observable is independent of the noise. As a result we can see the term that is inside the expectation has the following mathematical properties.
    \begin{enumerate}
        \item It is traceless because $\tr{\left(U_I^{\dagger} O U_I\right)} = \tr{O} = 0$.
        \item It is Hermitian because $\left(U_I^{\dagger} O U_I\right)^{\dagger} = \left(U_I^{\dagger} O U_I\right)$, assuming that $O^{\dagger}=O$ which is true for any quantum obervable.
        \item It is unitary because $\left(U_I^{\dagger} O U_I\right)^{\dagger}\left(U_I^{\dagger} O U_I\right)=I$. This property holds for the Pauli observables and will still hold for any operator such that $O^2 = I$.
        \item Consequently all its eigenvalues are real and lie on the unit circle, so the only possibility is that the half the eigenvalues are $+1$ and the other half is $-1$. 
    \end{enumerate}
    Now, if we look into the mathematical properties of the whole expectation term, we find that it is a sum of traceless Hermitian unitary operators. Consequently, the expectation should satisfy the following properties:
    \begin{enumerate}
        \item It is traceless $\tr{\braket{U_I^{\dagger} O U_I}_c}=0$.
        \item It is Hermitian $\braket{U_I^{\dagger} O U_I}_c^{\dagger} = \braket{U_I^{\dagger} O U_I}_c$.
        \item In general, it will not be unitary, however the eigenvalues should be real and satisfy that $-1 \le \lambda\left(\braket{U_I^{\dagger} O U_I}\right) \le 1$. This can be proved as follows. Suppose for the sake of convenience that the probability distribution of the $U_I^{\dagger} O U_I$ with respect to the noise is finite and discrete. That is it can only take values $\tilde{O}_i$ with probability $P_i$. Then  
        \begin{align}
            \lambda_{\text{max}}\left(\braket{U_I^{\dagger} O U_I}_c \right) &= \lambda_{\text{max}}\left(\sum_{i=1}^{i_{\text{max}}}{P_i \tilde{O}_i} \right)\\
                                                                           &= \lambda_{\text{max}}\left(P_1 \tilde{O}_1 + \sum_{i=2}^{i_{\text{max}}}{P_i \tilde{O}_i} \right)\\
                                                                           &\le \lambda_{\text{max}}\left(P_1 \tilde{O}_1\right) + \lambda_{\text{max}}\left( \sum_{i=2}^{i_{\text{max}}}{P_i \tilde{O}_i} \right) \\
                                                                           &= P_1 \lambda_{\text{max}}\left(\tilde{O}_1\right) + \lambda_{\text{max}}\left( \sum_{i=2}^{i_{\text{max}}}{P_i \tilde{O}_i} \right) \\
                                                                           &= P_1 +\lambda_{\text{max}}\left( \sum_{i=2}^{i_{\text{max}}}{P_i \tilde{O}_i} \right).
        \end{align}
        The third line follows from Weyl's inequality since all the terms of the form $P_i \tilde{O_i}$ are Hermitian. Now, if we repeat recursively the same steps on the second remaining term, we get
         \begin{align}
            \lambda_{\text{max}}\left(\braket{U_I^{\dagger} O U_I}_c \right) &= P_1 + P_2 + \cdots P_{i_{\text{max}}}\\
            &= 1,
        \end{align}
        as the $P_i$'s form a probability distribution. Similarly, we can show that
        \begin{align}
            \lambda_{\text{min}}\left(\braket{U_I^{\dagger} O U_I}_c \right) &\ge \lambda_{\text{min}}\left(\sum_{i=1}^{i_{\text{max}}}{P_i \tilde{O}_i} \right)\\
                                                                           &\ge -1.
        \end{align}
        and so by combining the two results we get
        \begin{align}
            -1 \le \lambda\left(\braket{U_I^{\dagger} O U_I}_c\right) \le 1
        \end{align}
        \end{enumerate}
        
    This proof can be extended to the more realistic situation when the noise distribution is continuous. Thus, for a $d\times d$ system, by specifying a diagonal matrix $D$ whose entries are real numbers in the interval $[-1,1]$ adding up to 0, and by choosing a general unitary matrix $Q$, we can reconstruct any $V_O$ operator in such a way that satisfies its mathematical properties, using the eigendecomposition 
    \begin{align}
        V_O = \braket{O^{-1} U_I^{\dagger} O U_I}_c = O^{-1}QDQ^{\dagger}.
    \end{align}
    In particular for the case of qubit presented in this paper (i.e. $d=2$), we can completely specify the $V_O$ operator using 4 parameters which we would refer to as $\psi$, $\theta$, $\Delta$, and $\mu$ such that
    \begin{align}
        Q = \begin{pmatrix} e^{i\psi} & 0 \\ 0 & e^{-i\psi} \end{pmatrix} \begin{pmatrix} \cos{\theta} & \sin{\theta} \\ -\sin{\theta} & \cos{\theta} \end{pmatrix} \begin{pmatrix} e^{i\Delta} & 0 \\ 0 & e^{-i\Delta} \end{pmatrix}, 
    \end{align}
    where we neglected a degree of freedom that represents an overall global phase shift, and
    \begin{align}
        D = \begin{pmatrix} \mu & 0 \\ 0 & -\mu \end{pmatrix}.
    \end{align}
    The parameters $\psi$, $\theta$, and $\Delta$ can take any real values as they are arguments of periodic functions. However, based on the previous discussion, the real parameter $\mu$ must lie in the interval $[0,1]$.
%%%%%%%%%%%%%%%%%%%%%%%%%%%%%%%%%%%%%%%%%%%%%%%%%%%%%%%%%
\subsection{Model architecture}\label{sec:architecture}
\begin{figure}
    \centering
    \includegraphics{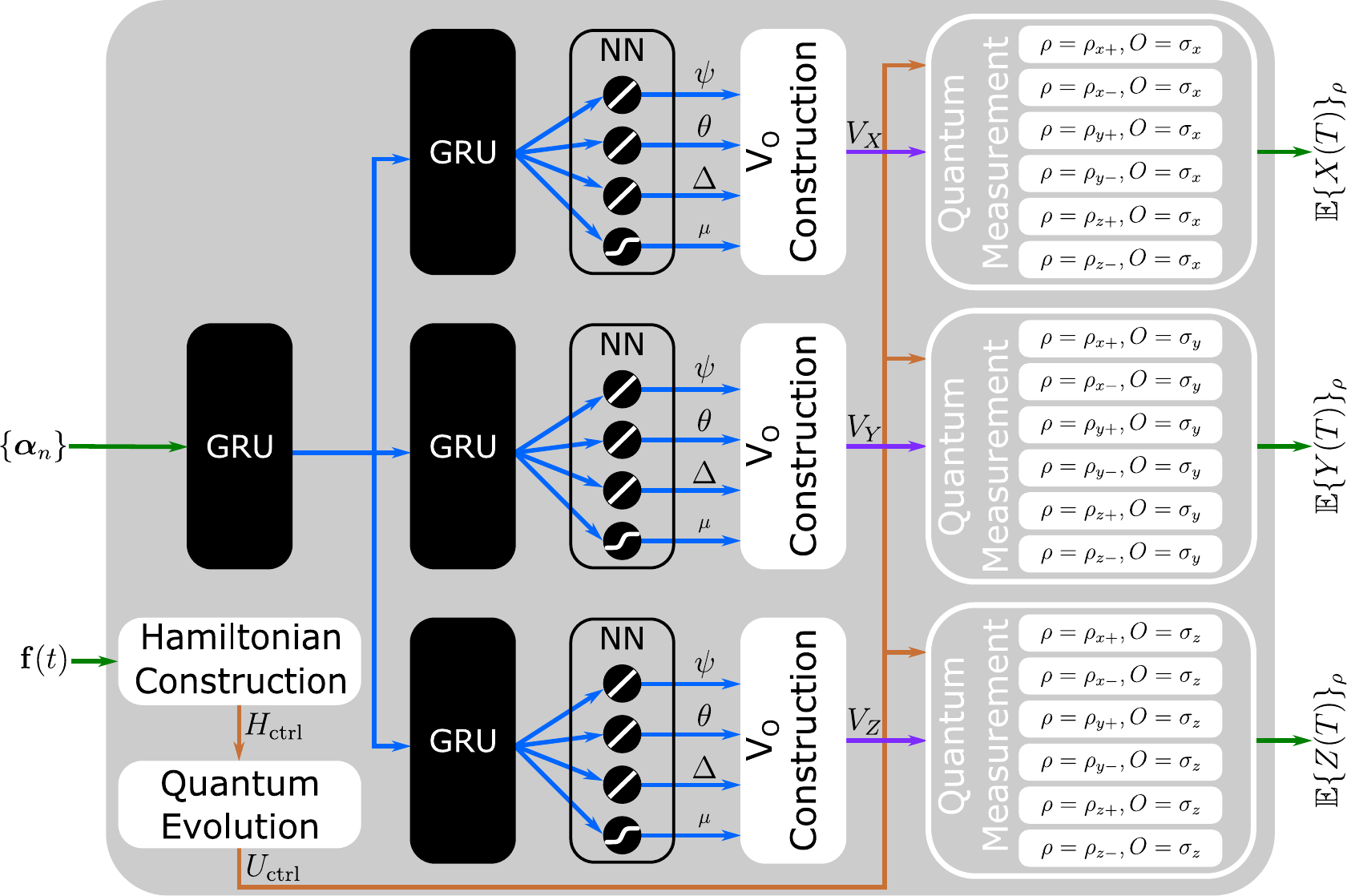}
    \caption{The proposed graybox architecture for modeling the noisy qubit following Equation \ref{equ:Vo}. The inputs of the model are the sequence of control signal parameters $\{\boldsymbol{\alpha}_n\}$, and the actual time-domain waveform $\mathbf{f}(t)$. The outputs of the model are the expectations over the noise for all the Pauli eigenstates as initial states, and all Pauli's as measurement operators. The black box part of the model consist of two layers of GRU followed by a neuron layer. The output of this layer represents the parameters that can be used to construct the ``$V_O$'' operator. There are three different branches corresponding to each of the three possible Pauli obersvables. The whitebox part of the model is formed from the layers that implement specific formulas known from quantum mechanics. This includes layers for constructing the $V_O$ operators from the parameters generated from the blackbox, constructing the control Hamiltonian from the time domain representation of the control pulse sequence, the time-ordered evolution to generate the control unitary, and the quantum measurement layer. The model is trained using a set of pairs of control pulse sequence and corresponding expectation values of the observables. After training, the model can be used to predict the measurements for new pulse sequences. It can also be probed to estimate one of the ``$V_O$" operators, and thus can be used as a part of a quantum control algorithm to achieve a desired quantum gate. }
    \label{fig:model}
\end{figure}

The proposed graybox ML model is shown in Figure \ref{fig:model}. We shall explain in detail the structure as follows.

\subsubsection{Model inputs and outputs}
The purpose of the proposed architecture is to have a model that relates the control pulses applied on the qubit (which we have control over in an experiment) to the classical average of the quantum observables (which we can physically measure). The model internal parameters will act as an abstract representation for the noise, as well as how it affects the measurement outcomes. The model has two main inputs which depend only on the control pulses. The first one represents a set of features extracted from the time domain representation of the pulse sequence. We assume in this paper that the control signal can be parameterized by a finite set of parameters. This still allows having infinitely large number of possible control signals since each of the parameters can take infinitely many values. For example a train of $N$ Gaussian pulses can be completely defined by $3N$ parameters: the amplitude, mean, and variance of each of the $N$ pulses. Similarly, a train of square pulses can be defined by each of the pulse positions, pulse widths, and pulse amplitudes. In the language of machine learning, these are called features. The process of evaluating those features is called feature extraction. However, particularly for the presented application, these features are pre-known, since we choose the control sequence in the first place. Now, these features have to be represented in a way that is suitable for the subsequent blocks (standard ML blackboxes) to process. So, the first step is to normalize each of signal parameters to be in the range of $[0,1]$ across all the examples. For instance, the pulse locations can be normalized with respect to the the total evolution time $T$ since there will be no pulses beyond this point in time. The pulse amplitudes could also be normalized such that the maximum amplitude for any pulse sequence is 1. The second is step is the proper formatting of these parameters. We choose to organize the signal parameters of an $n_{\text{max}}$ pulse train in the form of a sequence of vectors $\{\boldsymbol{\alpha}_n\}_{n=1}^{n_{\text{max}}}$, where each vector represents the $n^{\text{th}}$ pulse and has $r$ entries representing the normalized pulse parameters (example: Gaussian pulse train will have $r=3$). For the case of multi-axis control, we concatenate the parameterization along each direction into one vector assuming the controls are independent along each direction. We emphasize here that we take Gaussian and square pulses as examples to demonstrate our ideas, but in general any waveform with any suitable parameterization could be uses.

The second input to the model is the actual time domain representation of the pulse sequence, discretized into $M$ steps. This input is only processed only by customized whiteboxes. Although in principle we can calculate the time-domain representation from the signal parameters in the first input, it turns out that the overall algorithm performs better if we do not do this calculation directly. In other words treat both features as two ``independent inputs" to the model.

The output of the model should be the measurement outcomes. If we initialize our qubit to each of the eigenstates (``up/down") of each Pauli operators (that is a total of 6 states), and measure the three Pauli operators, then we have enough information (tomographically complete) to predict the dynamics for other configurations. So, we need to perform a total of 18 ``prepare-measure" experiments and collect their results. And so, our model will have 18 outputs corresponding to each of the measurement settings.
%%%%%%%%%%%%%%%%%%%%%%%%%%%%%%%%%%%%%%%%%%%
\subsubsection{Model whiteboxes}
As discussed previously, there are lots of known relations from quantum mechanics that we are certain of. It is better in terms of the overall performance to directly implement as much as possible of these relations in non-standard customized layers. This saves the machine from essentially having to learn everything about quantum mechanics from the data, which would be hard and could decrease the overall accuracy. Moreover, this allows us to evaluate physically significant quantities which is one of the most important general advantages of the graybox approach. In our proposed model, we make use of the following whiteboxes.
\begin{itemize}
    \item Hamiltonian Construction
    
    This layer takes the discretized time domain representation of the control pulses (which is exactly the second input to the model), and outputs the control Hamiltonian $H_{\text{ctrl}}$ evaluated at each of the $M$ time steps using Equation \ref{equ:Hctrl}. The layer is also parameterized by the energy gap $\Omega$ which we fix at the beginning. 
    
    \item Quantum Evolution
    
    This layer follows the ``Hamiltonian Construction" layer, and thus it takes the control Hamiltonian at each time step as input and evaluates the time-ordered quantum evolution as output (i.e. Equation \ref{equ:Uctrl}). Numerically, this is calculated using the approximation of an infinitesimal product of exponentials
    \begin{align}
        U_{\text{ctrl}} &= \mathcal{T}_+ e^{-i\int_0^T{H_{\text{ctrl}}(s) ds}} \\
                        &= \lim_{M\to\infty} e^{-i H_{\text{ctrl}}(t_M)\Delta T} e^{-i H_{\text{ctrl}}(t_{M-1})\Delta T} \cdots e^{-i H_{\text{ctrl}}(t_0)\Delta T} \\
                        & \approx e^{-i H_{\text{ctrl}}(t_M)\Delta T} e^{-i H_{\text{ctrl}}(t_{M-1})\Delta T} \cdots e^{-i H_{\text{ctrl}}(t_0)\Delta T} \label{equ:Texp}, 
    \end{align}
    where $t_k= k\Delta T$ and $\Delta T = \frac{T}{M}$. The last line follows if $M$ is large enough. 
    
    \item $V_O$ Construction
    
    This layer is responsible to reconstruct the $V_O$ operator. It takes the parameters $\psi$, $\theta$, $\Delta$, and $\mu$ as inputs and outputs the $V_O$ following the reconstruction discussed in Section \ref{sec:Vo}. The blackboxes of the overall model are responsible for estimating those parameters. The output of this layer can be probed to estimate the $V_O$ operator given a control pulse. This allows us to do further operations including noise spectroscopy and quantum control.

    \item Quantum Measurement
    
    This layer is essentially the implementation of Equation \ref{equ:Vo}. So, it takes the $V_O$ operator as input, together with the control unitary, and outputs the trace value. It is parameterized by the initial state of the qubit, as well as the observable to measure. Therefore, in order to calculate all possible 18 measurements, we need 18 of such layers in the model, each with the correct combination of inputs and parameterization. The outputs of all 18 layers are concatenated finally and they represent the model's output.  
\end{itemize}
%%%%%%%%%%%%%%%%%%%%%%%%%%%%%%%%%%%%%%%%%%
\subsubsection{Model blackboxes}

 The exact calculation of the measurement outcomes requires assumptions on both the noise and control pulse sequence. So, by using the standard ML blackbox layers, such as Neural Networks (NN) and Gated Reccurent Units (GRU) (see Appendix \ref{appx:ml} for an overview), we can have an abstract assumption-free representation of the noise and its interaction with the control. This would allow us to estimate the required parameters for reconstructing the $V_O$ operators using a whitebox. The power of such layers comes from their effectiveness in representing unknown maps due to their highly non-linear complex structure. In our proposed model we have three such layers explained as follows.

\begin{itemize}
    \item Initial GRU
    
    This layer is connected to the first input of the model, i.e. the parameters of the control pulse sequence. The purpose behind this layer is to have an initial pre-processing of the input features. Feature transformation is commonly used in ML algorithms, to provide a better feature space that would essentially enhance the learning capability of the model. In the modern deep learning paradigm, instead of doing feature transformation at the beginning, we actually integrate it within the overall algorithm. In this way, the algorithm learns the best optimal transformation of features that increases the overall accuracy. In our application, the intuition behind this layer is to have some sort of abstract representation of the interaction unitary $U_I$. This would depend on the noise as well as the control. In this sense, the input of the layer represents the control pulses, the output represents the interaction evolution operator, and the weights of the layer represent the noise. This does not mean that probing the output layer is exactly related to the actual $U_I$ as the algorithm might have a completely different abstract representation, which is a general feature of blackboxes. In the proposed model, we choose the GRU unit to have 10 hidden nodes.
    
    \item Final GRU
    
    This is another GRU layer that is connected to the output of the initial GRU layer. The purpose of this layer is to increase the complexity of the blackboxes so that the overall structure is complex enough to represent our relations. For our application, this layer serves as a way to estimate the operator $\braket{U_I^{\dagger} O U_I}_c$ in some abstract representation. And thus we need actually three of such layers to correspond to the three Pauli observables. We choose to have 60 hidden nodes for each of these layers.
    
    \item Neural Network
    
    This is a fully-connected single neural layer consisting of four nodes. The output of the final GRU  layer is connected to each of the nodes. The first three nodes have linear activation function and their output represent the actual parameters $\psi$, $\theta$, and $\Delta$ that are used to construct the $V_O$ operator. The last node has a sigmoid activation function and its output corresponds exactly to the $\mu$ parameter of the $V_O$ operator. As discused before the $\mu$ parameter has to be in the range $[0,1]$ which is exactly the range of the sigmoid function. Since the parameters will differ for each of the three observables, we need three of such layers each connected to one of the final GRU layers.  
\end{itemize}
%%%%%%%%%%%%%%%%%%%%%%%%%%%%%%%%%%%%%%%%%%%%%%%%%%%%%%%%%%
\subsection{Training and Testing}\label{sec:training}
In order to find the parameters of the blackboxes of the model, we have to train the model using some dataset. So, the first step is to prepare a training dataset that consists of pairs of inputs and corresponding outputs. In our application, this corresponds to performing an experiment in the lab where we prepare the qubit in an initial state, apply some control, then measure the observable. This is repeated for all 18 possible configurations of initial states/observable. This pair consisting of the control pulse parameterization and time-domain representation, and the value of the 18 measurements would correspond to one example in the dataset. To have more examples we need to choose a different control sequence and repeat the process. Now, we need the model to be able to generalize, i.e. predict the outcomes for new control pulses that were not in the training set. This is an important requirement for any machine learning algorithm. The way to do it is to make sure the training set is large enough to represent wide range of cases. For instance, consider constructing a training set of CPMG-like sequences~\cite{CPMG}, i.e., sequences composed of equally spaced pulses. Then, in order for the model to have the capability of predicting the correct outcomes if the pulses are shifted (maybe for some experimental errors), we have to provide training examples that have the control pulses randomly shifted. Similarly, if we want good enough predictions for control pulses that have powers other than $\pi$, then we need to include such examples in the training set. Also, the prediction would work for the same pulse shape. This means we need to prepare different training sets and train different models if we have different pulse shapes. However, usually the pulse shapes are fixed in experiment, for example Gaussian or square. In this case, we might be interested in training only one model with one training set. One could also consider training the same model with different datasets, but this is out of scope of this paper.

The second step is to choose a loss function for the model. This a function that measures how accurate the outputs predicted by the model compared to the true outputs. This choice depends on the application under consideration. In our case, we shall use the Mean-Square-Error (MSE), averaged over all 18 measurement outcomes. The weights of the model are chosen such that the loss function is minimized. Ideally, we seek a global minimum of the loss function but in practice this might be hard and we probably end up with a local minimum. However, practically, this usually provides sufficient performance. 

The third step is to choose an optimization algorithm. The optimization is for finding the weights of the model that minimizes the loss function averaged over all training examples. The standard method used in ML is backpropagation which is essentially a gradient descent based method combined with an efficient way of calculating the gradients of the loss function with respect to the weights. There are many variants of the backprogation method in the literature, the one we choose to use in this paper is the Adam algorithm \cite{Adam}. There exist also other gradient-free approaches such as Genetic Algorithm (GA) based optimization \cite{bausch2020quantum}.  

The fourth step is to actually perform the training. In this case, we initialize the weights of the model to some random values, then apply enough iterations of the optimization algorithm till the loss function reaches a sufficiently small value. In the case of MSE, we would like it ideally to be as close as possible to $0$, but this could require infinite number of iterations. So, practically we stop either when we reach sufficient accuracy or we exceed a maximum number of steps. 

A final thing to mention is that because the whiteboxes do not have any trainable parameters, the blackboxes are enforced through the training to generate outputs that are compatible with the whiteboxes, so that we end up with the correct physical quantities.  

\subsection{Performance analysis}\label{sec:performance}
After executing the aforementioned steps for training the model, it will be able to predict accurately the outputs of the training examples. In our application, this by itself is useful because we can easily probe the output of the $V_O$ layers and use that prediction for various purposes. However, we have to ensure the model is also capable of generalizing to new examples. The way to assess this ability is to prepare another dataset for testing. It is similar to the training set, just a different set of pairs of inputs and true outputs. We can then evaluate the MSE of the testing dataset and compare it with the MSE of training dataset. If the MSE of the testing set is sufficiently small then this indicates the model has good predictive power. Ideally, we need the MSE of the testing set to be as close as possible to the MSE of the training set. Sometimes this does not happen and we end up with MSE of testing set that is significantly higher than that of the training set. This is referred to as overfitting. In order to diagnose this behaviour we usually plot the MSE over the training as well as the testing sets on the same axes,  versus the iteration number. If both curves decrease with increasing the number of iterations until reaching a sufficiently low level, then the model has good fit. If the testing dastaset MSE saturates eventually or worse starts increasing again then there is overfitting. There are many methods proposed in the classical ML literature to overcome overfitting including decreasing the model complexity, increasing the number of training examples, and early stopping (i.e. stop the iterations before the MSE of the testing set starts to increase). On the other hand, the significance of overfitting on the performance of a model depends on the application, and the required level of accuracy. This means that a model might be experiencing some overfitting behaviour, but the prediction accuracy is still sufficient. Finally, it is worth emphasizing that when we do such performance analysis, the MSE evaluated over the testing set is never used for updating the weights during training.  

%%%%%%%%%%%%%%%%%%%%%%%%%%%%%%%%%%%%%%%%%%%%%%%%%%%%%%%%%
\section{Simulation Results}\label{sec:simulations}
In this section, we describe the numerical simulations we performed in order to verify the proposed method. We chose to create six datasets of different pulse configurations to train and test the ML structure. This is described in Section \ref{sec:implementation}. Next, in Section \ref{sec:results} we present the performance analysis results regarding the accuracy of trained models for each of the datasets. In Section \ref{sec:applications}, we show the applicability of using our trained model to do standard tasks such as decoherence suppression and quantum control. Finally, we discuss the significance of these results in \ref{sec:discussion}

\subsection{Implementation}\label{sec:implementation}
We implemented the proposed protocol using the ``Tensorflow" Python package \cite{tensorflow}, and its high-level API package ``Keras" \cite{keras}. The code is publicly available\footnote{https://github.com/akramyoussry/BQNS}. We also implemented a noisy qubit simulator, to generate the datasets for training and testing. It simulates the dynamics of the qubit using Monte Carlo method rather than solving a master equation, to be general enough to simulate any type of noise. The details of the design and implementation of this simulator are presented in Appendix \ref{appx:simulator}. We chose the simulation parameters as shown in Table \ref{tab:simulation}.

\begin{table}[h]
    \centering
    \caption{The different simulation parameters used for generating the datasets.}
    \begin{tabular}{|l|l|l|}
    \hline
    \textbf{Parameter}            & \textbf{Description}                      & \textbf{Value}                                                                    \\ \hline
    $T$                           & Evolution time                            & 1                                                                                 \\ \hline
    $M$                           & Number of discrete time steps             & 4096                                                                              \\ \hline
    $K$                           & Number of noise process realizations      & 1000                                                                              \\ \hline
    $\Omega$                      & Energy gap                                & 10                                                                                \\ \hline
    \end{tabular}
    \label{tab:simulation}
\end{table}

We created three categories of datasets using the simulator, summarized in Table \ref{tab:datasets}, as follows. 
\begin{enumerate}
    \item Qubit with noise on a single-axis and control pulses on an orthogonal axis. 
    
    The Hamiltonian in this case takes the form   
    \begin{align}
        H = \frac{1}{2}\left(\Omega + \beta_z(t)\right)\sigma_z + \frac{1}{2}f_x(t) \sigma_x.
    \end{align}
    We chose the noise to have a the following power spectral density (single-side band representation, i.e. the frequency $f$ is non-negative)
    \begin{align}
        S_Z(f) = \begin{cases}
        \frac{1}{f+1}+0.8e^{-\frac{(f-20)^2}{10}}             & 0 < f \le 50 \\
        0.25+0.8e^{-\frac{(f-20)^2}{10}}                      & f > 50
        \end{cases}
    \label{equ:PSD_Z}
    \end{align}
    Figure \ref{fig:PSD_Z} shows the plot of this power spectral density. The reason for choosing such a form is to ensure that the result noise is general enough but also covers some special cases (such as $1/f$ noise). Also, the total power of the noise is chosen such the effect of noise is evident on the dynamics (i.e. having coherence $<1$). In this category, we generated two datasets. The first one is for CPMG pulse sequences with Gaussian pulses instead of the ideal delta pulses. So, the control function takes the form
    \begin{align}
        f(t) = \sum_{n=1}^{n_{\text{max}}}{ A e^{-\frac{(t-\tau_n)^2}{2\sigma^2}} },
    \end{align}
    where $\sigma = \frac{6T}{M}$, and $A=\frac{\pi}{\sqrt{2\pi\sigma^2}}$, and $\tau_n = \left(\frac{n-0.5}{n_{\text{max}}}\right)T$. The highest order of the sequence was chosen to be $n_{\text{max}}=28$. Now, this means we have a set of 28 examples only in the dataset. In order to introduce more examples in the dataset, we randomize the parameters of the signal as follows. The position of the $n^{\text{th}}$ pulse of a given sequence is randomly shifted by a small amount $\delta_\tau$ chosen at uniform from the interval $[-6\sigma, 6\sigma]$. As a result, we lose the CPMG property that all pulses are equally spaced. However, this can be useful experimentally when there is jitter noise on the pulses. Additionally, we also randomize the power of the pulse. In this case, we vary the amplitude $A$ by scaling it with randomly by amount $\delta_A$ chosen at uniform from the interval $[0,2]$. For this randomization, we scale all the pulses in the same sequence with the same amount. Again we lose the property of CPMG sequences that they are $\pi-$ pulses, but this is needed so that the algorithm can have sufficient generalization power. With these two sources of randomness, we generate 100 instances of the same order resulting in a total of 2800 examples. Finally, we split randomly the dataset following the 75:25 ratio convention into training and testing sets.
    
    The second dataset in this category is very similar with the only difference being the shapes of the pulses. Instead of Gaussian pulses we have square pulses with finite width. The control function takes the form
    \begin{align}
        f(t) = \sum_{n=1}^{n_{\text{max}}}{ A u(t-\tau_n-0.5\sigma) u(\tau_n+0.5\sigma-t) },
    \end{align}
    where $u(\cdot)$ is the Heaviside unit step function, $\sigma=\frac{6T}{M}$, and $A=\frac{\pi}{\sigma}$. The same scheme for randomization and splitting is used in this dataset.

%%%%%%%%%%%%%%%%%%%%%%
    \item Qubit with multi-axis noise, and control pulses on two orthogonal directions.
    
    The Hamiltonian in that category takes the form
    \begin{align}
        H = \frac{1}{2} \left(\Omega + \beta_z(t)\right) \sigma_z + \frac{1}{2} \left( f_x(t) + \beta_x(t) \right) \sigma_x + \frac{1}{2} f_y(t) \sigma_y
    \end{align}
    We chose the noise along $z-$ axis to have the same power spectral density as in Equation \ref{equ:PSD_X}, while the noise along the x-axis has the power spectral density 
    \begin{align}
        S_X(f) = \begin{cases}
        \frac{1}{(f+1)^{1.5}}+0.5e^{-\frac{(f-15)^2}{10}}          & 0 < f \le 20 \\
        (5/48)+0.5e^{-\frac{(f-15)^2}{10}}                       & f > 20
        \end{cases}
    \label{equ:PSD_X}
    \end{align}
    Figure \ref{fig:PSD_X} shows the plot of this power spectral density. This category consists of two datasets. The first one consists of CPMG sequences of maxmimum order of $7$ for the $x-$ and $y-$ directions. We take all possible combinations of orders along each direction. This leaves us with 49 possible configurations. We follow the same randomization scheme discussed before applied to the pulses along the $x-$ and $y-$ directions separately. We generate 100 examples per each configuration and then split into training and testing sets. The second dataset is similar with the only difference that the we do not randomize over the pulse power, we just randomize over the pulse positions.
    
\begin{figure}
    \centering
    \subfloat[Z-axis noise]{\includegraphics[scale=0.75]{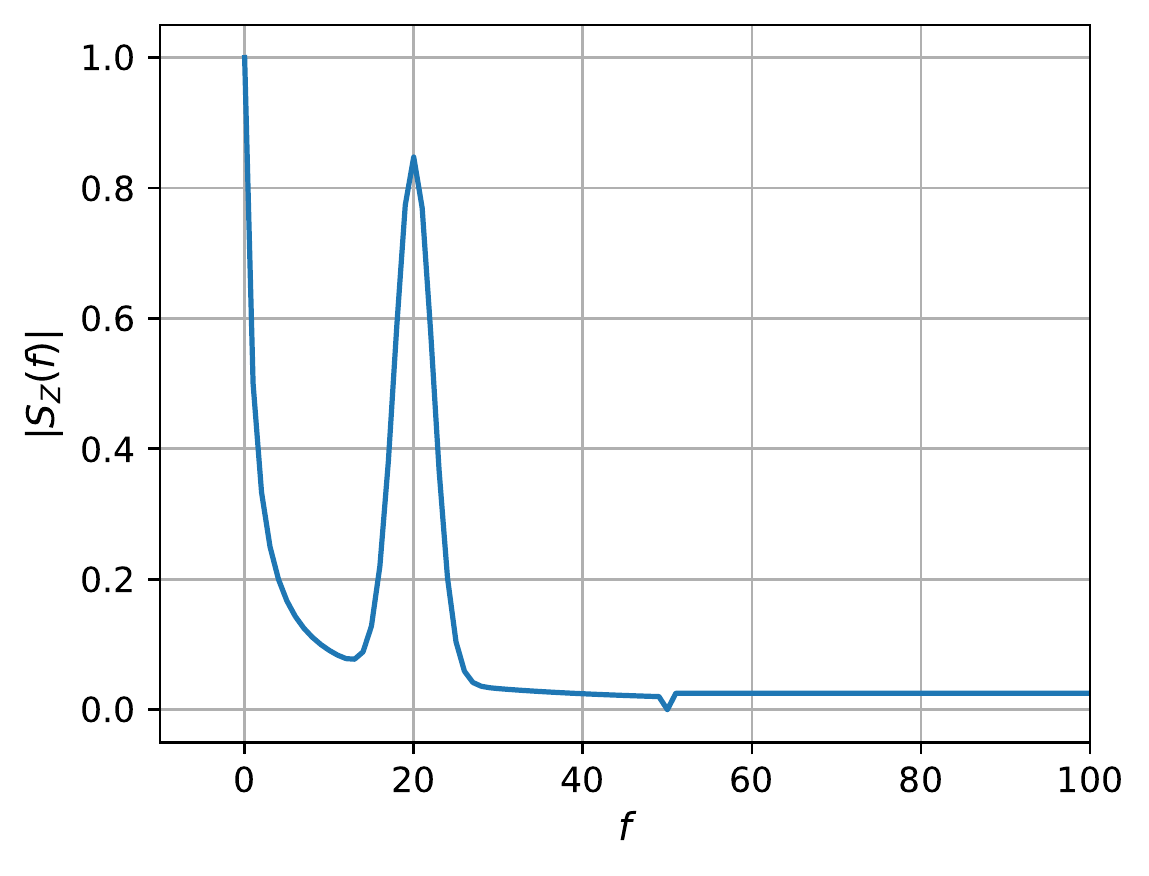}\label{fig:PSD_Z}}
    \subfloat[X-axis noise]{\includegraphics[scale=0.75]{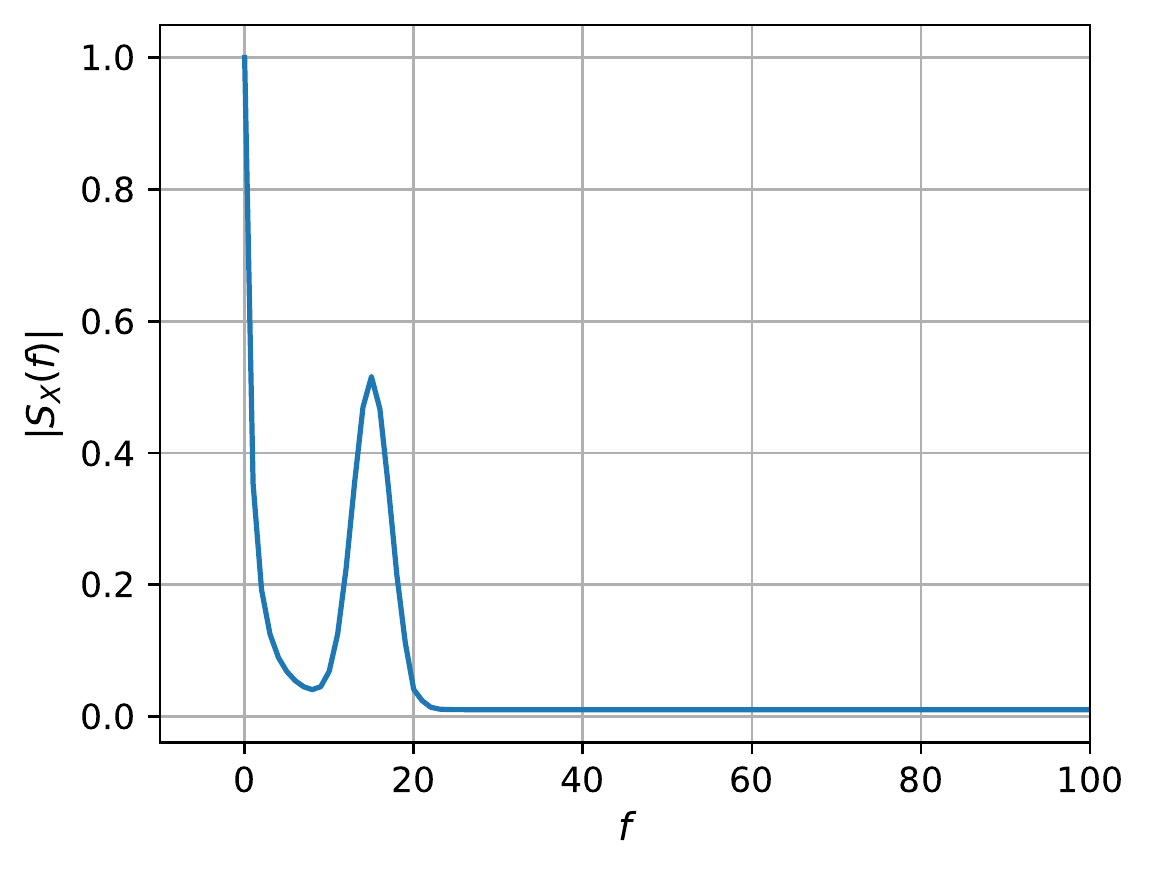}\label{fig:PSD_X}}
    \caption{Powers Spectral Density of the noise that was used to generate the datasets in categories 1 and 2.}
    \label{fig:my_label}
\end{figure}

%%%%%%%%%%%%%%%%%%%%%
    \item Qubit without noise (i.e. a closed quantum system), and pulses on two orthogonal directions.
     
    The Hamiltonian takes the form
    \begin{align}
        H = \frac{1}{2} \Omega \sigma_z + \frac{1}{2} f_x(t) \sigma_x + \frac{1}{2} f_y(t) \sigma_y
    \end{align}
    This category has only datasets as well which follow the same scheme of pulse configuration and randomization as the second category dataset. The only difference is that the absence of noise.
%%%%%%%%%%%%%%%%%%%%%
\end{enumerate}

\begin{table}[h]
    \centering
   \begin{tabular}{|l|l|l|l|l|l|l|}
    \hline
    \textbf{Category}   & \textbf{Name}                 & \textbf{Pulse Shape}   & \textbf{Noise}   & \textbf{control}     & \textbf{\# Training}        & \textbf{\# Testing} \\ \hline
    1                   & CPMG\_G\_X\_28                & Gaussian               & $(z)$            & $(x)$                & 2100                        & 700                 \\ \hline
    1                   & CPMG\_S\_X\_28                & Square                 & $(z)$            & $(x)$                & 2100                        & 700                 \\ \hline
    2                   & CPMG\_G\_XY\_7                & Gaussian               & $(x,z)$          & $(x,y)$              & 3625                        & 1225                \\ \hline
    2                   & CPMG\_G\_XY\_pi\_7            & Gaussian               & $(x,z)$          & $(x,y)$              & 3625                        & 1225                \\ \hline
    3                   & CPMG\_G\_XY\_7\_nl            & Gaussian               & $-$              & $(x,y)$              & 3625                        & 1225                \\ \hline
    3                   & CPMG\_G\_XY\_pi\_7\_nl        & Gaussian               & $-$              & $(x,y)$              & 3625                        & 1225                \\ \hline
    \end{tabular}
    \caption{The three different categories of datasets generated for doing the training and testing of the proposed algorithm. The first category is for qubits with noise along $z-$axis, and control pulses along $x-$ axis. The second category is for qubits with noise along $z-$ and $x-$ axes, and control pulses along $x-$ and $y-$ axes. The final category is for noiseless qubits with pulses along $x-$ and $y-$ axes.}
    \label{tab:datasets}
\end{table}

It is worth emphasizing here that since we are generating the datasets by simulation, we had to arbitrarily chose some noise models and pulse configurations. In a physical experiment however, we do not assume any noise models and just directly measure the different outcomes. Moreover, the pulse configurations should be chosen according to the capability of the available experimental setup. 
%%%%%%%%%%%%%%%%%%%%%%%%%%%%%%%%%%%%%%%%%%%%%%%%%%%%%%
\subsection{Results}\label{sec:results}
\begin{figure}[h]
    \centering
    \subfloat[CPMG\_G\_X\_28]{\includegraphics[scale=0.5]{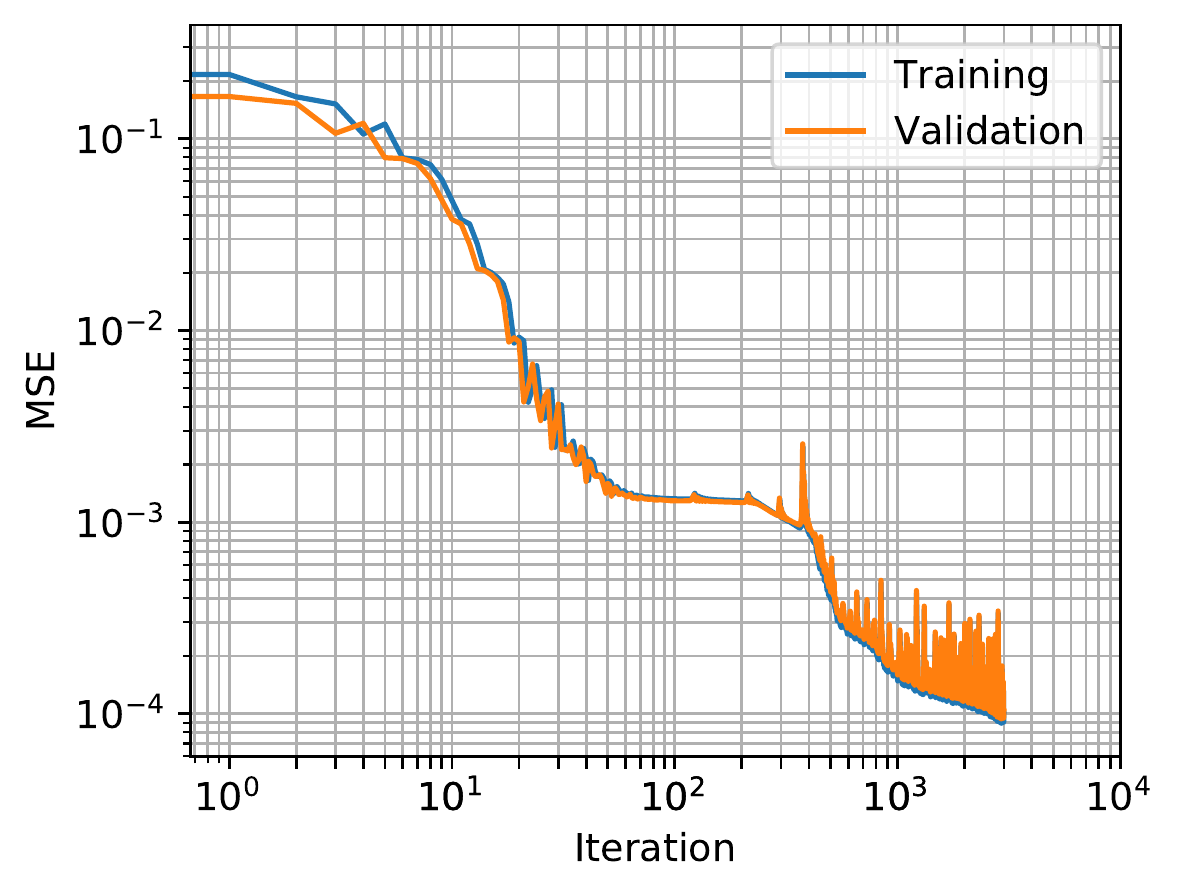}}
    \subfloat[CPMG\_S\_X\_28]{\includegraphics[scale=0.5]{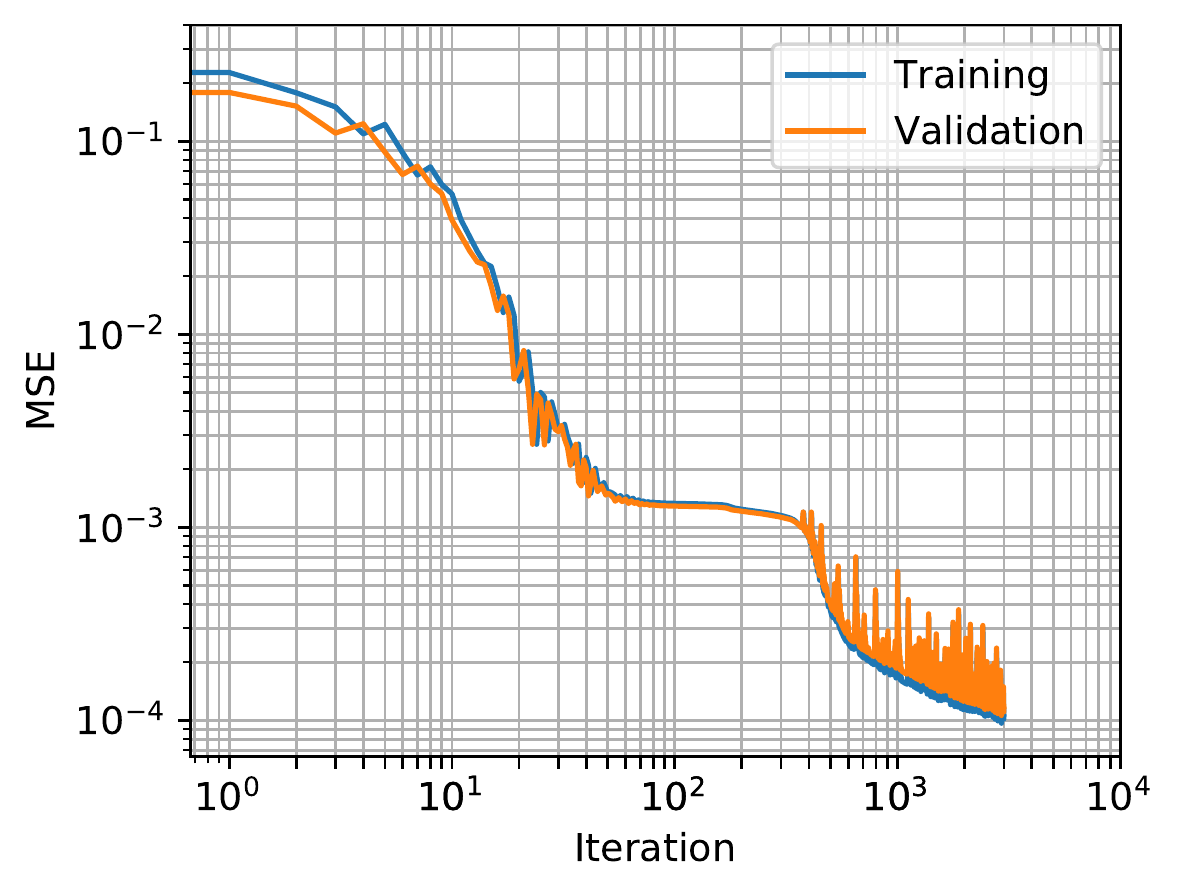}} \\ 
    \subfloat[CPMG\_G\_XY\_7]{\includegraphics[scale=0.5]{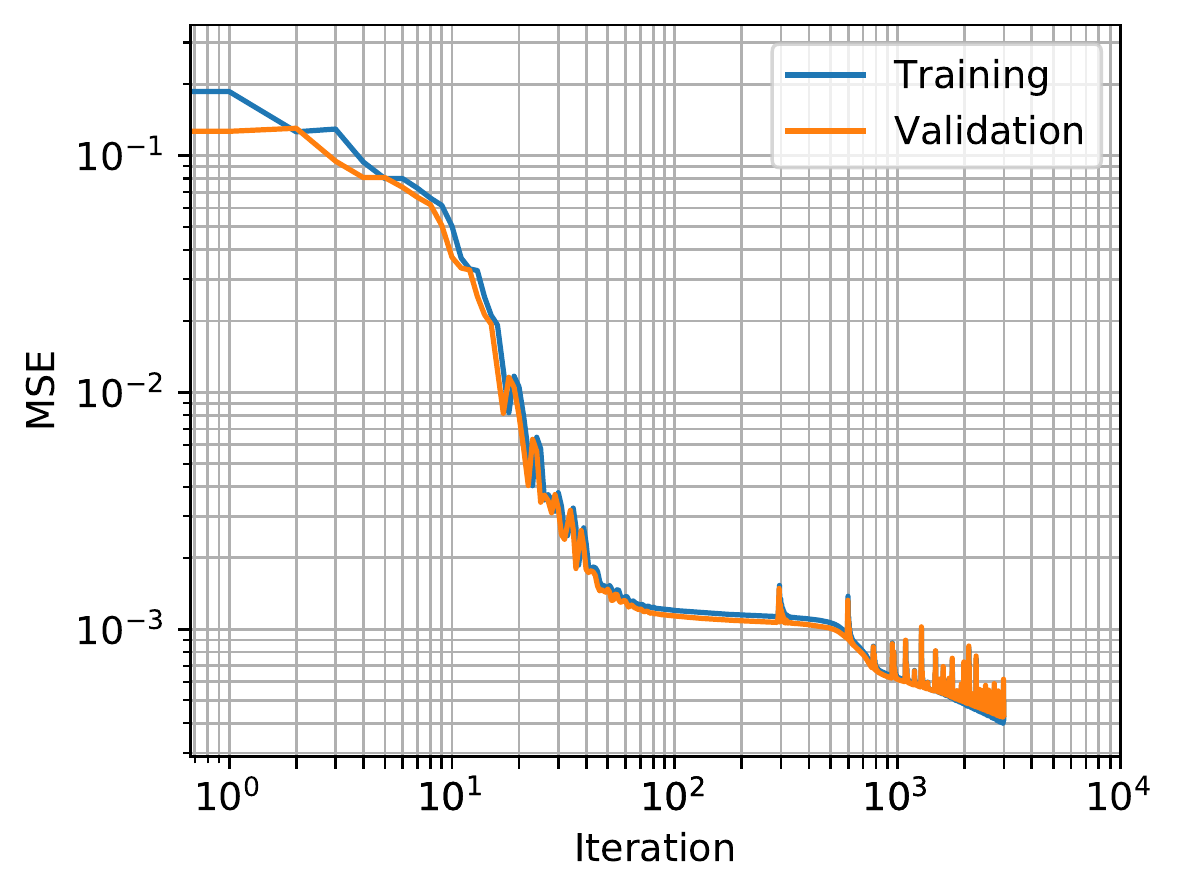}}
    \subfloat[CPMG\_G\_XY\_pi\_7]{\includegraphics[scale=0.5]{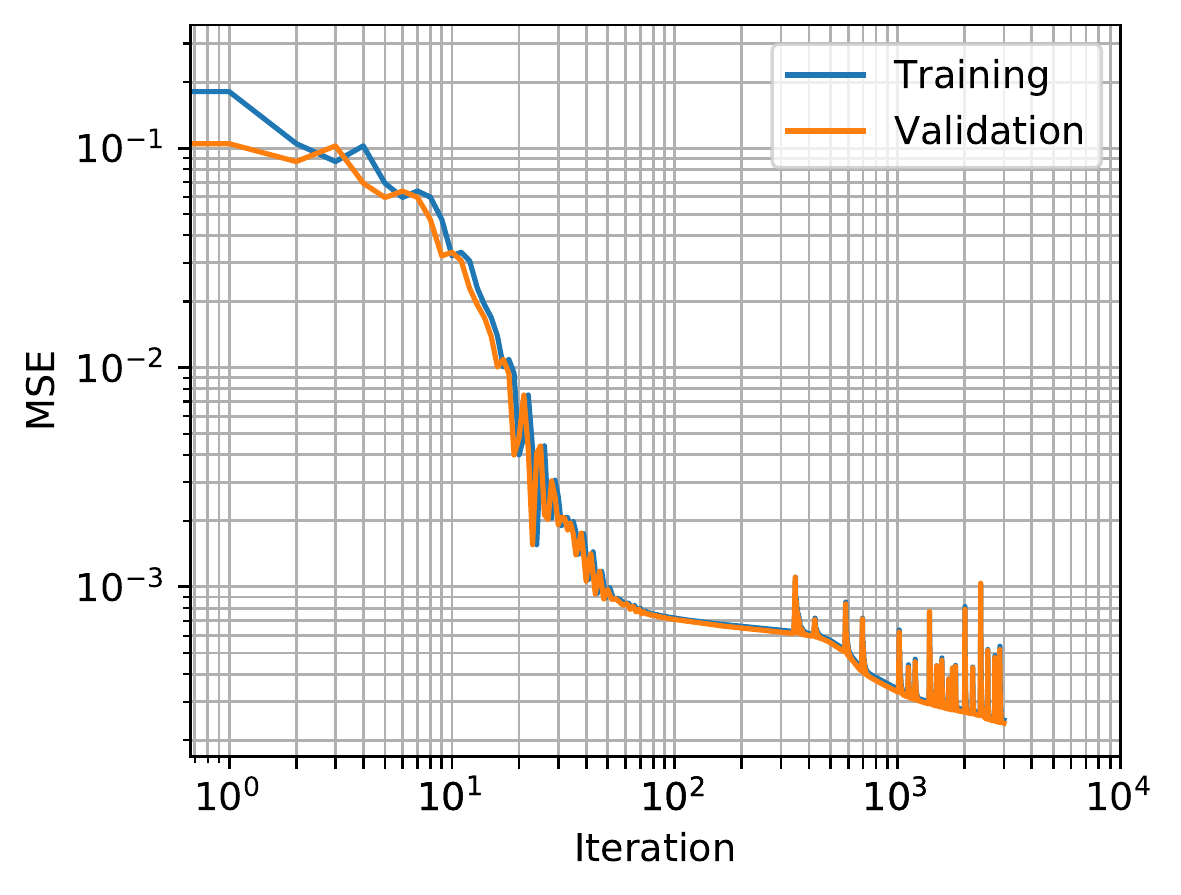}}\\
    \subfloat[CPMG\_G\_XY\_7\_nl]{\includegraphics[scale=0.5]{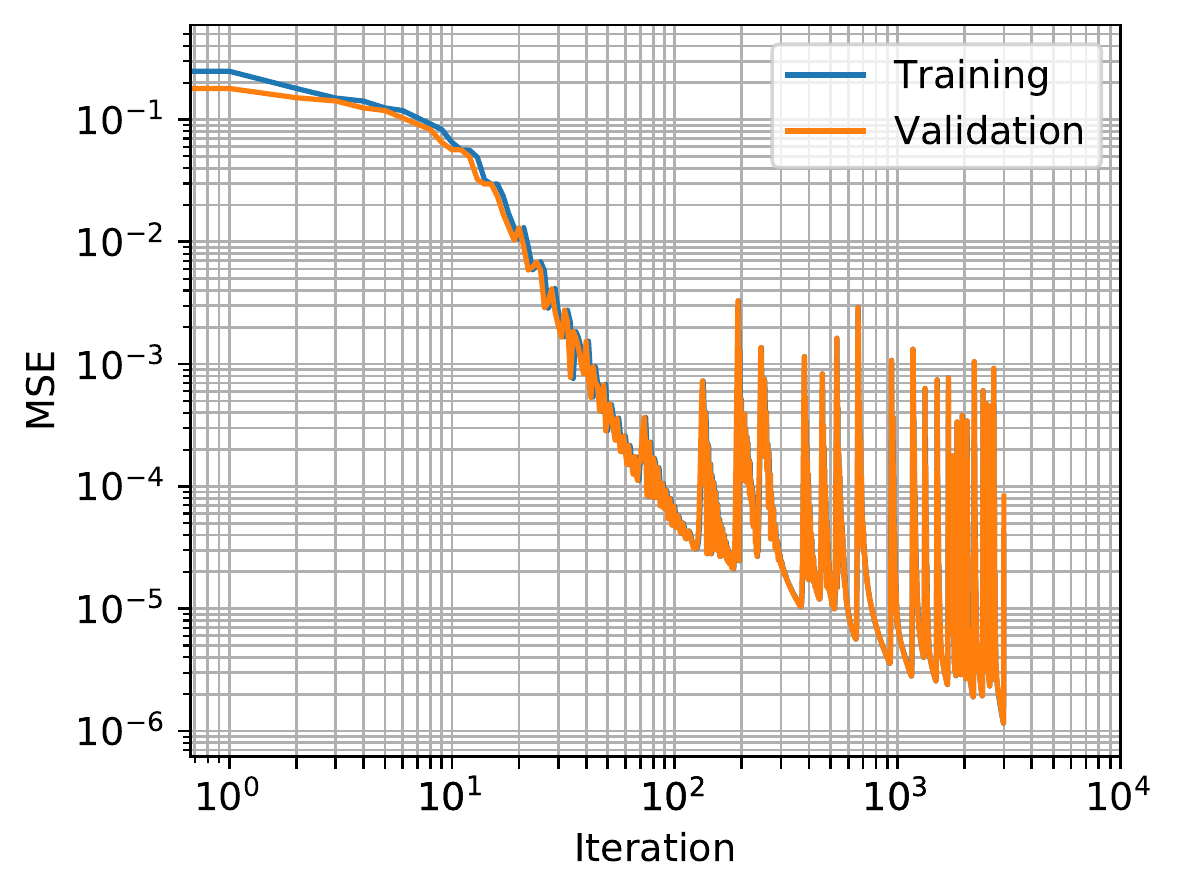}}
    \subfloat[CPMG\_G\_XY\_pi\_7\_nl]{\includegraphics[scale=0.5]{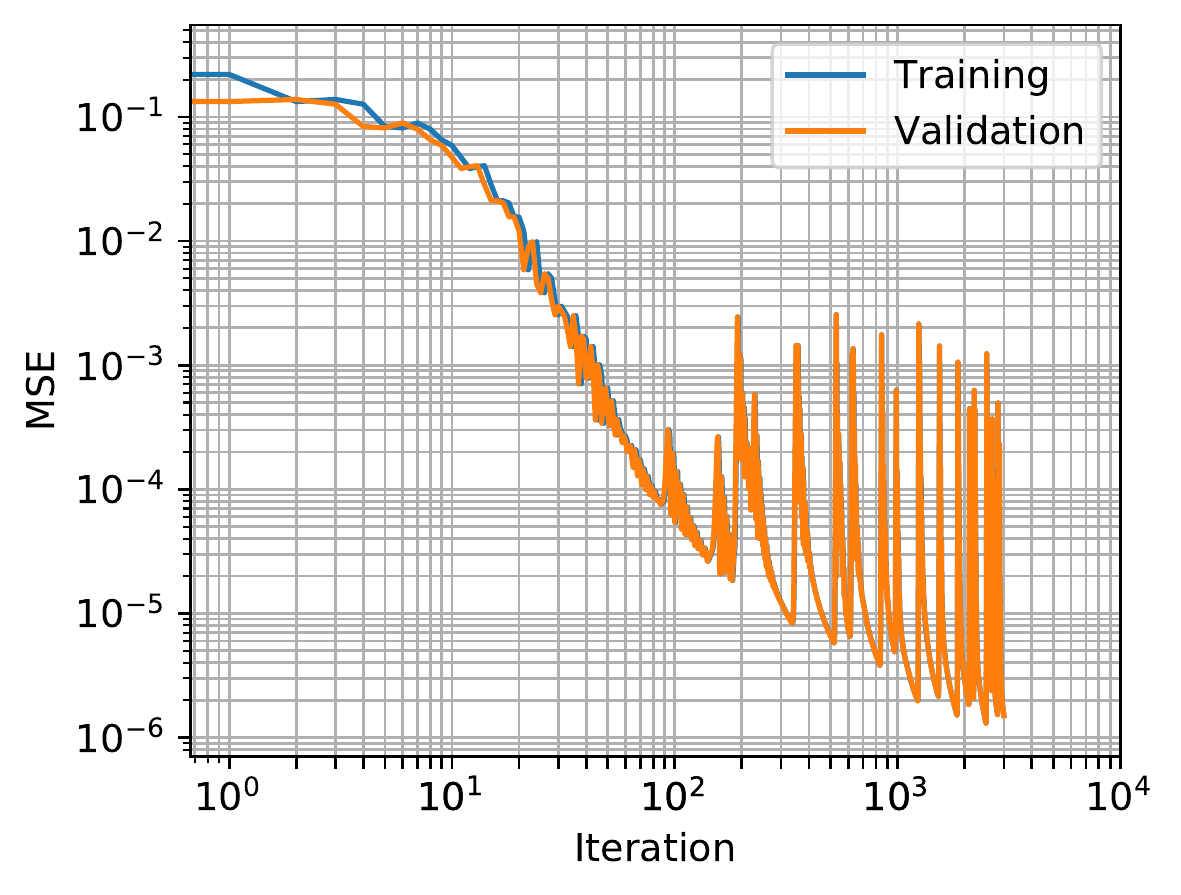}}
    \caption{MSE evaluated for the training and testing examples versus the iteration number for the various datasets.}
    \label{fig:MSE}
\end{figure}

\begin{figure}
    \centering
    \includegraphics[scale=0.75]{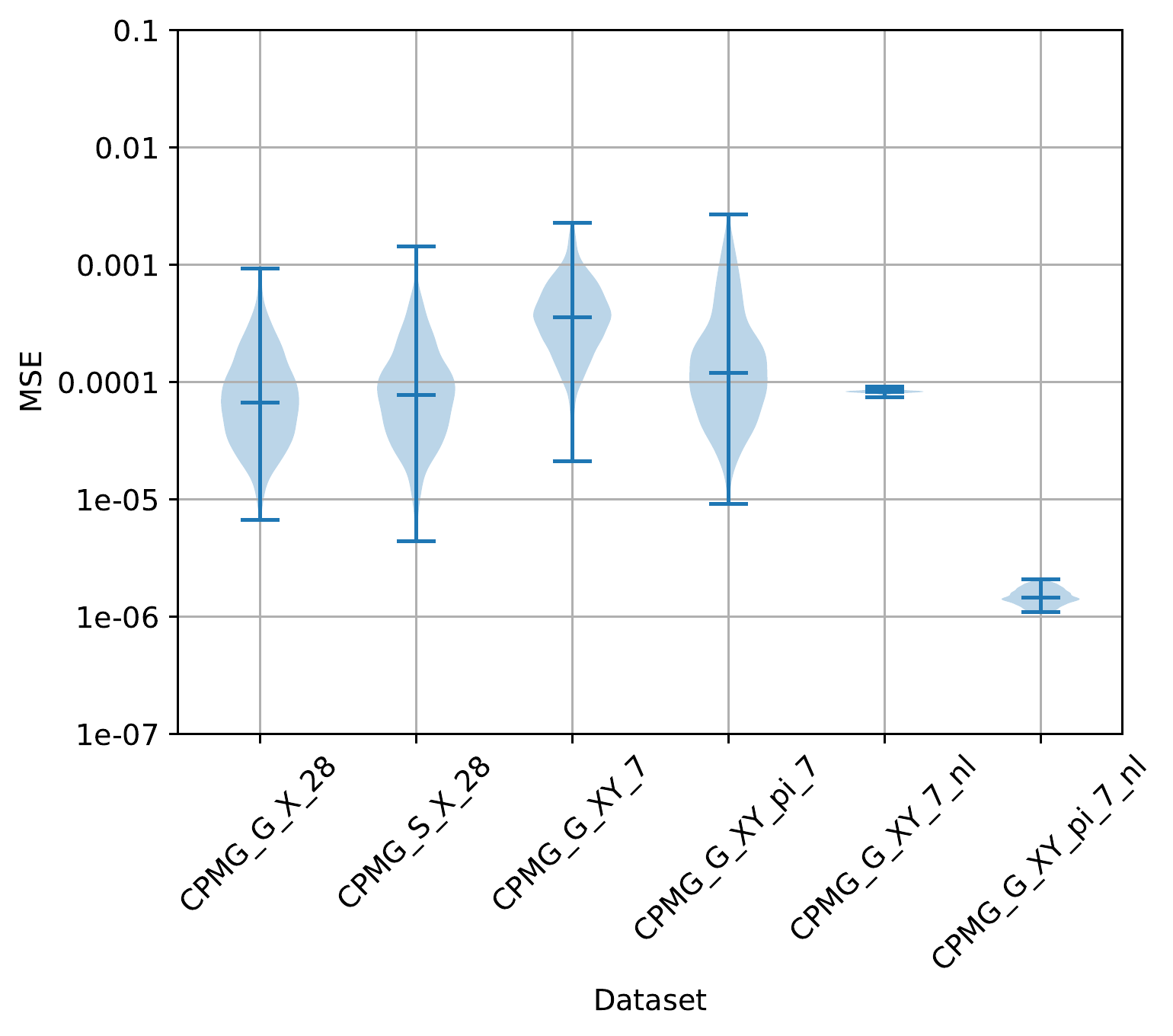}
    \caption{Violin plot of the MSE compared over all testing datasets.}
    \label{fig:violinplot}
\end{figure}

The proposed algorithm was trained on each of the different datasets to assess its performance in different situations. The number of iterations is chosen to be $3000$. Table \ref{tab:MSE} summarizes the MSE evaluated at the end of the training stage for both training and testing examples. Figure \ref{fig:MSE} shows the history of the training procedure for each of the datasets. The plot shows the MSE evaluated after each iteration for both the training and testing examples. For the testing examples, the MSE evaluated is just recorded and does not contribute to the calculation of the gradients for updating the weights. Figure \ref{fig:violinplot} shows a violin plot of the MSE compared across the different datasets; while Appendix Figure \ref{fig:boxplot} shows the boxplot. Appendix Figures \ref{fig:ex_cat1a} to  \ref{fig:ex_cat3b} show the square of the prediction errors for measurement outcome in the best case, average case, and worst case examples of the testing datasets. 

\begin{table}[]
    \centering
    \begin{tabular}{|c|c|c|c|}
    \hline
         \textbf{Dataset}         &  \textbf{\# Iterations} & \textbf{MSE training}             &  \textbf{MSE Testing}            \\ \hline
          CPMG\_G\_X\_28          &  3000                   & $9.86 \times 10^{-5}$                & $1.03 \times 10^{-4}$               \\ \hline
          CPMG\_S\_X\_28          &  3000                   & $1.05 \times 10^{-4}$                & $1.14 \times 10^{-4}$               \\ \hline
          CPMG\_G\_XY\_7          &  3000                   & $4.12 \times 10^{-4}$                & $4.36 \times 10^{-4}$               \\ \hline
          CPMG\_G\_XY\_pi\_7      &  3000                   & $2.47 \times 10^{-4}$                & $2.38 \times 10^{-4}$               \\ \hline
          CPMG\_G\_XY\_7\_nl      &  3000                   & $8.32 \times 10^{-5}$                & $8.31 \times 10^{-5}$               \\ \hline
          CPMG\_G\_XY\_pi\_7\_nl  &  3000                   & $1.50\times 10^{-6}$                & $1.52 \times 10^{-6}$               \\ \hline
    \end{tabular}
    \caption{MSE Evaluated for each of the datsaets at the end of the training process.}
    \label{tab:MSE}
\end{table}

%%%%%%%%%%%%%%%%%%%%%%%%%%%%%%%%%%%%%%%
\subsection{Applications using the trained model}\label{sec:applications}

\subsubsection{Dynamical decoupling and quantum control}
\begin{figure}[h]
    \centering
    \subfloat[$I$]{\includegraphics[scale=0.5]{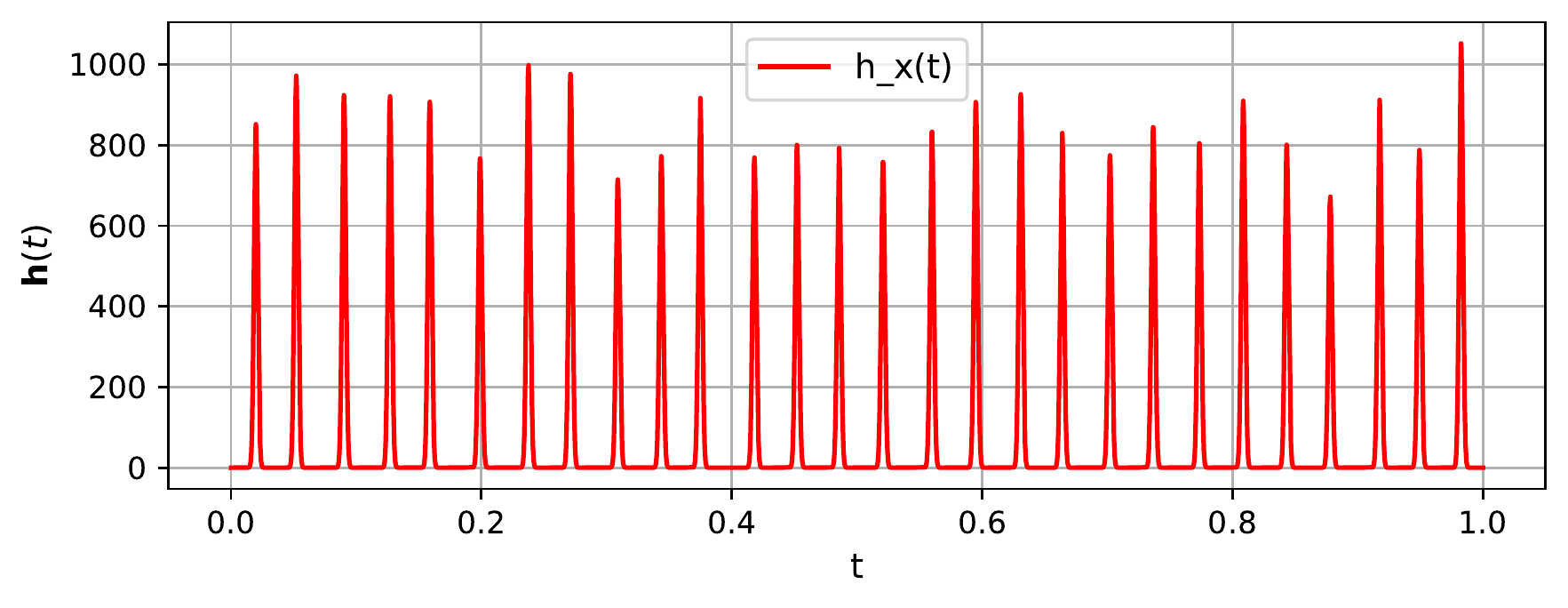}}
    \subfloat[$X$]{\includegraphics[scale=0.5]{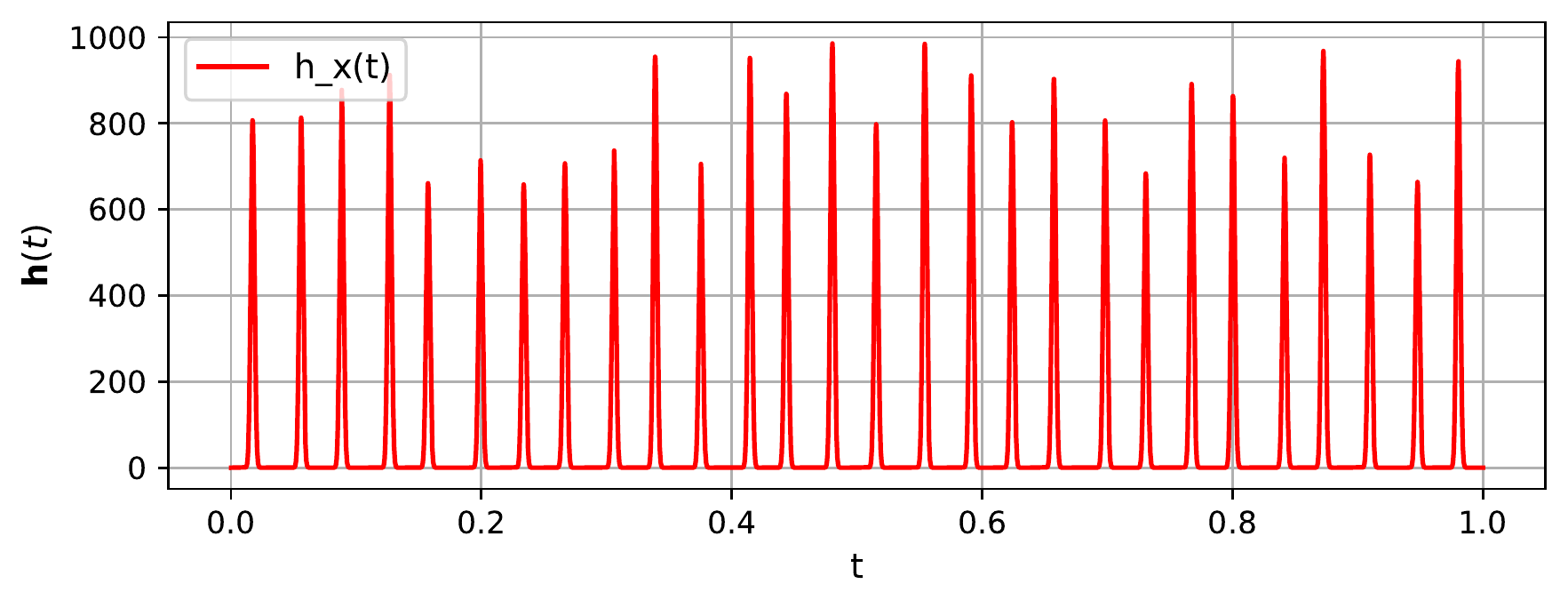}} \\ 
    \subfloat[$Y$]{\includegraphics[scale=0.5]{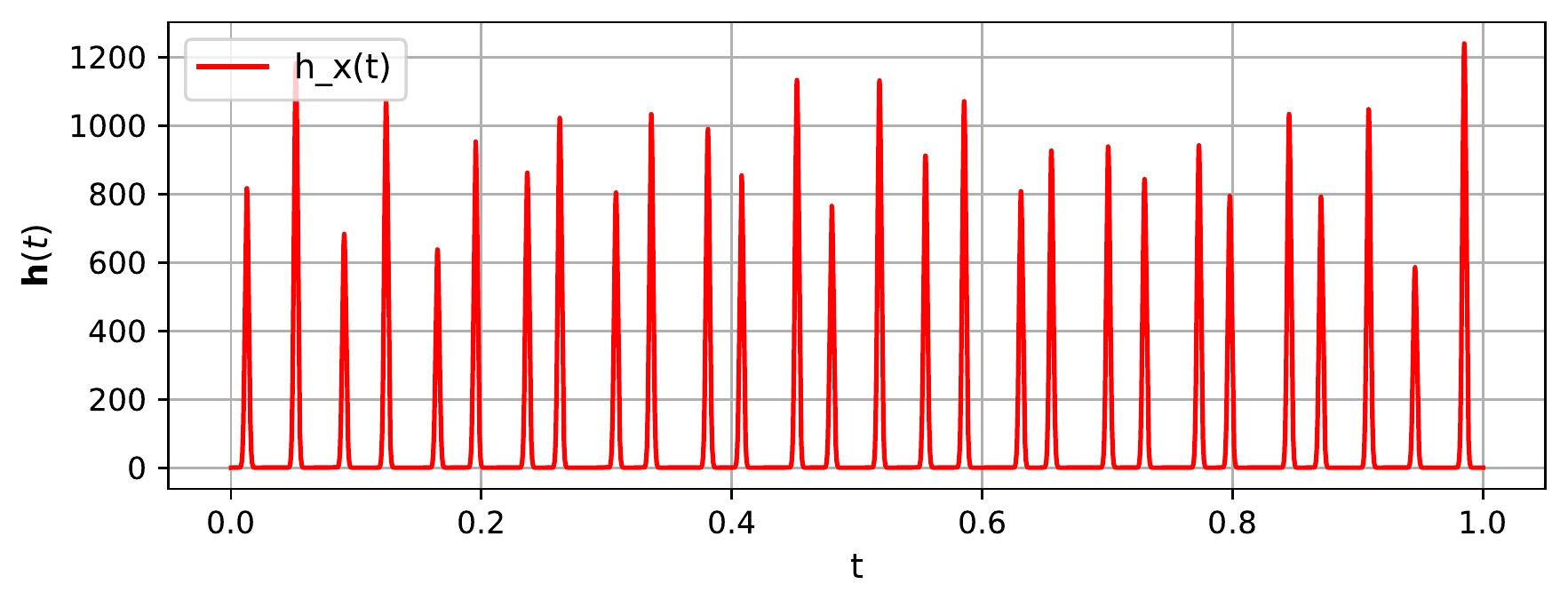}}
    \subfloat[$Z$]{\includegraphics[scale=0.5]{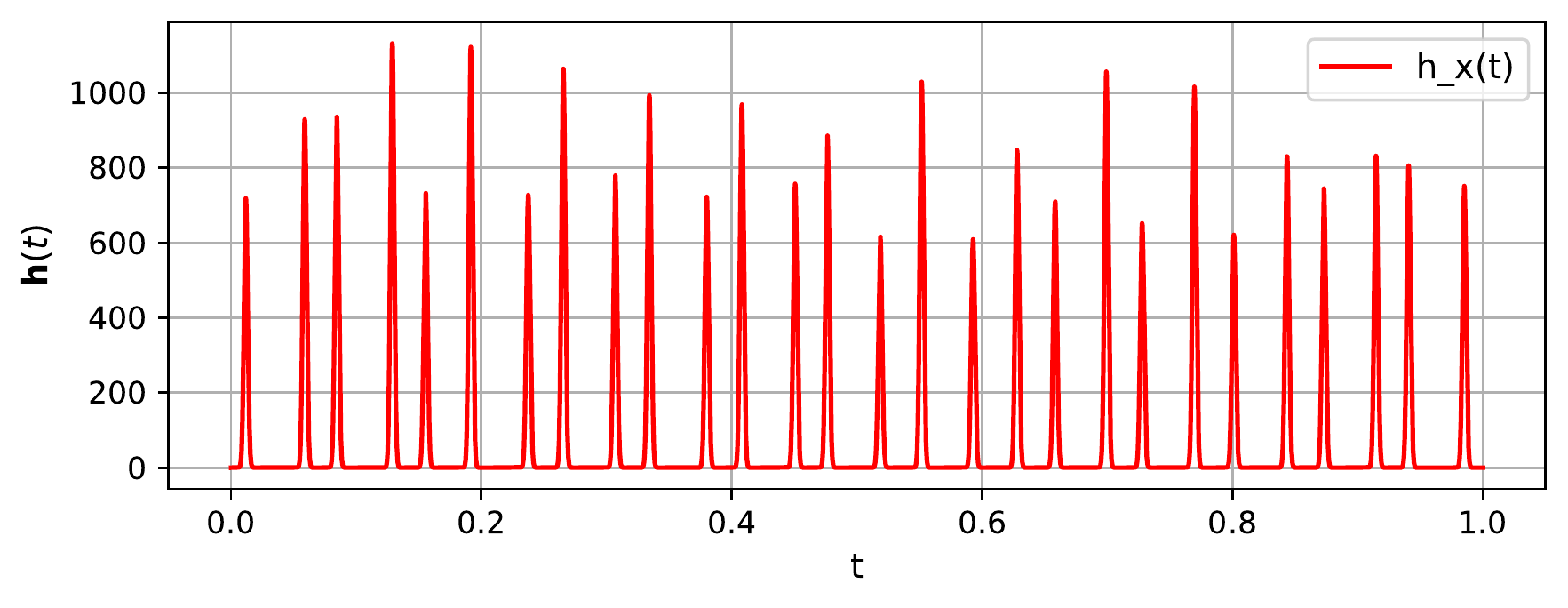}}\\
    \subfloat[$H$]{\includegraphics[scale=0.5]{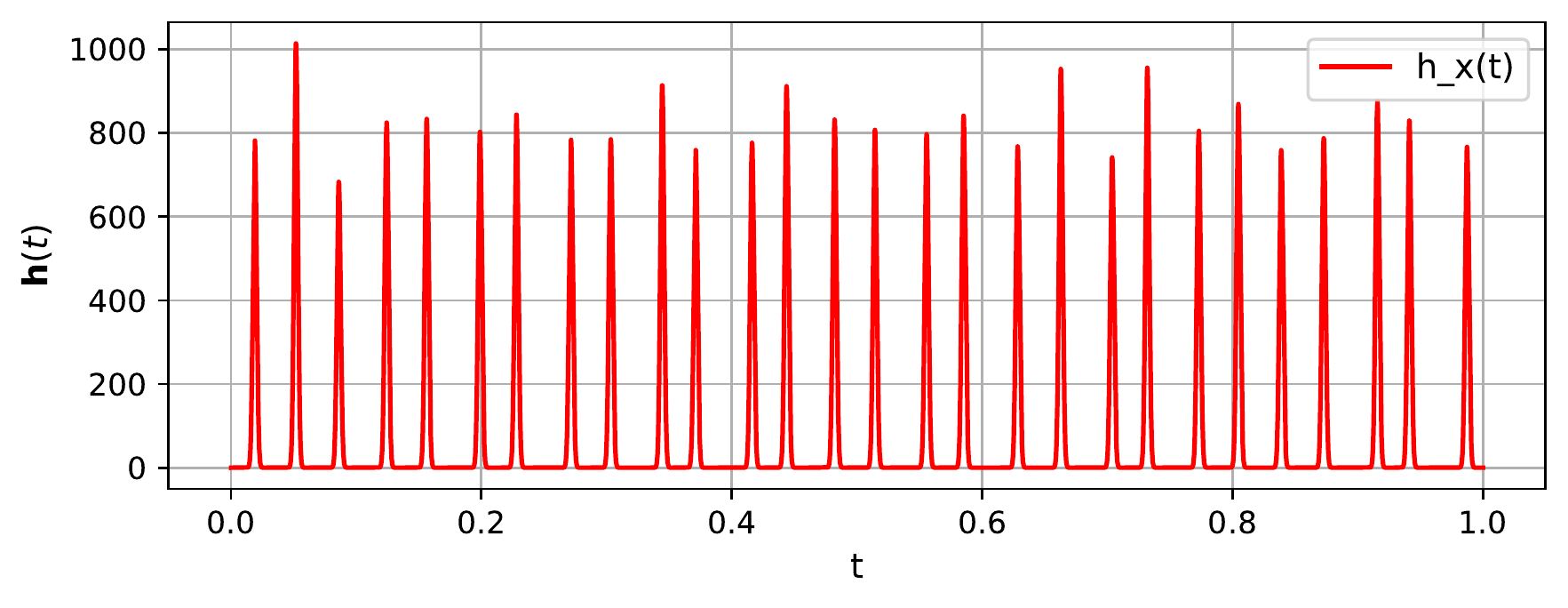}} 
    \subfloat[$R_X\left(\frac{\pi}{4}\right)$]{\includegraphics[scale=0.5]{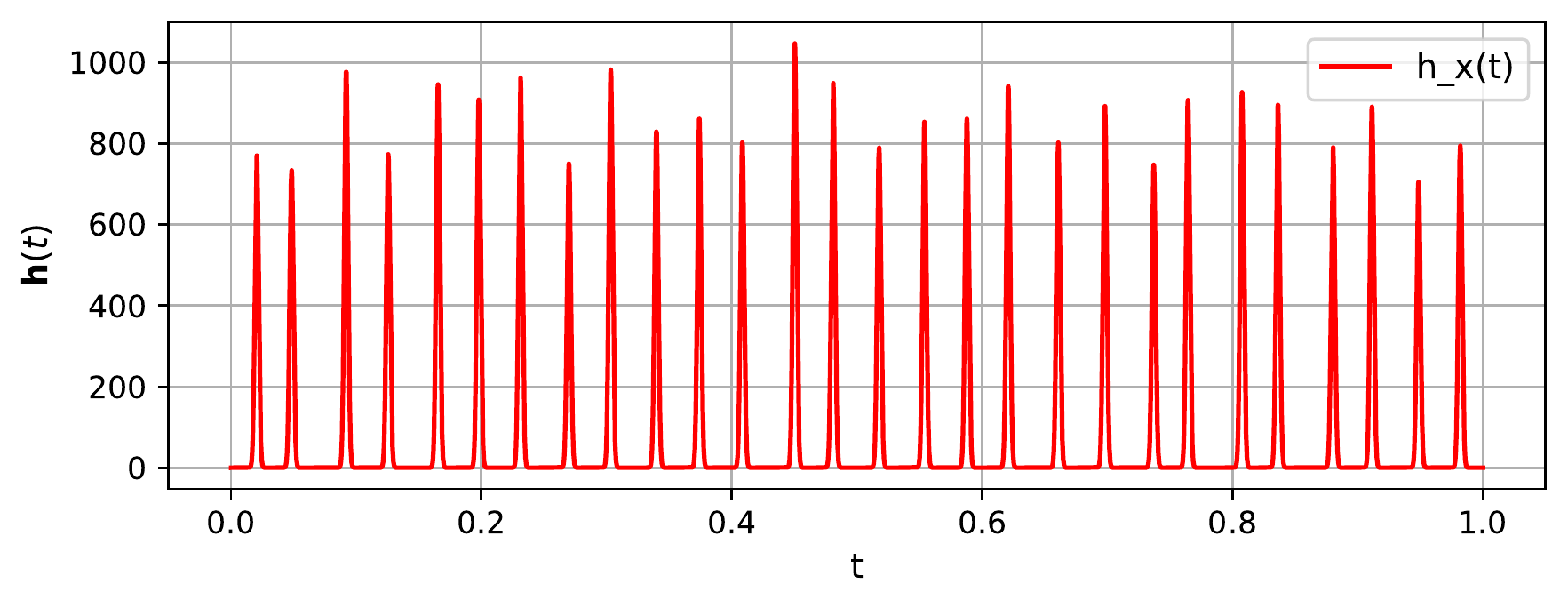}}
    \caption{Control Pulses to implement various quantum gates obtained from optimizing over the trained model of the ``CPMG\_G\_X\_28" dataset. }
    \label{fig:control_pulses_G}
\end{figure}

For the model trained on the single-axis Gaussian dataset ``CPMG\_G\_X\_28", we tested the possibility of using it perform some quantum control tasks. Particularly, we implemented a simple numerical optimization-based controller that aims to find the optimal set of signal parameters to achieve some target quantum gate $G$. We used the fidelity as an objective function, which is defined for two $d\times d$ matrices $U$ and $V$ as
\begin{align}
    F(U,V) = \frac{1}{d^2}|\tr{\left(U^{\dagger}V\right)}|^2.
\end{align}
Ideally, we target four objectives listed as follows:
\begin{align}
    F(V_O,I) &= 1, \quad \forall O \in \{X, Y, Z\} \\
    F(U_c,G) &= 1,
\end{align}
where $V_O$ and $U_c$ are estimated from the trained model. The first three conditions are equivalent to getting rid of the effects of noise, while the last one is equivalent to having achieve evolution described by quantum gate $G$. Practically, It is impossible to completely remove the noise effects, so what we want to do is to find the set of optimal pulse parameters $\{\boldsymbol{\alpha}^*_n\}$ such that
\begin{align}
    \boldsymbol{\alpha}^* = \argmin_{\boldsymbol{\alpha}} (F(V_X[\boldsymbol{\alpha}],I)-1)^2 + (F(V_Y[\boldsymbol{\alpha}],I)-1)^2 + (F(V_Z[\boldsymbol{\alpha}],I)-)^2 + (F(U_c[\boldsymbol{\alpha}],G)-1)^2.
\end{align}
Then using this objective function we can numerically find the optimal pulse sequence. Utilizing this formulation allows us to treat the problem of dynamical decoupling exactly the same, with $G=I$. It is important to mention that this is just one method to do quantum control which might have some drawbacks because of its multi-objective nature. For instance, the optimization could result in one or more of the objectives having sufficient performance, while the others are not. An example of this case is where $U_c$ becomes so close to $G$, while the $V_O$ operators are still far from the identity. This means that the overall evolution will not be equivalent to $G$. There are ways to overcome this problem. For example, we can optimize over the observables instead of the operators or optimize over the overall noisy unitary $U$. However, this is by itself is a separate issue, and we defer it to the future work of this paper. We present these results as a proof of concept that it is possible to use the trained model as a part of a quantum control algorithm. We tested this idea to implement a set of universal quantum gates for a qubit. The resulting fidelities are shown in Table \ref{tab:control}. The control pulses obtained from the optimization procedures are shown in Figures \ref{fig:control_pulses_G}

\begin{table}[]
    \centering
    \begin{tabular}{|c|c|c|c|c|}
    \hline
         $G$         &  $F(V_X,I)$                      & $F(V_Y,I)$                 &  $F(V_Z,I)$          & $F(U_c,G)$          \\ \hline
          I          &  $99.613962$                     & $99.873708$                & $99.873708$          & $99.999960$         \\ \hline
          X          &  $99.604923$                     & $99.868495$                & $99.903021$          & $99.999824$         \\ \hline
          Y          &  $99.604802$                     & $99.870700$                & $99.891827$          & $99.997947$         \\ \hline
          Z          &  $99.594505$                     & $99.857868$                & $99.910219$          & $99.997244$         \\ \hline
          H          &  $99.596911$                     & $99.867958$                & $99.917101$          & $99.999393$         \\ \hline
          $R_X\left(\frac{\pi}{4}\right)$ & $99.596116$ & $99.869907$                & $99.907704$          & $99.999935$         \\ \hline 
          
    \end{tabular}
    \caption{The resulting fidelity between the predicted $V_X$, $V_Y$, $V_Z$, and $U_c$ from the machine learning model trained on the ``CPMG\_G\_X\_28" dataset, and the corresponding targets, after optimizing the control pulses using the trained model.}
    \label{tab:control}
\end{table}

\subsubsection{Quantum noise spectroscopy}
\begin{figure}
    \centering
    \subfloat[Coherences]{\includegraphics[scale=0.5]{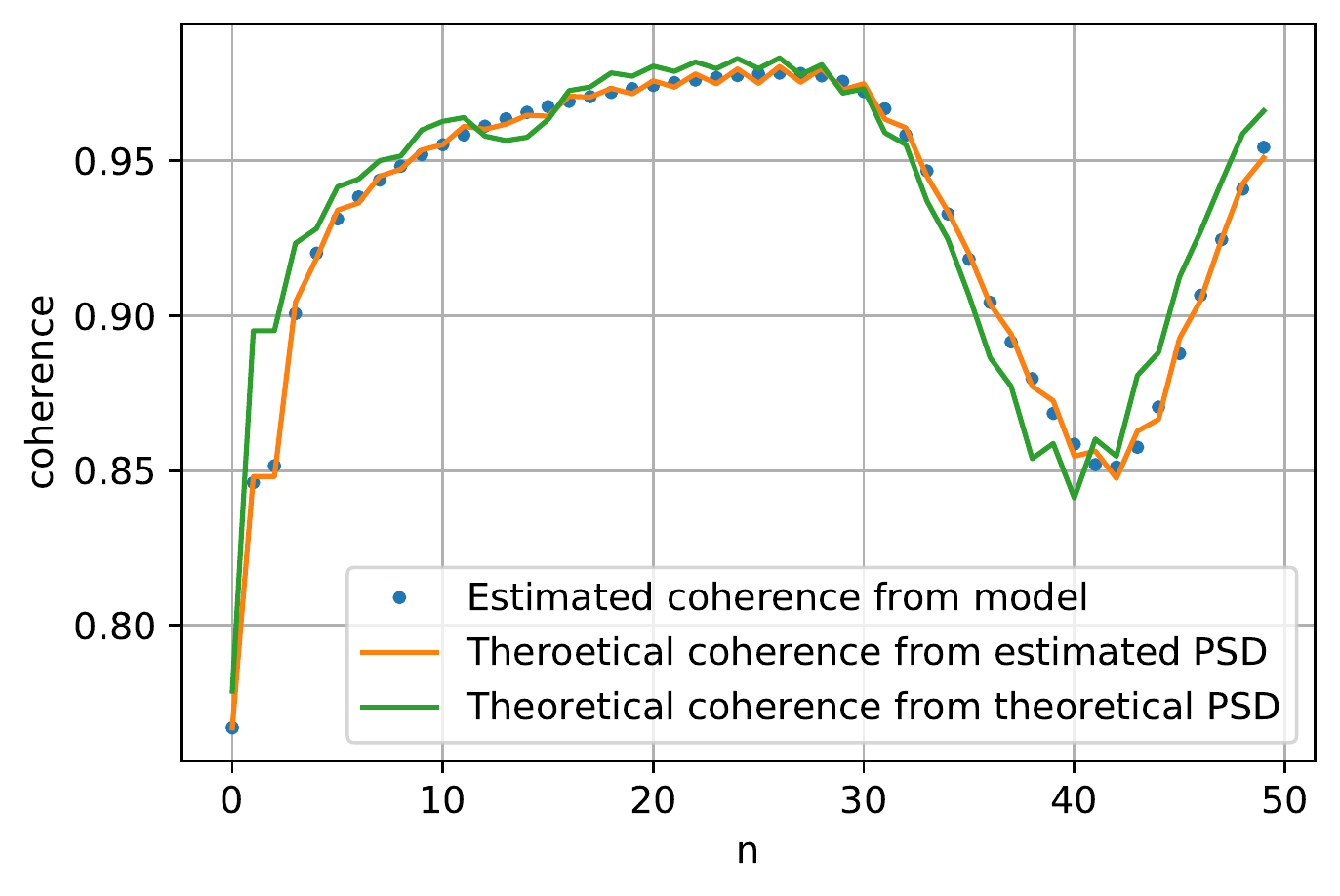}}
    \subfloat[Power Spectrum]{\includegraphics[scale=0.5]{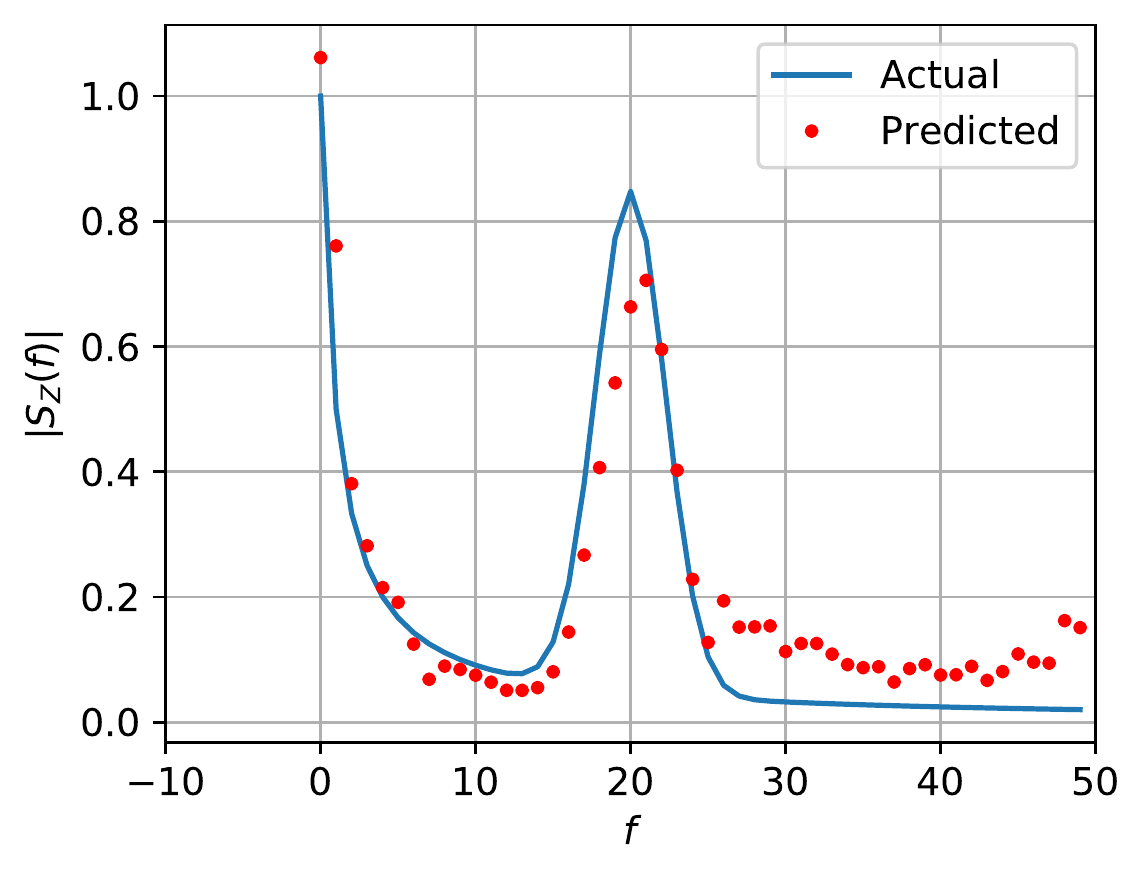}}
    \caption{The estimated and theoretical coherence measurements and noise power spectrum using a trained model on 100 CPMG sequences with finite widths.}
    \label{fig:QNS}
\end{figure}

It is also possible to use the trained model to estimate the power spectral density of the noise using the standard Alvarez-Suter (AS) method \cite{PhysRevLett.107.230501}. In this case, we use the trained model to predict the coherence of the qubit (that is the expectation of the $X$ observable for the $X+$ initial state $\mathbb{E}\{X(T)\}_{\rho=X+}$) for a set of CPMG sequences at the correct locations and powers. Then, from the predicted coherence we can find the power spectrum that theoretically produces these values. In order to do so we have to assume the noise is stationary and Gaussian. Here, we have trained a separate model with CPMG sequences up to order 50. Since the evolution time $T$ is fixed, the higher the order of the sequence is, the higher the accuracy of the estimated spectrum would be specially at high frequencies. On the other hand, because the pulses still have finite width, there is a maximum we could apply during the evolution time and thus we can only probe the spectrum up to some frequency. Figure \ref{fig:QNS} shows the plot of the estimated PSD of the noise versus the theoretical one, as well as the coherences obtained from  predictions of the model as well as the theoretical ones.  We emphasize here that the point of presenting this work is to develop a method that is more general than the standard QNS method. However, we show in this application that we can still utilize the conventional methods combined with our proposed one. Also, in this experiment the focus was on showing the possibility of doing spectrum estimation. We did not use the trained models discussed in the previous section as they are limited to 28 pulses which prevents the probing of the spectrum using the (AS) method to high frequencies.    

%%%%%%%%%%%%%%%%%%%%%%%%%%%%%%%%%%%%%%%%%%%%%%%%%%%%
\subsection{Discussion}\label{sec:discussion}
The plots presented in the previous subsection are useful to assess the performance of the proposed method. First, we can see that for all training sets, the MSE curve versus iterations decreases on average with the number of iterations. This means that the structure is able to learn some abstract representation of the each of the $V_O$ operators as function of the input pulses. For the testing sets, we see that the MSE curves goes down following the training dataset MSE curves. The violin and boxplots show that there exist some minor outliers which we would expect anyway from a machine learning based algorithm. If we look into the worst-case examples, we see that they are actually performing well in terms of accuracy for the different datasets. The overall conclusion from this analysis is that proposed model is able to learn how to predict the measurement outcomes with high accuracy. The noisy multi-axis datasets had slightly less accuracy than the single-axis and the noiseless datasets, which might be worth investigation and is subject to the future work. 

The results of the applications of the trained model are also very promising. The fidelities obtained for the different quantum gates are above 99\% including the identity gate which equivalent to dynamical decoupling. This indicates that we can use numerical quantum control methods combined with our proposed one. We were also able to show the possibility of estimating the spectrum of the noise using the AS method. These results could be enhanced by including longer pulse sequences which requires increasing the overall time of evolution. Thus, the proposed framework is general enough to be used for different tasks in quantum control.   

%%%%%%%%%%%%%%%%%%%%%%%%%%%%%%%%%%%%%%%%%%%%%%%%%%%%%%%%%
\section{Conclusion}\label{sec:conclusions}
In this paper, we presented a machine-learning based method for characterizing and predicting the dynamics of an open quantum system based on measurable information. We followed a graybox approach that allows us to estimate the $V_O$ operators, which are generally difficult to calculate analytically without assumptions on noise and control signals. The numerical results show good performance in terms of prediction accuracy of the measurement outcomes. However, this is not the end of the story. There are lots of points to explore as an extension of this work.

In terms of our control problem, there are two direct extensions. The first relates to that generality of the model used in the whiteboxes. We chose here to study a classical bath because its dynamics was amenable to simulations. However, the learning algorithm itself does not depend on the details of the noise Hamiltonian, and indeed one only needs that Eq.~\eqref{equ:Vo} holds---essentially that what we know about about open quantum systems holds---and that given $\{f_\alpha(t)\}$ one can write $U_{\rm ctrl}(T)$.  
A more important direction our results allow, however, is related to the observation that we are characterizing the open quantum system {\it relative to the given set of control capabilities}. Here we chose a simple set to demonstrate how our algorithm learned about the open quantum system dynamics {\it relative to them} (see also ~\cite{Behnam20} for a formal analysis of this observation) and is capable of predicting the dynamics under control functions $\{f_\alpha(t)\}$ achievable by those capabilities. However, our algorithm is generic, in the sense that it can be applied to a set of control capabilities that is relevant to a specific experimental setup, e.g., when convex sums of Gaussian or Slepian pulses are available or when certain timing constraints must be obeyed, etc. Similar to what we showed here in the sample applications, one can then implement optimal control routines tailored to a specific platform. We are currently working on such targeted result. 

On the ML side, the blackboxes can be further optimized to increase the accuracy specially for the noisy multi-axis datasets. There are lots of architectures that one could exploit. We tried to present the problem in a way that that would facilitate for researchers from the machine learning community to explore and contribute to this field. we emphasize on the importance of understanding the theory so one can know limitations and assumptions of various tools. This is the essence of using the graybox approach, as opposed to using only a blackbox (which is often criticized in the physics community). Moreover, one could make use of existing results in machine learning that deals with incomplete training data~\cite{doi:10.1080/713827181, 10.1007/978-3-642-17103-1_60, garcia2010pattern, raykar2010learning}. These could be leveraged to reduce the number of required experiments which would be useful particularly for higher dimensional systems. 

In summary, we have established a general ML tool which integrates the concept of a graybox with the problem of characterizing (and eventually controlling) an open quantum system relative to a set of given control capabilities. We made every effort to present the result in a way that is palatable for both the physics and the ML community, with the hopes of establishing a bridge between the two communities. We believe this interaction to be necessary in order to achieve efficient and robust protocols that can tackle the extremely relevant problem of high quality control of multiple qubits in NISQ era machines and beyond.

\paragraph*{Acknowledgements}
The authors thank Holger Haas and Daniel Puzzuoli for helpful discussions. Funding for this work was provided by the Australian Government via the AUSMURI grant AUSMURI000002. This research is also supported in part by the iHPC facility at UTS. AY is supported by an Australian Government Research Training Program Scholarship. CF acknowledges Australian Research Council Discovery Early Career Researcher Awards project No. DE170100421. GPS is pleased to acknowledge support from Australian Research Council Discovery Early Career Researcher Awards project No. DE170100088.
%%%%%%%%%%%%%%%%%%%%%%%%%%%%%%%%%%%%%%%%%%%%%%%%%%%%%%%%
\bibliographystyle{apsrev4-1}
\bibliography{library} 

%%%%%%%%%%%%%%%%%%%%%%%%%%%%%%%%%%%%%%%%%%%%%%%%%%%%%%%%
\appendix
\section{Simulator Design}\label{appx:simulator}
The basic idea behind the simulator we implemented is to generate different realizations of the noise process, evaluate the Hamiltonian for each realization, simulate the time-ordered evolution to calculate the observables, and finally average over all realizations. As a result of the central limit theorem, the more noise realizations we average over, the more the sample average converges to the population average. This procedure is repeated for each input state and measurement operator. In this paper, we assume that the noise realizations are the same for all measurements of the same pulse input (i.e. same example in the dataset), but differ from one example to another.

\begin{algorithm}[H]
	\caption{Monte Carlo simulation of a noisy qubit}
	\label{alg:simulation}
	\begin{algorithmic}
	\Function{Evolve}{$H$, $\delta$}
	    \State $U \gets I$
	    \For{$t \gets 0, M-1$}
	        \State $U_t \gets e^{-iH_t\delta}$
	        \State $U \gets U_t U$
        \EndFor
        \State \Return $U$
	\EndFunction
	\Function{GenerateNoise}{$S$, $T$, $M$}
	    \State $N \gets \frac{M}{2}$
        \For{$j \gets 0, N-1$}
            \State $\phi \gets \text{Random}(0,1)$  
	        \State $P_j \gets \frac{M}{\sqrt{T}}\sqrt{S_j}e^{2\pi i\phi}$
	        \State $Q_{N-j} \gets \bar{P}$ 
	   \EndFor
	   \State $P \gets$ \Call{Concatenate}{$P$, $Q$}
	   \State $\beta \gets \text{Re}\{\text{ifft}(P)\}$
	   \State \Return $P$
	\EndFunction
	\Function{simulate}{$\rho$, $O$, $T$, $M$, $f_x$, $f_y$, $f_x$, $S_X$, $S_Y$, $S_Z$ }
	    \State $\delta \gets \frac{T}{M}$
	    \State $E \gets 0$
	    \For{$k \gets 0, K-1$} 
	        \State $\beta_x \gets$ \Call{GenerateNoise}{$S_X$, $T$, $M$}
	        \State $\beta_y \gets$ \Call{GenerateNoise}{$S_Y$, $T$, $M$}
	        \State $\beta_z \gets$ \Call{GenerateNoise}{$S_Z$, $T$, $M$}
	        \For{$j \gets 0, M-1$}
	            \State $t \gets (0.5+j)\delta $ 
	            \State $H_j \gets \frac{1}{2}\left(\Omega + \beta_z(t)\right)\sigma_z + \frac{1}{2}\left(f_x(t) + \beta_x(t)\right) \sigma_x + \frac{1}{2}\left(f_y(t) + \beta_y(t)\right) \sigma_y$
	        \EndFor
	        \State $U \gets$ \Call{Evolve}{$H$, $\delta$}
	        \State $E \gets E + \tr{\left(U \rho U^{\dagger} O\right)}$
	    \EndFor
	    \State $E \gets \frac{E}{K}$
        \State \Return $E$
	\EndFunction
	\end{algorithmic}
\end{algorithm}

There are three basic components in this simulator. The first is a function that calculates the time-ordered evolution of a Hamiltonian to generate a unitary. This is based on approximating the calculation using Equation \ref{equ:Texp}. The second component is a simulator that generates random realizations of the noise given its power spectral density (PSD). The algorithm consists of three steps. First, a random phase is added to each sample of the normalized desired PSD. Second, the complex-valued PSD is concatenated with a flipped version that is also complex conjugated. This step is done to ensure that the signal is symmetric around the middle (i.e the sample at $M/2$). Finally, we take the inverse Fourier transform of the signal and this will be real-valued as a result of the symmetry. We assume here that the desired PSD is single-side band, which means that the total power of the signal is obtained by integrating over positive frequencies only. The third component of the simulator is the main loop that simulates the quantum measurement. Inside the loop, we calculate the observables for each realization, and after that we average over all realizations. A pseudocode of the simulator implementation is shown in Algorithm \ref{alg:simulation}.

\begin{figure}[H]
    \centering
    \subfloat[Control Pulses]{\includegraphics[scale=0.5]{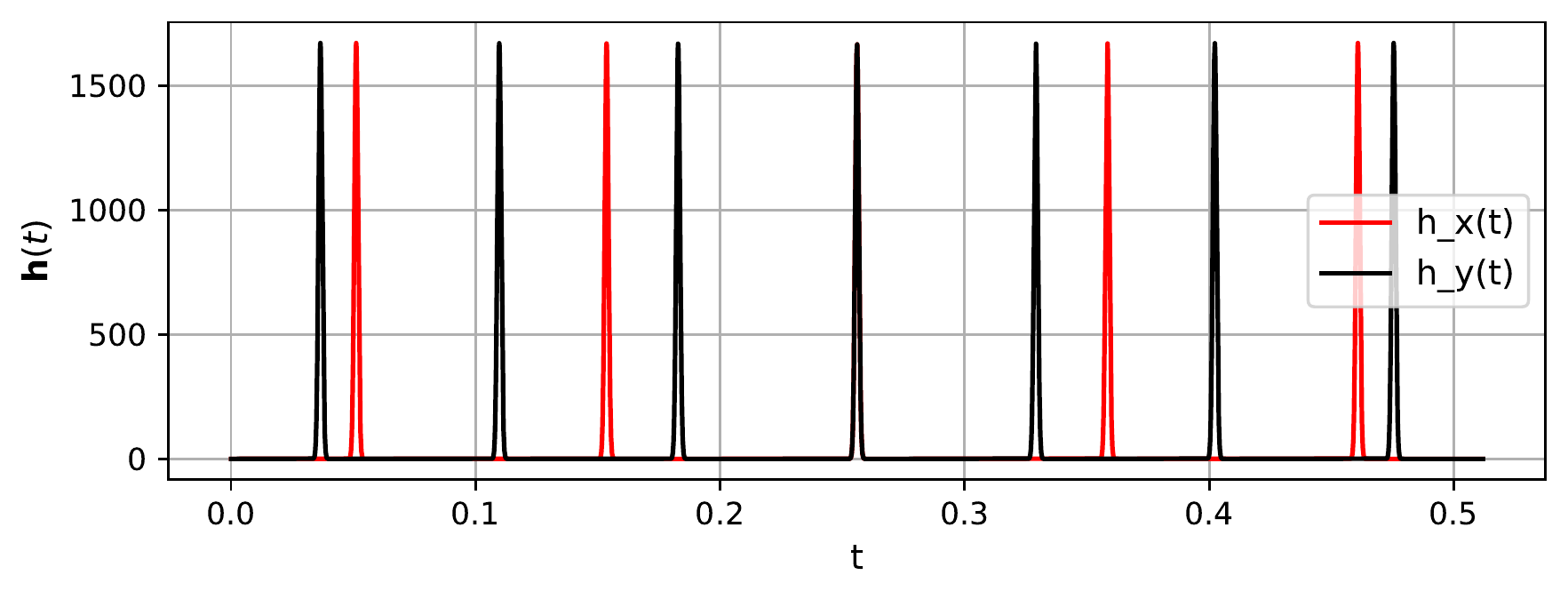}} \\
    \subfloat[$\rho=X+$]{\includegraphics[scale=0.5]{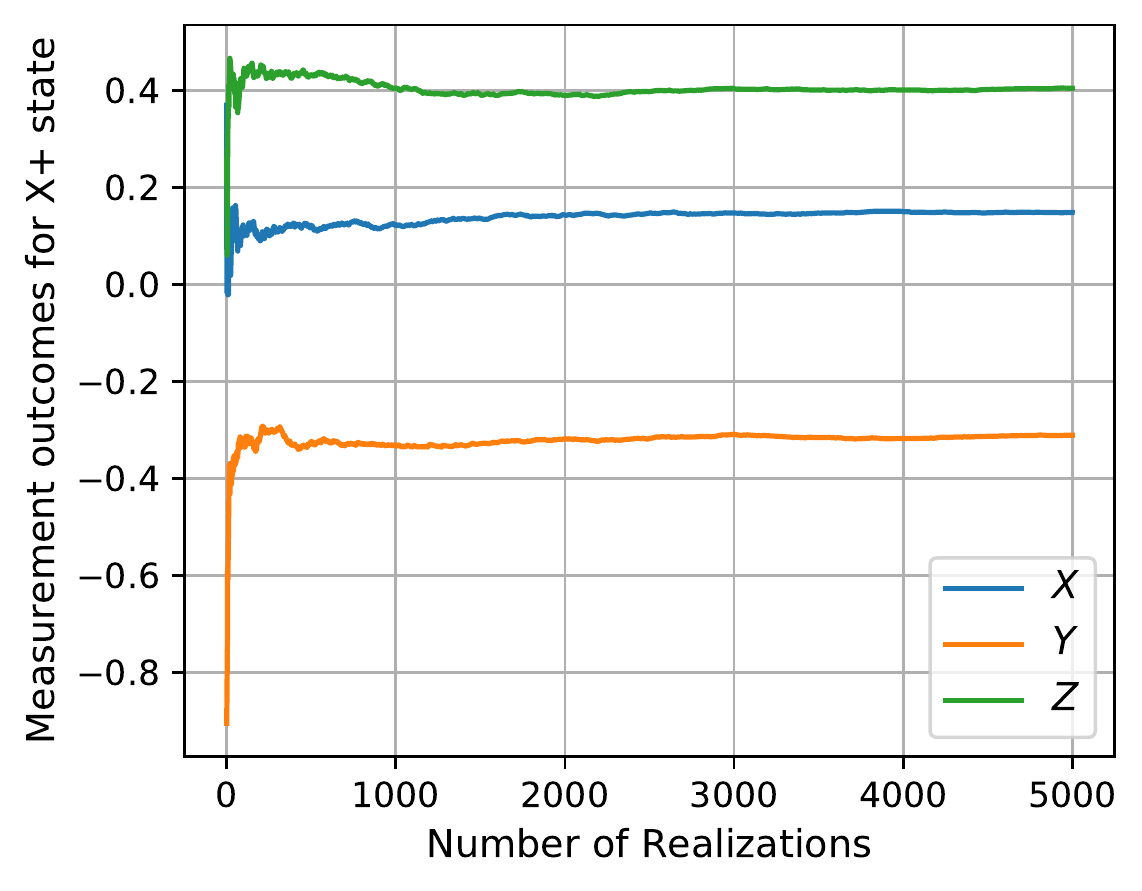}}
    \subfloat[$\rho=Y+$]{\includegraphics[scale=0.5]{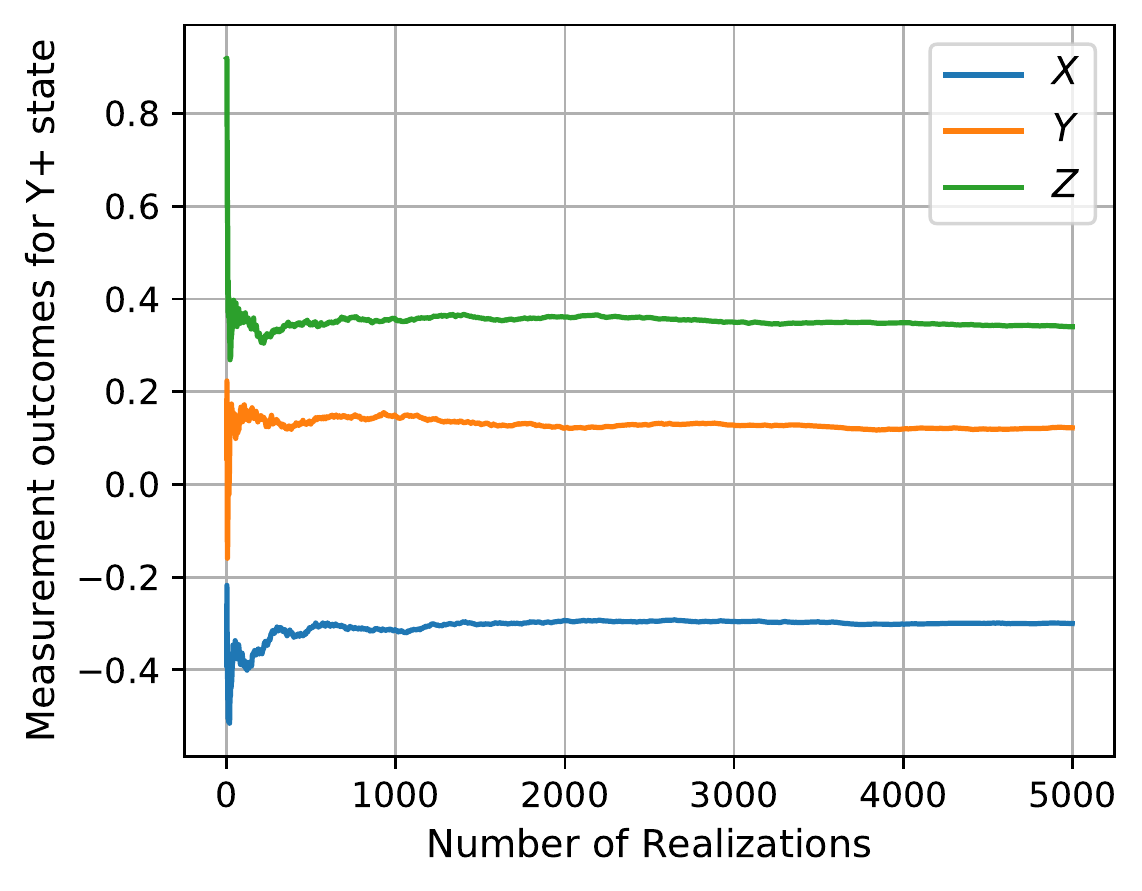}}
    \subfloat[$\rho=Z+$]{\includegraphics[scale=0.5]{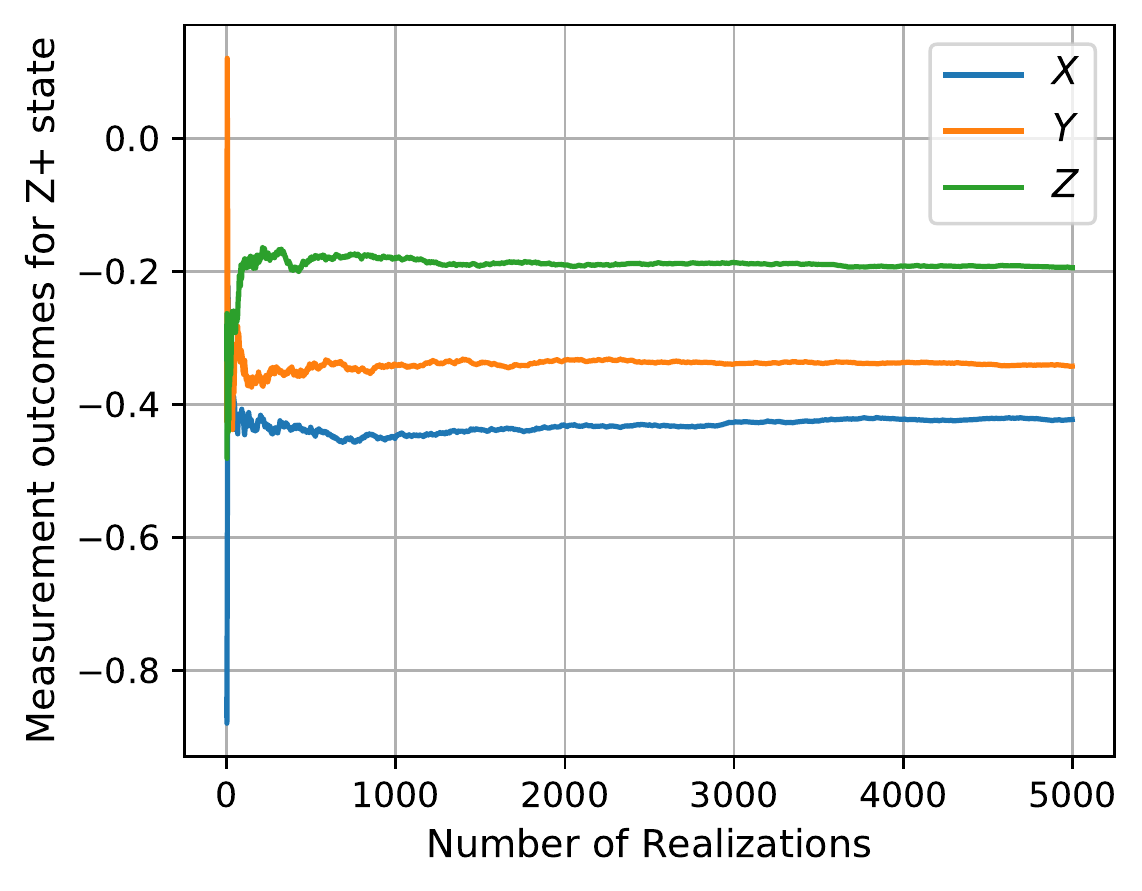}}
    \caption{The effect of the number of realization on the performance of the Monte Carlo simulation of the measurement outcomes. }
    \label{fig:montecarlo}
\end{figure}

In this paper, we selected the number of noise realizations based on doing the Monte Carlo simulation of a random pulse sequence, and then observing how much the expectation values change by increasing the number of realizations. As shown in Figure \ref{fig:montecarlo}, the values start to stabilize around 500 realizations, so we chose $K=1000$ for generating all the datasets.

%%%%%%%%%%%%%%%%%%%%%%%%%%%%%%%%%%%%%%%%%%%%%%%%%%%%%%%%%
\section{Derivation of Equation \ref{equ:Vo} }\label{appx:Vo}
In this Appendix we give some details on deriving Equation \ref{equ:Vo}, based on a modified interaction picture. We will start with the usual interaction picture, then see why it will not be suitable for our purpose, and finally present the modified interaction picture that will yield the desired form. 

As it is known the Schr\"odinger picture is where the states are time-independent while the operator are time-dependent, which is the more famous picture. The Heisenberg picture is the opposite, so the states are constants while the operators are time-dependent. Finally, there is the interaction (Dirac) picture, where both states and operators are time-dependent. The three pictures are equivalent to each other in the sense that they all yield the same expectation values of quantum observables, which is what can be measured physically. We can also move from one picture to another. The use of particular picture depends on the application, and one picture can make calculations easier more than another picture. 

If we start in the ``usual"  Schr\"odinger picture with a qubit in state $\rho(0)$, and the system has the Hamiltonian 
\begin{align}
     H(t) &= \frac{1}{2}\left(\Omega + \beta_z(t) + f_z(t) \right)\sigma_z + \frac{1}{2}\left( \beta_x(t) + f_x(t) \right) \sigma_x + \frac{1}{2}\left(\beta_y(t) +  f_y(t)\right) \sigma_y
\end{align}
then we can express the evolved state $\rho(t)$ at time $t$ as
\begin{align}
    \rho(t) = U(t) \rho(0) U(t)^{\dagger},
\end{align}
where the total unitary can be expressed as the time-ordered evolution of the total Hamiltonian $H(t)$
\begin{align}
    U(t) = \mathcal{T}_+ e^{-i\int_0^t{H(s) ds}},
\end{align}
and the classical expectation over the noise of a quantum observable $O$ for the evolved state will be in the form
\begin{align}
    \mathbb{E}\{O(T)\}_\rho= \left\langle \tr{\left(U(t) \rho(0)  U^{\dagger}(t) O\right)} \right \rangle_c.
\end{align}
The problem with this form is that $U(t)$ depends on both the noise and the control pulses. We are interested to find a form where we can separate both so that we can design then protocols for dynamical decoupling, quantum control and quantum noise spectroscopy. So, we are going to move to the interaction picture to help us separate the noise and control. We start by separating the Hamiltonian into two parts, the first is $H_1(t)$ which has all the noise terms
\begin{align}
    H_1(t) = \frac{1}{2} \beta_z(t) \sigma_z + \frac{1}{2} \beta_x(t) \sigma_x + \frac{1}{2} \beta_y(t) \sigma_y,
\end{align}
and the other part $H_0(t)$ collects all the remaining terms which includes the free evolution term and the control terms
\begin{align}
    H_0(t) = \frac{1}{2}\left(\Omega + f_z(t) \right)\sigma_z + \frac{1}{2} + f_x(t) \sigma_x + \frac{1}{2} +  f_y(t) \sigma_y.
\end{align}
So the total Hamiltonian now is  
\begin{align}
    H(t) &= H_{0}(t) + H_1(t).
\end{align}
Now, we will apply the standard procedures of moving from the Schr\"odinger to the interaction picture. First we define the unitary 
\begin{align}
    U_0(t) = \mathcal{T}_+ e^{-i\int_0^t{H_0(s) ds}}.
\end{align}
Then we use it to transform the states and operators as follows. First, the state becomes
\begin{align}
    \rho_I(t) &= U_0^{\dagger}(t)\rho(t) U_0(t) \\
              &= U_0^{\dagger}(t) U(t) \rho(0) U^{\dagger}(t) U_0(t).
    \label{equ:rho_I}
\end{align}
The operator $H_0(t)$ does not change between the two pictures (because $U_0(t)$ commutes with $H_0(t)$, conjugating $H_0(t)$ by $U_0(t)$ has no effect). Next, we transform $H_1(t)$ to become $H_I(t)$ as
\begin{align}
    H_I(t) = U_0^{\dagger}(t) H_1(t) U_0(t),       
\end{align}
and consequently the interaction unitary is
\begin{align}
    U_I(t) &= \mathcal{T}_+ e^{-i\int_0^t{H_I(s) ds}}. 
\end{align}
On the other hand, we know that the time-evolution of the state in the interaction picture is given by
\begin{align}
    \frac{d}{dt} \rho_I(t) = [H_I(t), \rho_I(t)],
\end{align}
which is equivalent to 
\begin{align}
    \rho_I(t) = U_I(t) \rho_I(0) U_I^{\dagger}(t).
\end{align}
By comparing this form with that in Equation \ref{equ:rho_I} (and noticing that $\rho_I(0)= \rho(0)$) we find that
\begin{align}
    U(t) = U_0(t) U_I(t) 
\end{align}
which means we separated the noise and control parts. Thus, the expectation becomes
\begin{align}
   \mathbb{E}\{O(T)\}_\rho  &= \left\langle\tr{\left(U(t) \rho(0)  U^{\dagger}(t) O\right)}\right\rangle_c \\
                            &= \left\langle\tr{\left(U_0(t) U_I(t) \rho(0) U_I{^\dagger}(t)U_0^{\dagger}(t)   O\right)}\right\rangle_c
\end{align}
Now, the problem with that form is that the initial state gets conjugated by the noise unitary first. This will result in the dependence of the classical expectation on the initial quantum state, and it will not facilitate the expressing the optimal control problem. This is why the usual conventional interaction picture does not solve the problem completely. The form we need is the one where the state is conjugated with the control unitary first. So, we are going to modify the interaction picture as follows.  
\begin{align}
    U(t) &= U_0(t) U_I(t) \\
         &= U_0(t) U_I(t) U_0{^\dagger}(t) U_0(t) \\
         &= U_0(t) \mathcal{T}_+ e^{-i\int_0^t{ H_I(s)  ds}} U_0{^\dagger}(t) U_0(t) \\
         &= U_0(t) \left( 1 -i \int_0^t H_I(t_1)  dt_1 +\frac{(-i)^2}{2!} \int_0^t \int_0^t  H_I(t_1)  H_I(t_2) dt_1 dt_2 + \cdots \right) U_0{^\dagger}(t) U_0(t) \\
         &= \Bigg(\, 1 -i \int_0^t U_0(t)  H_I(t_1) U_0^{\dagger}(t) dt_1 \nonumber \\
         &+\frac{(-i)^2}{2!} \int_0^t \int_0^t U_0(t)  H_I(t_1) \left(U_0^{\dagger}(t) U_0(t) \right) H_I(t_2) U_0^{\dagger}(t) dt_1 dt_2 + \cdots \,\Bigg)U_0(t) \\
         &= \left(\mathcal{T}_+ e^{-i\int_0^t{U_0(t) H_I(s) U_0^{\dagger}(t) ds}} \right)U_0(t) \\
         &= \left(\mathcal{T}_+ e^{-i\int_0^{t}{H_I(s-t) ds}} \right)U_0(t)
\end{align}
Notice, the second line is just multiplying the identity from left. The fifth line and sixth we multiplied multiplied the $U_0(t)$ from left to all terms in the infinite series and $U_I^{\dagger}$ from right. We also resolve the identity between each $H_I$ term. This means effectively the $H_I$ terms gets conjugated by $U_0(t)$ inside the time-ordered exponential as in the seventh line. In the second last line, the effect of this conjugation is the time-evolving of $H_I(s)$ backwards in time to $H_I(s-t)$. Remember in the interaction picture, states evolve according to $U_I$, while operators evolve according to $U_0^{\dagger}$. Also, for a fact, the evolution operator depends only on the time-interval of evolution and not the end points, so $U_0(t) = U_0(s+t, s)$. Finally, in the last line we performed a change of variables for the integration. So, now we can finally write
\begin{align}
    U(t) &= \left(\mathcal{T}_- e^{-i\int_0^{t}{H_I(s) ds}} \right)U_0(t)\\
         &\equiv \tilde{U}_I U_0, \\
\end{align}
Where $\mathcal{T}_-$ is the reverse time-ordering operator. Now, we can express the classical expectation of the observable as
\begin{align}
   \mathbb{E}\{O(T)\}_\rho  &= \left\langle \tr{\left(U(t) \rho(0)  U^{\dagger}(t) O\right)} \right\rangle_c \\
                            &= \left\langle \tr{\left( \tilde{U}_I(t) U_0(t) \rho(0) U_0^{\dagger}(t) \tilde{U}_I{^\dagger}(t)  O\right)} \right\rangle_c\\
                            &= \left\langle \tr{\left(\tilde{U}_I{^\dagger}(t) O\tilde{U}_I(t) U_0(t) \rho(0) U_0^{\dagger}(t)   \right)} \right\rangle_c \\
                            &= \left\langle \tr{\left(\tilde{U}_I{^\dagger}(t) O\tilde{U}_I(t) U_0(t) \rho(0) U_0^{\dagger}(t) O O^{-1}  \right)} \right\rangle_c \\
                            &= \tr{\left( \braket{ O^{-1} \tilde{U}_I{^\dagger}(t) O\tilde{U}_I(t)}_c U_0(t) \rho(0) U_0^{\dagger}(t) O  \right)}\\
                            &\equiv \tr{\left( V_O U_0(t) \rho(0) U_0^{\dagger}(t) O  \right)}.
\end{align}
In the third line we applied the cyclic property of the trace twice. In the fourth line we multiplied by identity from left under the assumption. The second last line we applied the cyclic property again, and moved the classical expectation inside and it acts only on the first part that depends on the noise (i.e. the $\tilde{U}_I$). Now, this is exactly the form we want, because we can recover the closed system dynamics ($H_1 =0$, and thus $U_I = I$, and so $V_O=I$). The initial state is now conjugated with $U_0$ which is just the control part of the Hamiltonian which we have access to. Thus, we can formulate different quantum control problems utilizing this form, and the $V_O$ operator becomes encodes everything about the noise and its interaction with the control, independent on the initial quantum state of the qubit.
%%%%%%%%%%%%%%%%%%%%%%%%%%%%%%%%%%%%%%%%%%%%%%%%%%%%%%%%%
\section{Overview on Neural Networks (NN) and Gated-Recurrent Units (GRU)} \label{appx:ml}

In this Appendix, we give a brief overview on some of the commonly used classical machine learning blackboxes. The first blackbox is the neural network, which is a non-linear modular structure composed of basic computational units called neurons. A neuron calculates the weighted-average of its inputs and then applies a non-linear transformation, generating a single output. If we denote the set of inputs as $\mathbf{x} = [x_1, x_2 , \cdots x_n]^T$, then the output would be
\begin{align}
    y &= f\left(w_0 + \sum_{i=1}^{n}{w_i x_i}\right) \\
    &= f\left(\mathbf{W}\mathbf{x} + w_0 \right)
\end{align}
where $w_i$ are called the weights of a neuron, $w_0$ is called the bias and $f(\cdot)$ is a non-linear function called the activation function. The most common activation functions used are the linear activation $f(x)=x$, the sigmoid activation $f(x) = \frac{1}{1+e^{-x}}$, and hyperbolic tangent $f(x) = \tanh(x)$. The nice property about those three functions is that their gradients are easy to evaluate ($1$, $f(x)(1-f(x))$, and $1-f^2(x)$ respectively). Nonetheless other functions can be used. The weights and the bias are chosen through the training process to generate some desired output. For instance if the neuron output is denoted by $y$, and the desired output is $y_d$, then by constructing the loss function
\begin{align}
    L = (y-y_d)^2,
\end{align}
one can use steepest descent method to find the optimal weights as
\begin{align}
    w_i^{(t+1)} = w_i^{(t)} - \eta \frac{\partial L}{\partial w_i^{(t)}},
\end{align}
where $\eta$ is the learning rate. So, one starts with some random weights $w_i^{(0)}$ and applies these iterations until convergence. Now, this single neuron generates one output only. If we want to multiple outputs then we could have a layer of neurons who act on the same inputs. In many applications, this structure might not be complex enough to model our data. So, we can connect multiple layers where the output of one layer is fed as input to the next layer. This structure is what is commonly known as an Artificial Neural Network (ANN). The last layer is called the output layer, and the number of neurons there should match the number of desired outputs. The other layers are called hidden layers and they can have arbitrary number of neurons. One can also derive the update rule in such case which is commonly referred to as the backpropagation rule. There are also lots of variants that enhance the basic update rule. ANNs turn out to be very useful in lots of applications such as classification and regression. 

Another type of machine learning structures is the Recursive Neural Network (RNN). This is a structure that allows processing of sequences. Besides its input $\mathbf{x}_t$ and output $\mathbf{y}_t$, at time instant $t$, it has an internal hidden state denoted by $h_t$. At each time instant, the RNN processes the inputs to update the hidden state from the previous time instant, as well as generates the new output. So, it works like a feedback system. Generally, the new hidden state $\mathbf{h}_{t+1}$ and the output $\mathbf{y}_t$ can both depend on $\mathbf{x}_t$ and $\mathbf{h}_t$. The relations between different variables would depend on some weights which are adjusted during training to produce some desired output. Based on this idea, there are lots of such ``update rules" resulting in various kinds of structures. In this paper, we make use of the Gated Recurrent Unit (GRU) \cite{cho2014learning}. It consists of the following structure. First, there is a ``reset gate", which is essentially a neural network the operates on the concatenation of current input at time instant $t$ and the previous hidden state $\mathbf{h}_t$, to produce the output $\mathbf{r}_t$ defined using the sigmoid activation function. In other words,
\begin{align}
    \mathbf{r}_t = \sigma\left(\mathbf{W}_r\mathbf{x}_t + \mathbf{U}_r \mathbf{h}_{t-1} + \mathbf{b}_r \right),
\end{align}
where $\mathbf{W}_r$, $\mathbf{U}_r$, and $\mathbf{b}_r$ are the weights and the bias of the neural network, and $\sigma(\cdot)$ is the sigmoid function. The second component of a GRU is the update gate, which is also a similar neural network,
\begin{align}
    \mathbf{z}_t = \sigma\left(\mathbf{W}_z\mathbf{x}_t + \mathbf{U}_z \mathbf{h}_{t-1} + \mathbf{b}_z \right).
\end{align}
After calculating the outputs from the reset and update gates we can now calculate $\tilde{\mathbf{h}}_t$ which represents the new information we need to add to our hidden state,
\begin{align}
    \tilde{\mathbf{h}}_t = \tanh{\left(\mathbf{W}_h\mathbf{x}_t + \mathbf{U}_h \left(\mathbf{r_t} \odot \mathbf{h}_{t-1}\right) + \mathbf{b}_h \right)},
\end{align}
where $\odot$ is the Hadamard product (i.e. element-wise multiplication $(A \odot B)_{ij} = (A)_{ij} (B)_{ij}$). The final step is to calculate the new hidden state which would be a weighted average between the existing state $\mathbf{h}_t$ and the new information $\tilde{\mathbf{h}}_t$, using the output of the update gate $\mathbf{z}_t$
\begin{align}
    \mathbf{h}_t = \mathbf{z}_t \odot \mathbf{h}_{t-1} + \left(1 - \mathbf{z}_t\right)\odot \tilde{\mathbf{h}}_t.
\end{align}
The output of the GRU at time $t$ is simply $\mathbf{y}_t = \mathbf{h}_t$. The training will involve updating all the weight matrices and bias vectors, such that we obtain a target sequence of vectors $\mathbf{y}_t$ at every time instant $t$. This structure has lots of variations, but essentially they have the same overall structure where there is a hidden state that gets updated following some update rule. The GRU is a special class of a more general structure called the Long Short Term Memory (LSTM) \cite{hochreiter1997long}. In an LSTM, there there is a third gate that calculates the output given the hidden state rather than just outputting it as in a GRU (i.e.  an identity output gate). The name LSTM comes from the fact that the hidden state gets updated at every time without neither completely neglecting the new information $\tilde{\mathbf{h}}_t$ nor forgetting completely the old information $\mathbf{h}_{t-1}$. In this sense, it retains both a long and a short memory. These recurrent networks turn out to be very successful in application of time series analysis and natural language processing. However, they are generic enough for any application that involves sequence processing.
%%%%%%%%%%%%%%%%%%%%%%%%%%%%%%%%%%%%%%%%%%%%%%%%%%%%%%%%%%
\section{Supplementary figures}\label{appx:figures}
\begin{figure}[h]
    \centering
    \includegraphics[scale=0.75]{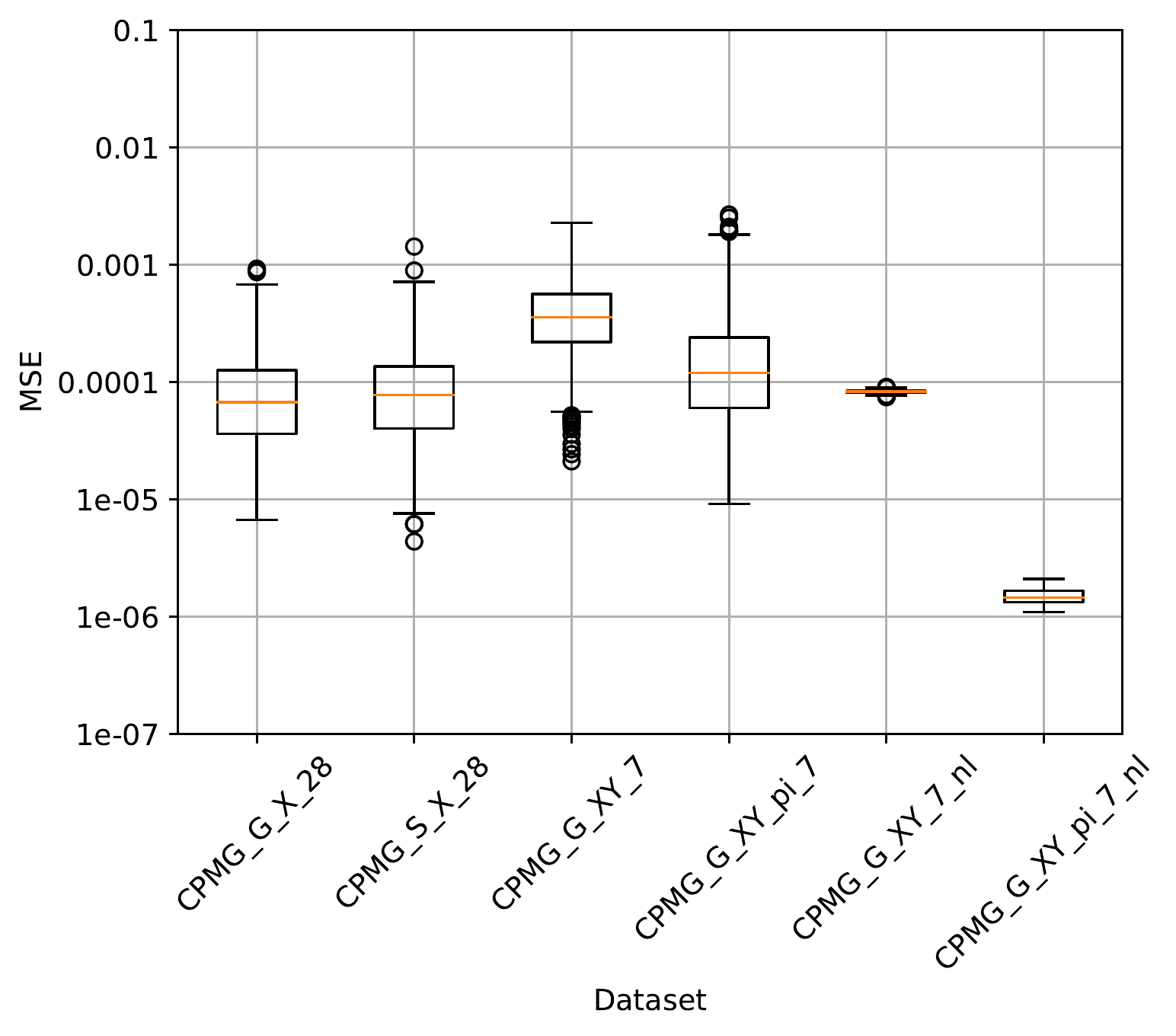}
    \caption{Violin plot of the MSE compared over all testing datasets.}
    \label{fig:boxplot}
\end{figure}

\begin{figure}[h]
    \centering
    \subfloat[Worst case]{\includegraphics[scale=0.5]{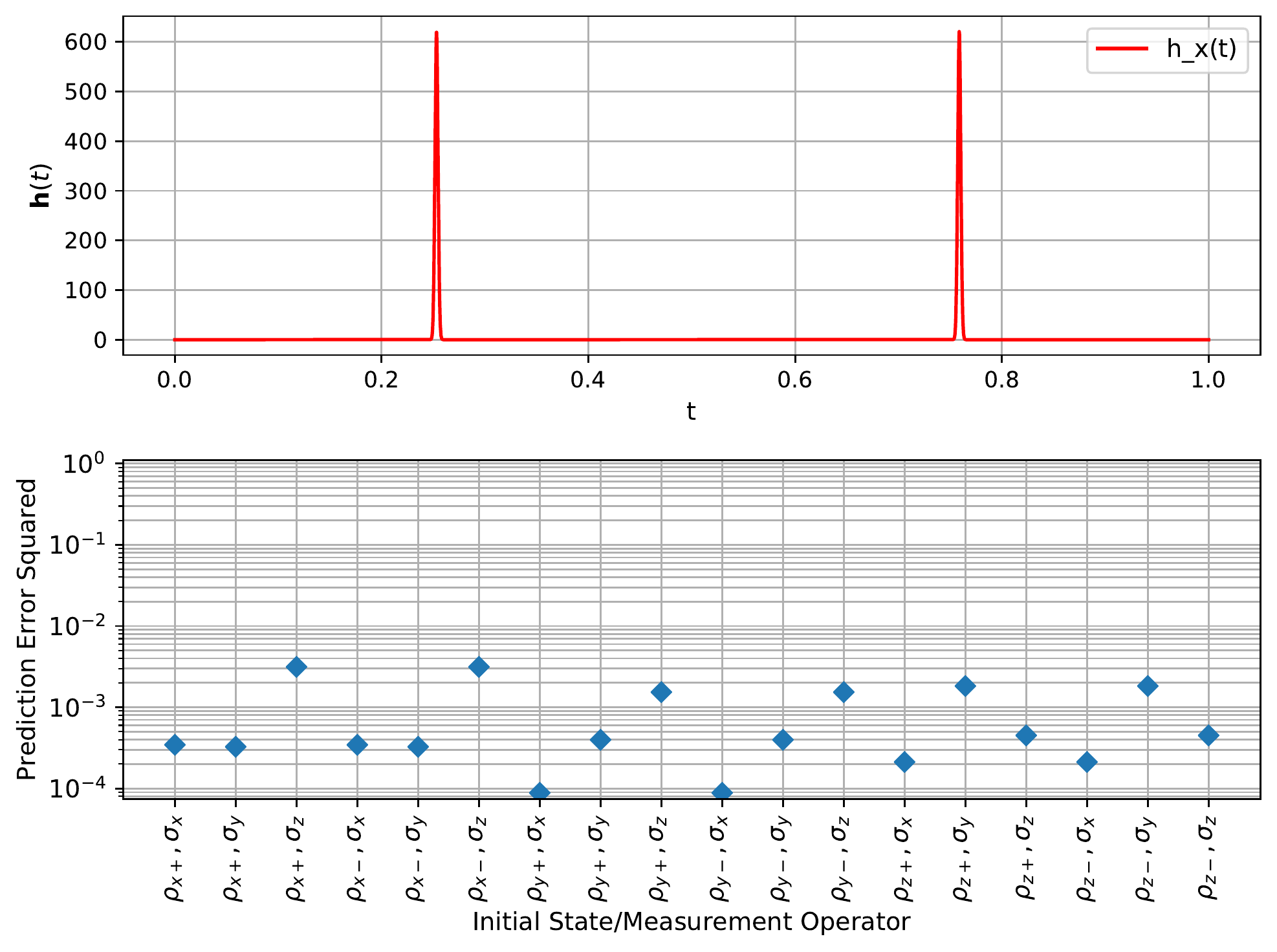}}\\
    \subfloat[Average Case]{\includegraphics[scale=0.5]{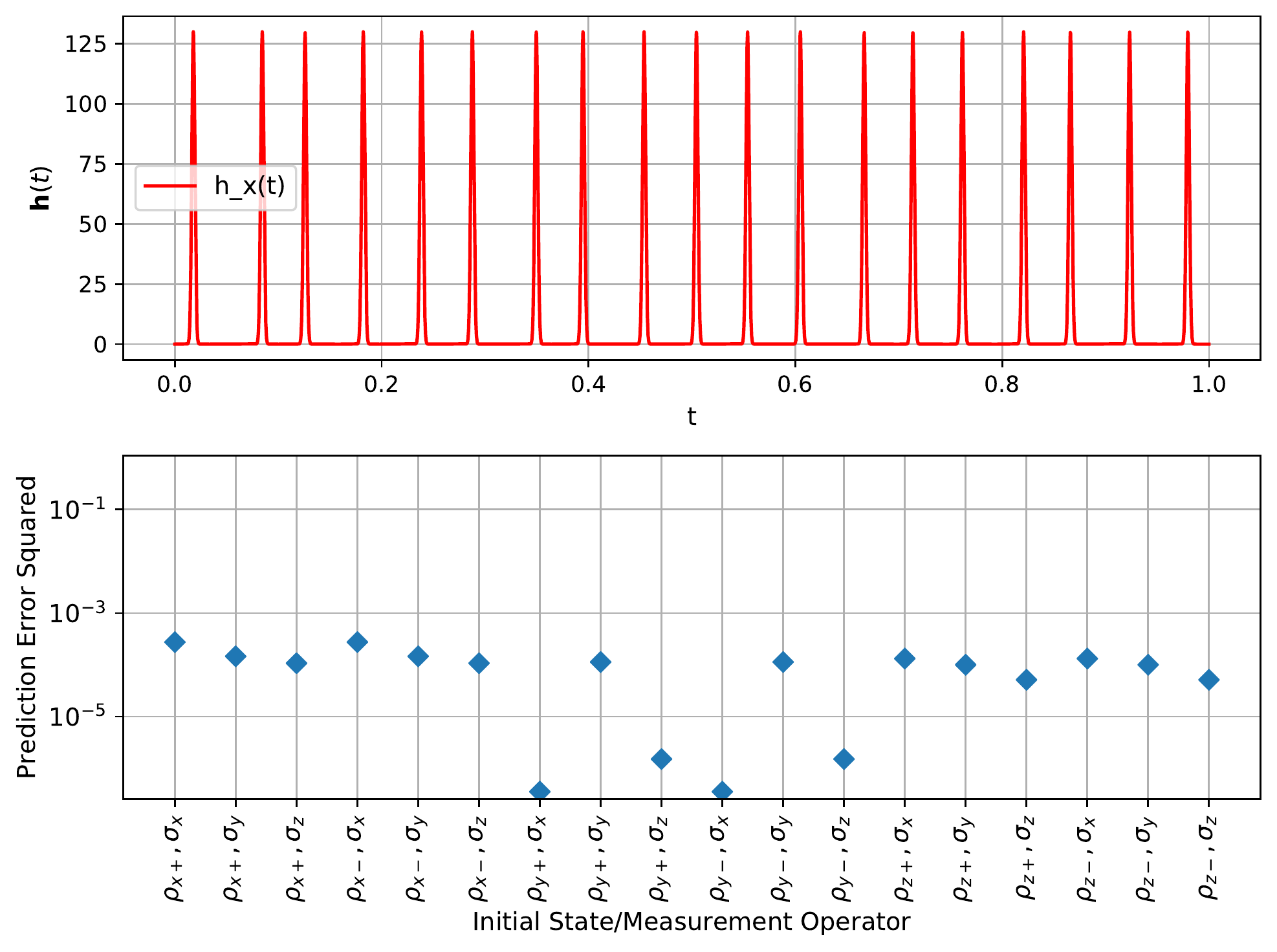}} \\
    \subfloat[Best Case]{\includegraphics[scale=0.5]{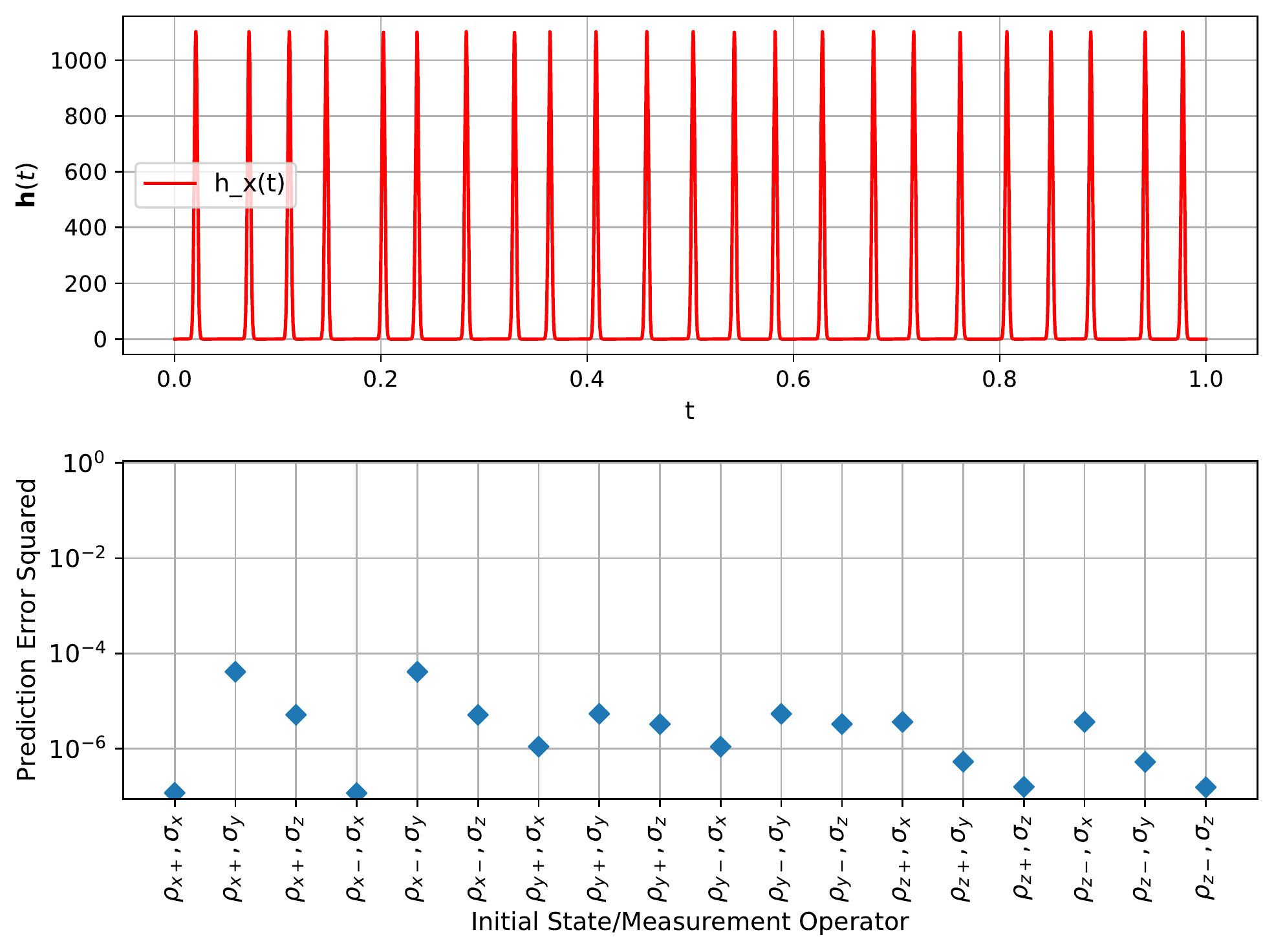}}
    \caption{The worst, average, and best case examples for the CPMG\_G\_X\_28 testing dataset.}
    \label{fig:ex_cat1a}
\end{figure}

\begin{figure}[h]
    \centering
    \subfloat[Worst case]{\includegraphics[scale=0.5]{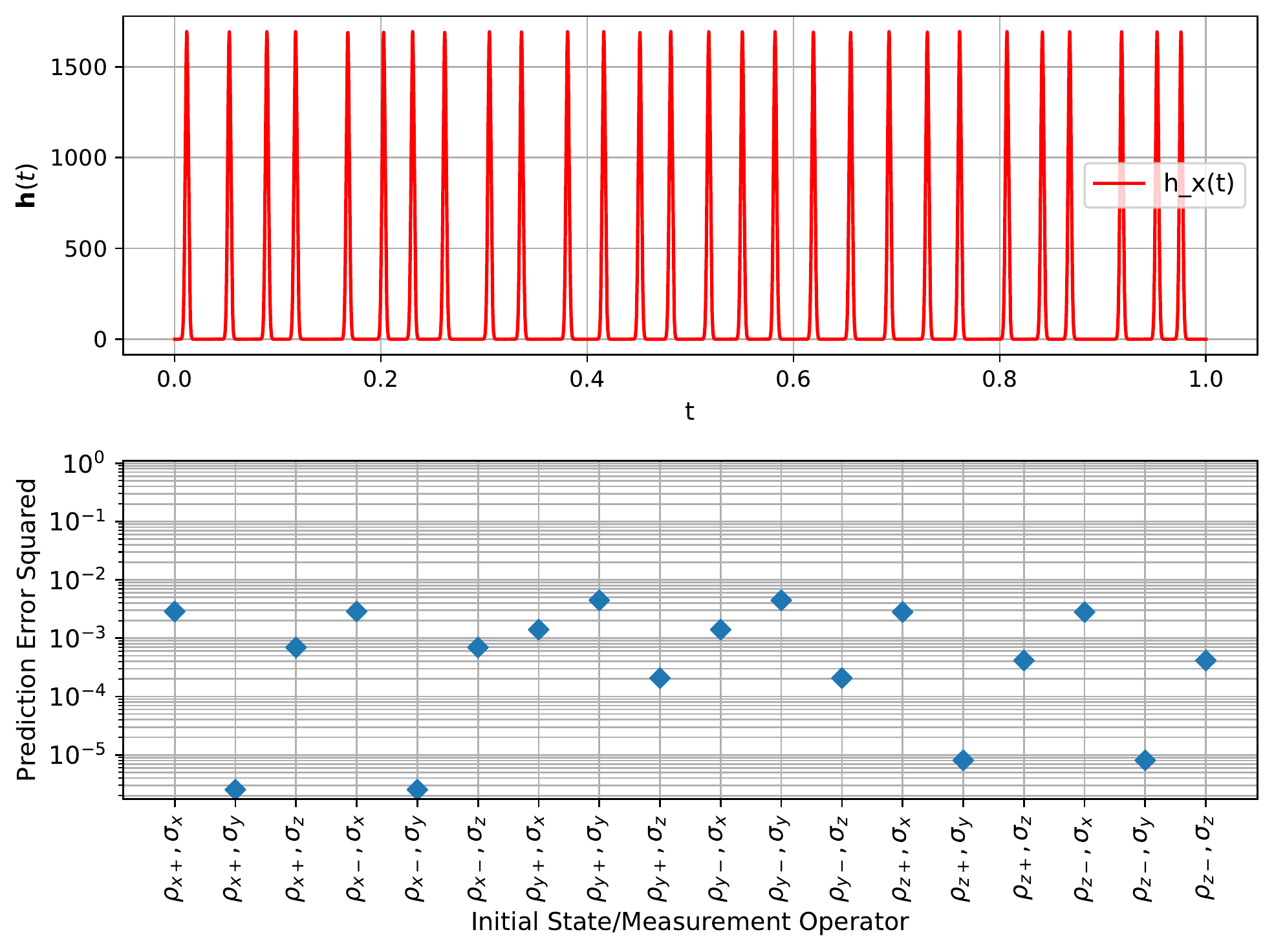}}\\
    \subfloat[Average Case]{\includegraphics[scale=0.5]{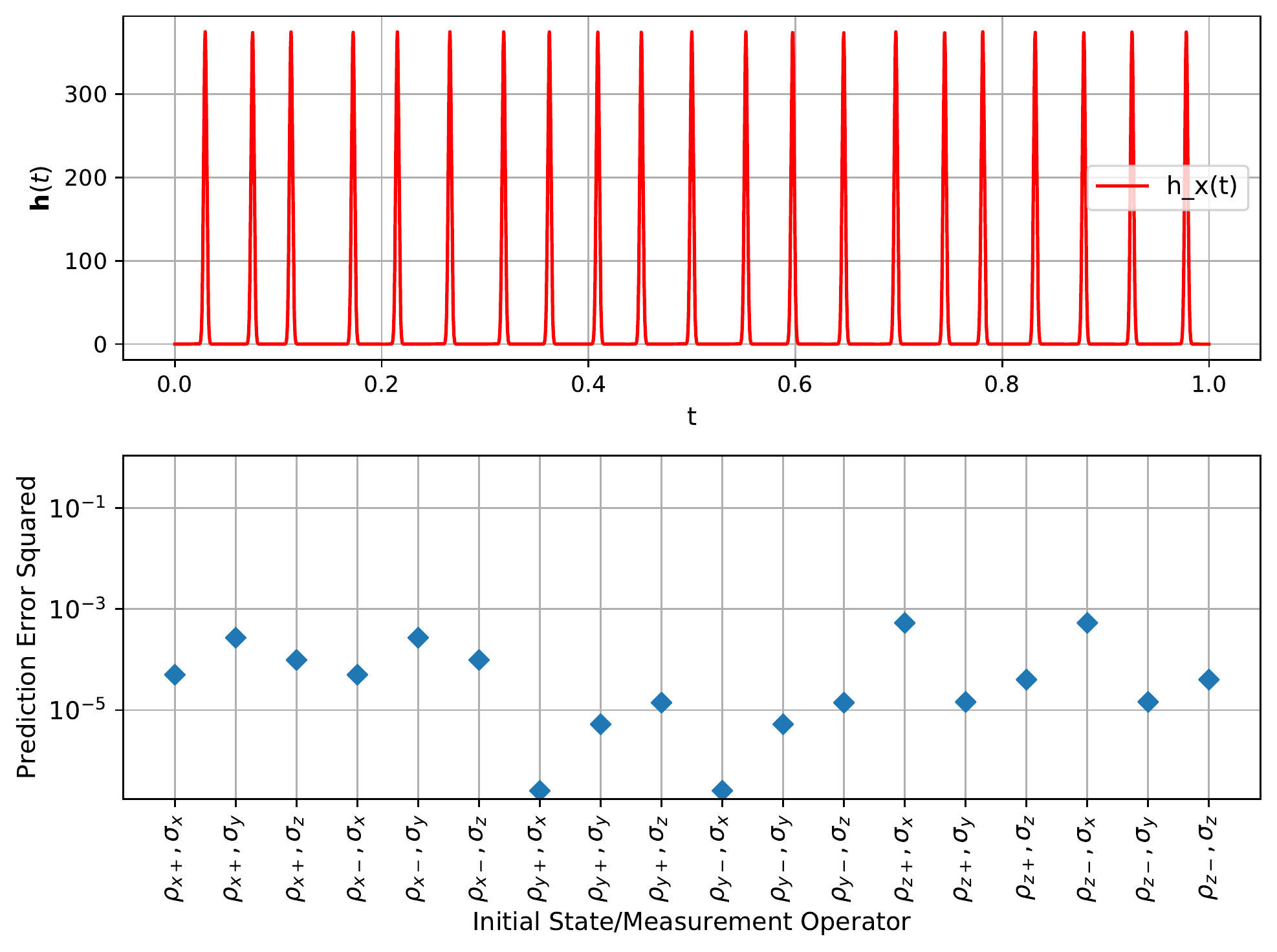}} \\
    \subfloat[Best Case]{\includegraphics[scale=0.5]{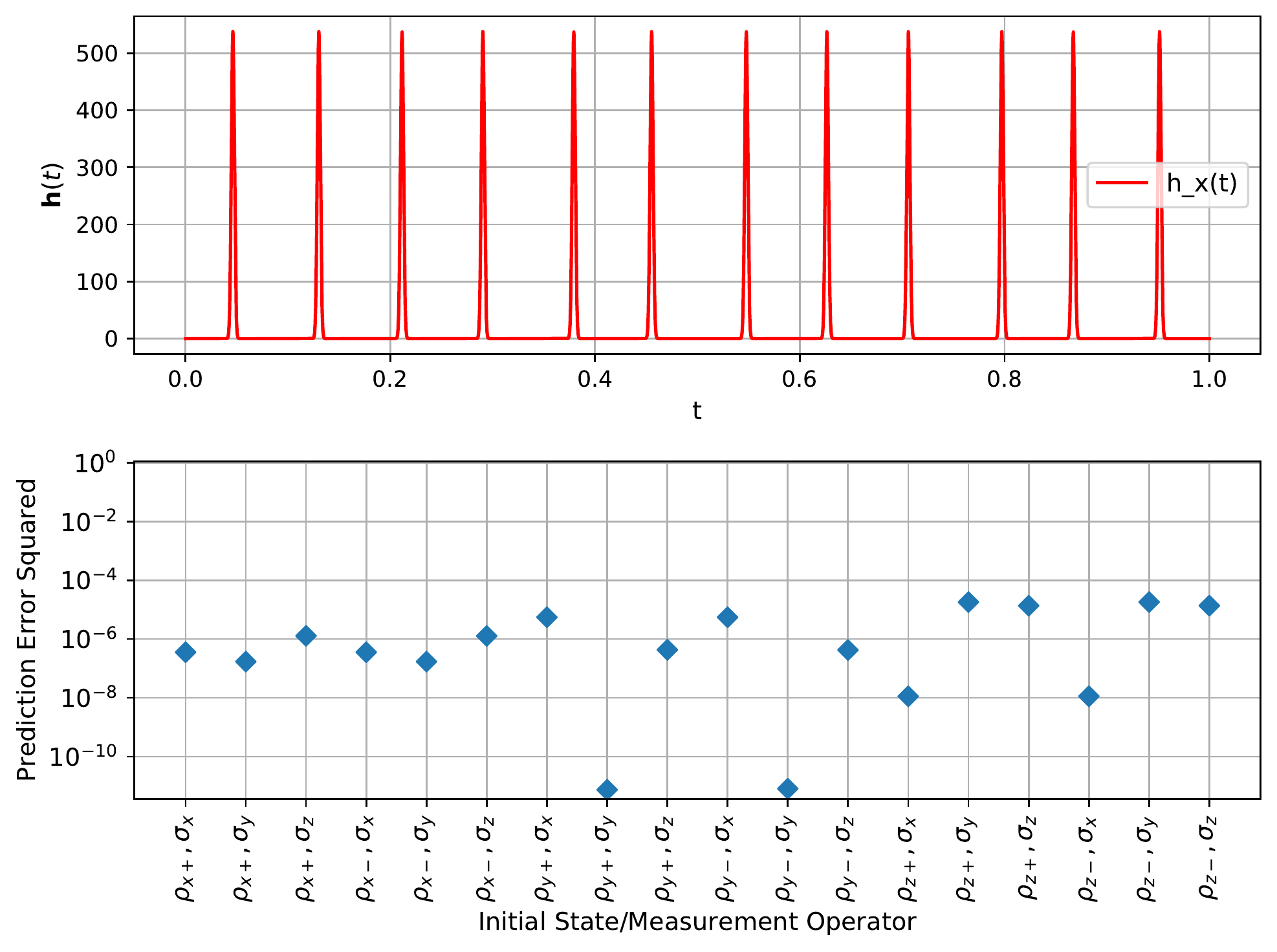}}
    \caption{The worst, average, and best case examples for the CPMG\_S\_X\_28 testing dataset.}
    \label{fig:ex_cat1b}
\end{figure}

\begin{figure}[h]
    \centering
    \subfloat[Worst case]{\includegraphics[scale=0.5]{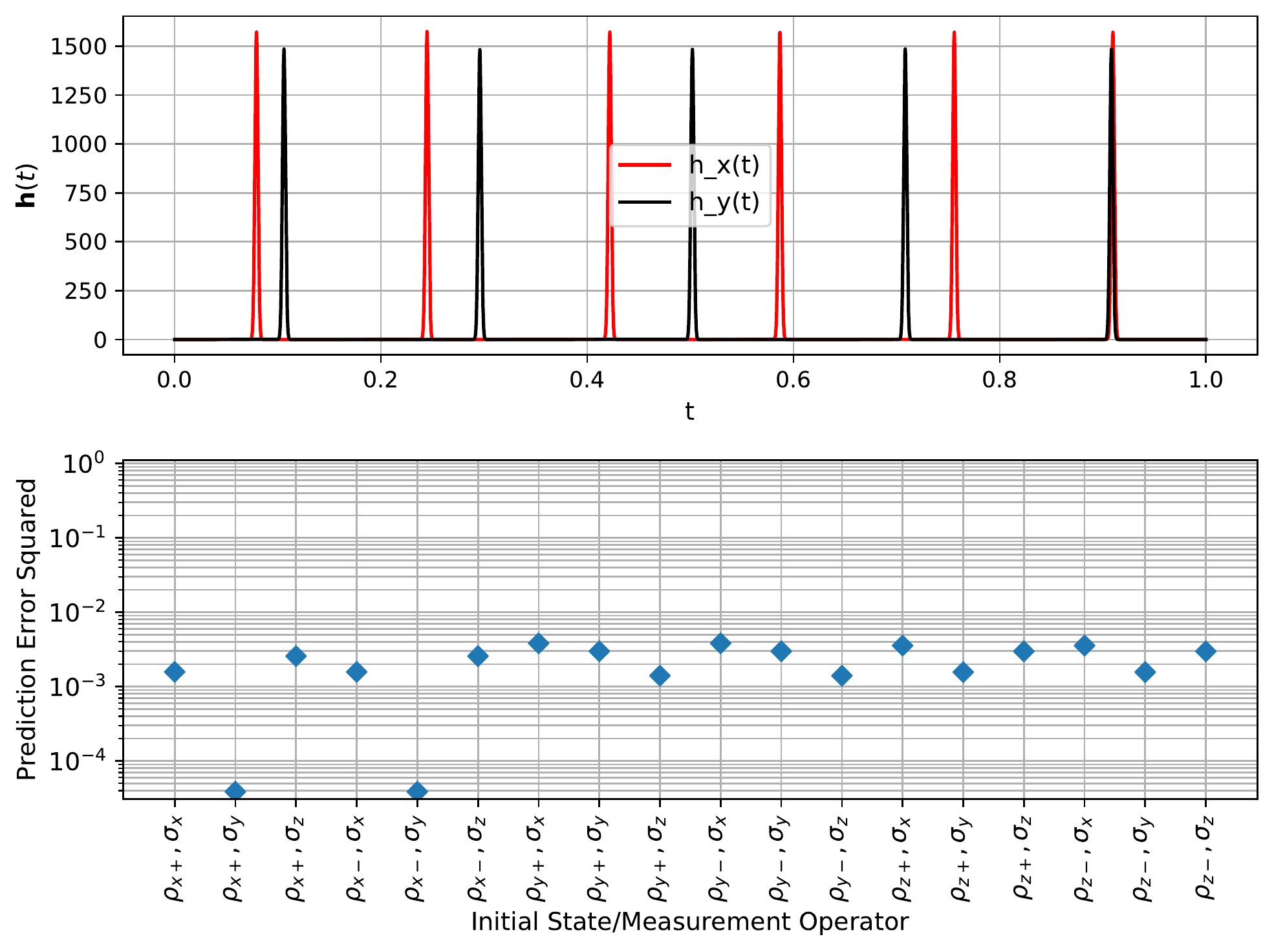}}\\
    \subfloat[Average Case]{\includegraphics[scale=0.5]{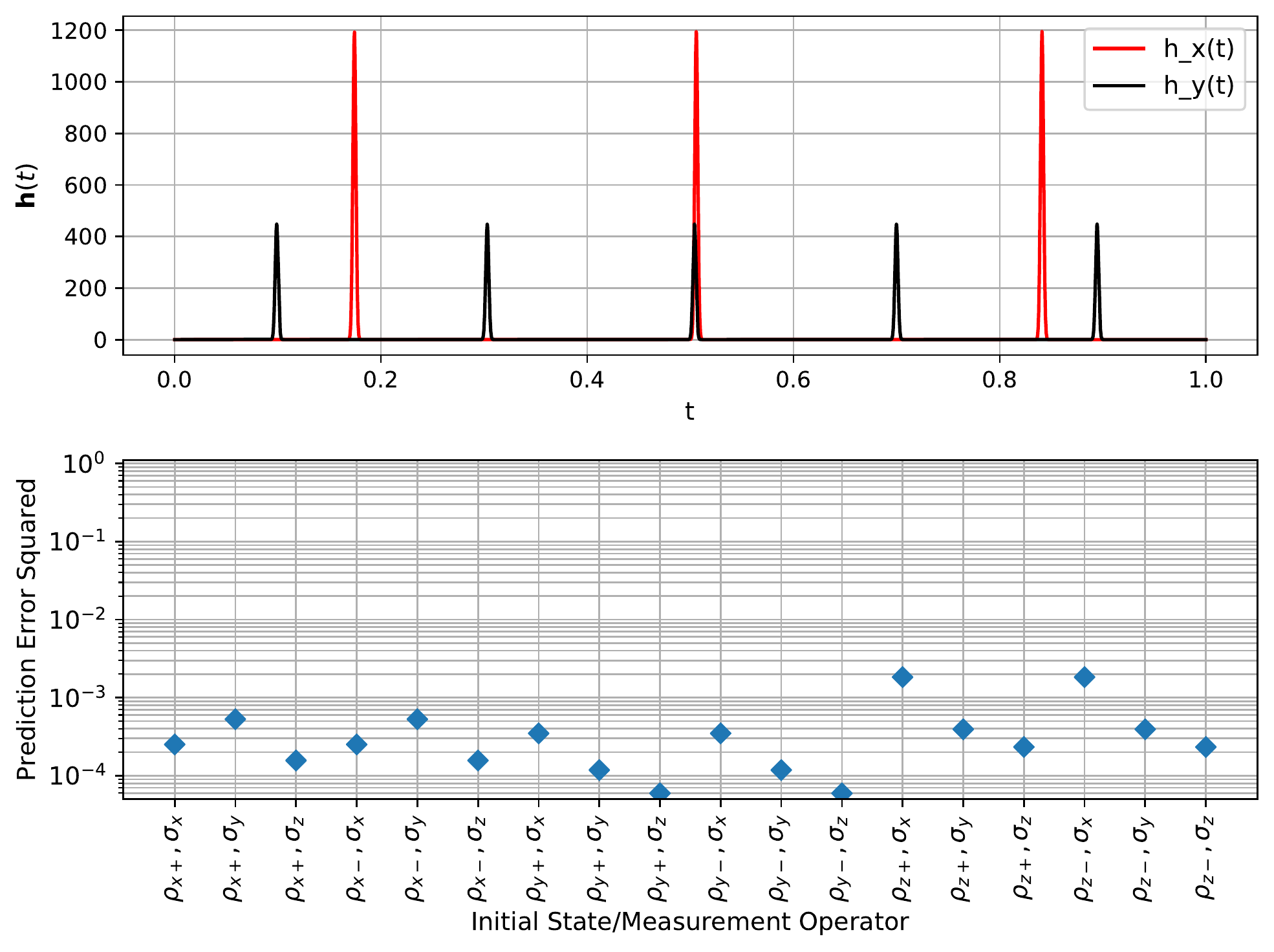}} \\
    \subfloat[Best Case]{\includegraphics[scale=0.5]{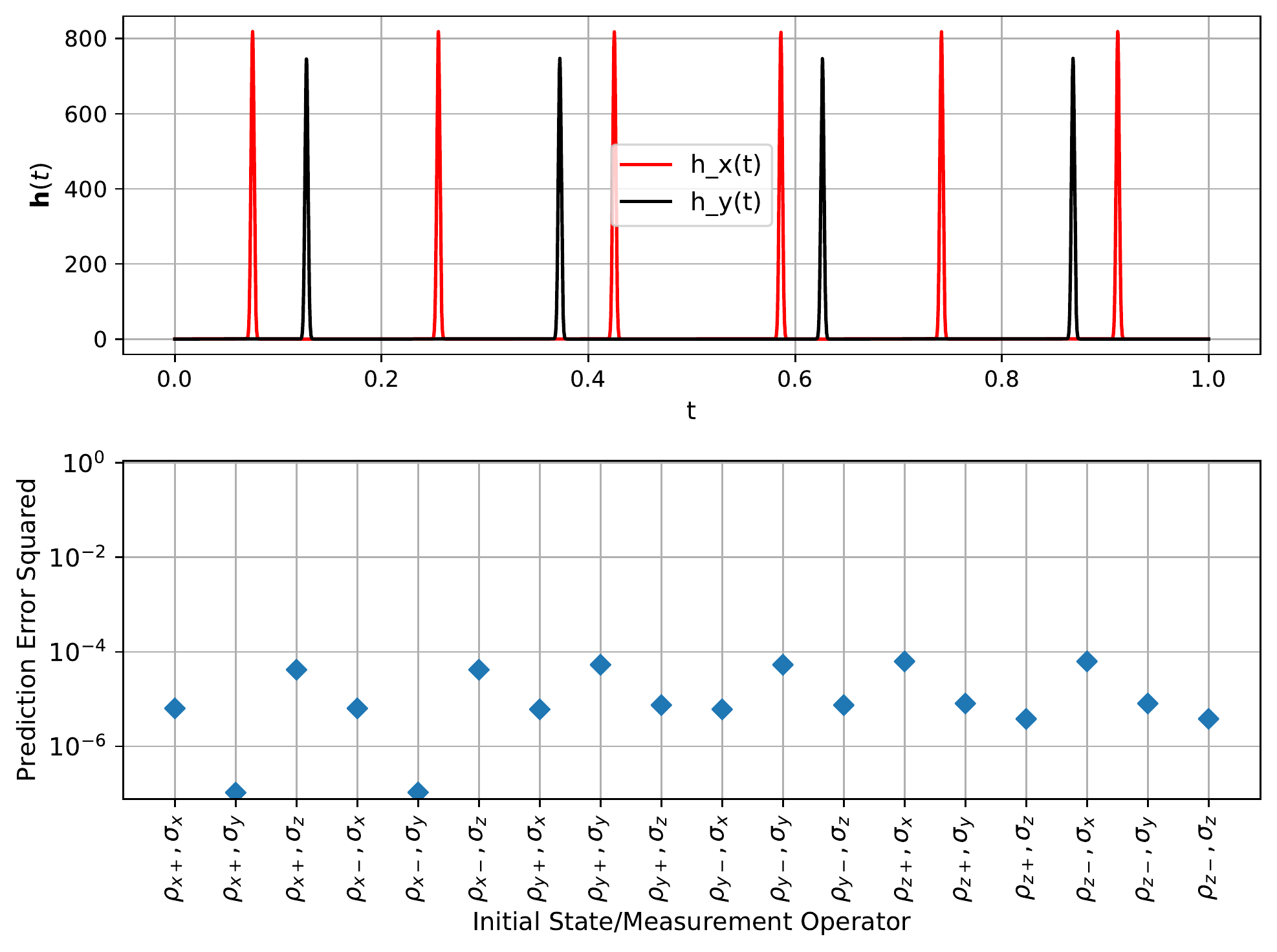}}
    \caption{The worst, average, and best case examples for the CPMG\_G\_XY\_7 testing dataset.}
    \label{fig:ex_cat2a}
\end{figure}

\begin{figure}[h]
    \centering
    \subfloat[Worst case]{\includegraphics[scale=0.5]{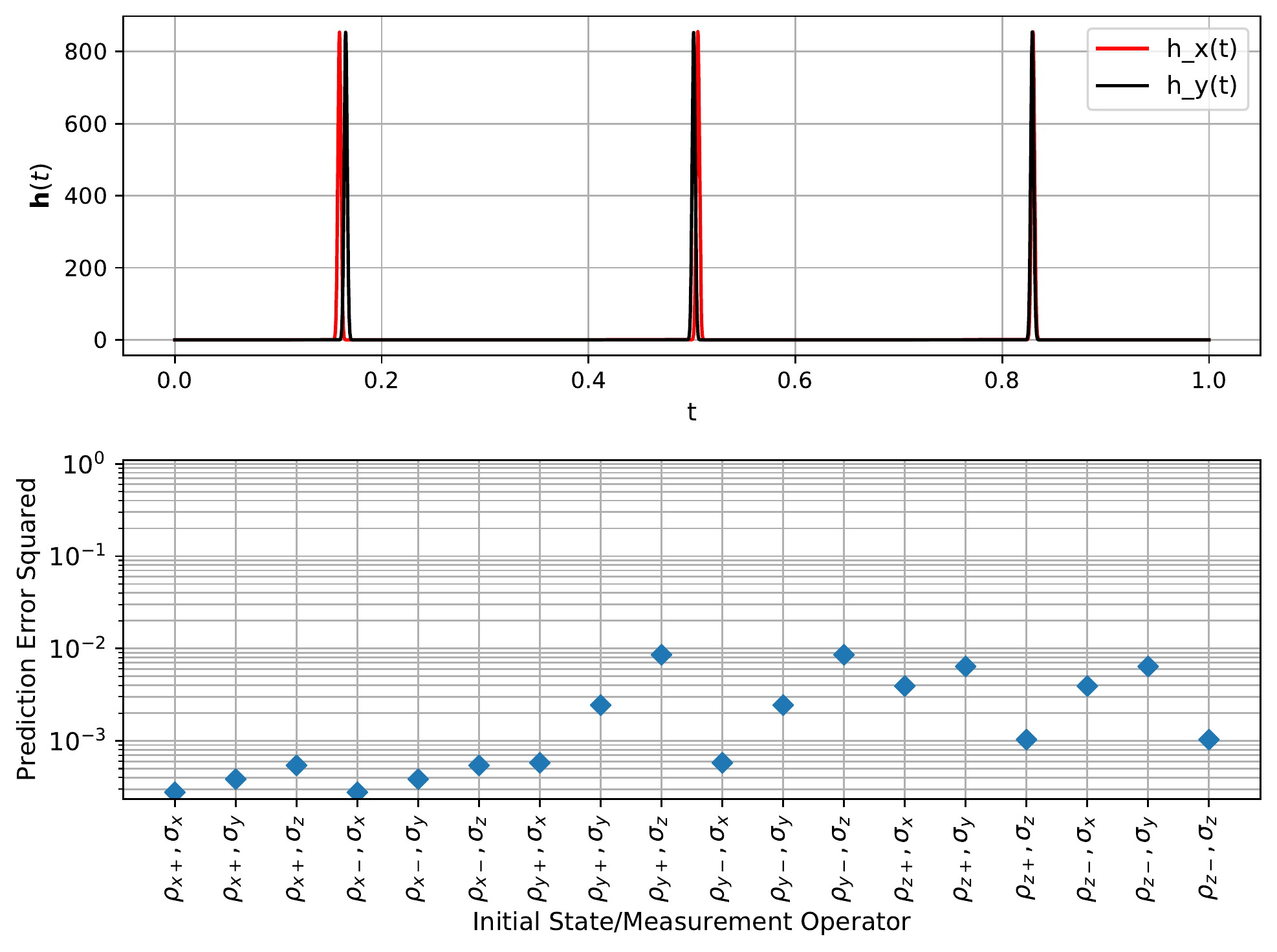}}\\
    \subfloat[Average Case]{\includegraphics[scale=0.5]{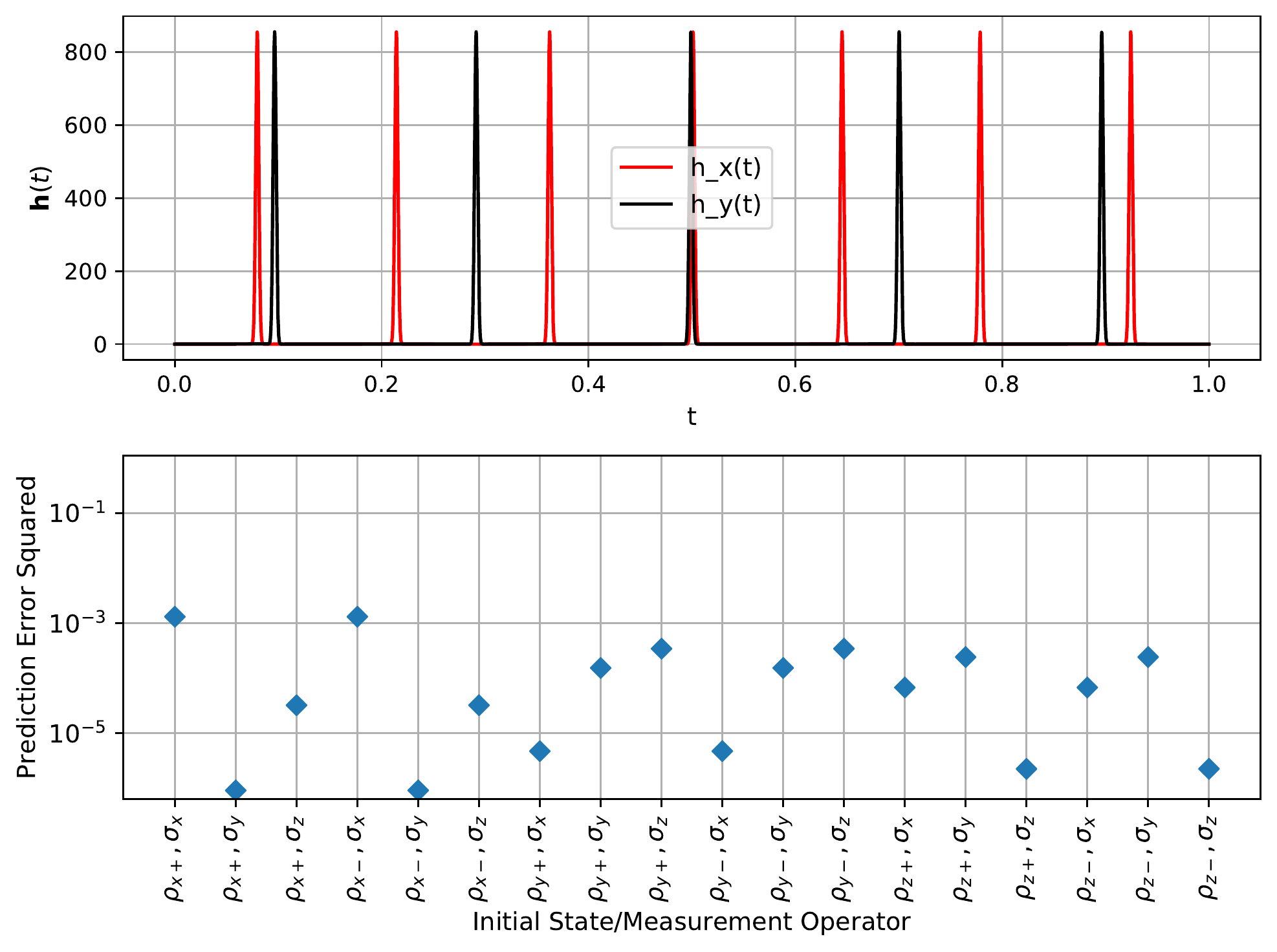}} \\
    \subfloat[Best Case]{\includegraphics[scale=0.5]{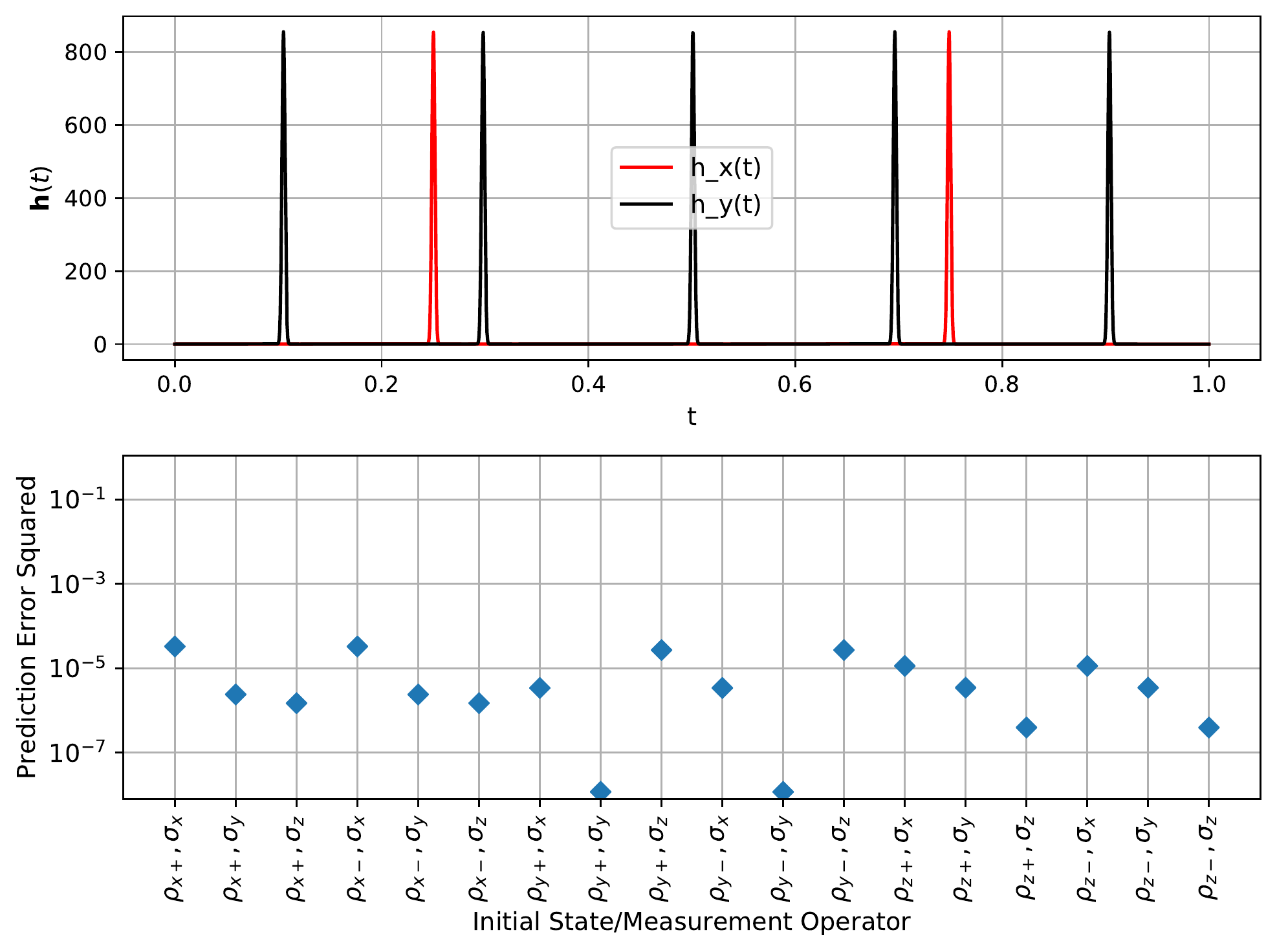}}
    \caption{The worst, average, and best case examples for the CPMG\_G\_XY\_pi\_7 testing dataset.}
    \label{fig:ex_cat2b}
\end{figure}

\begin{figure}[h]
    \centering
    \subfloat[Worst case]{\includegraphics[scale=0.5]{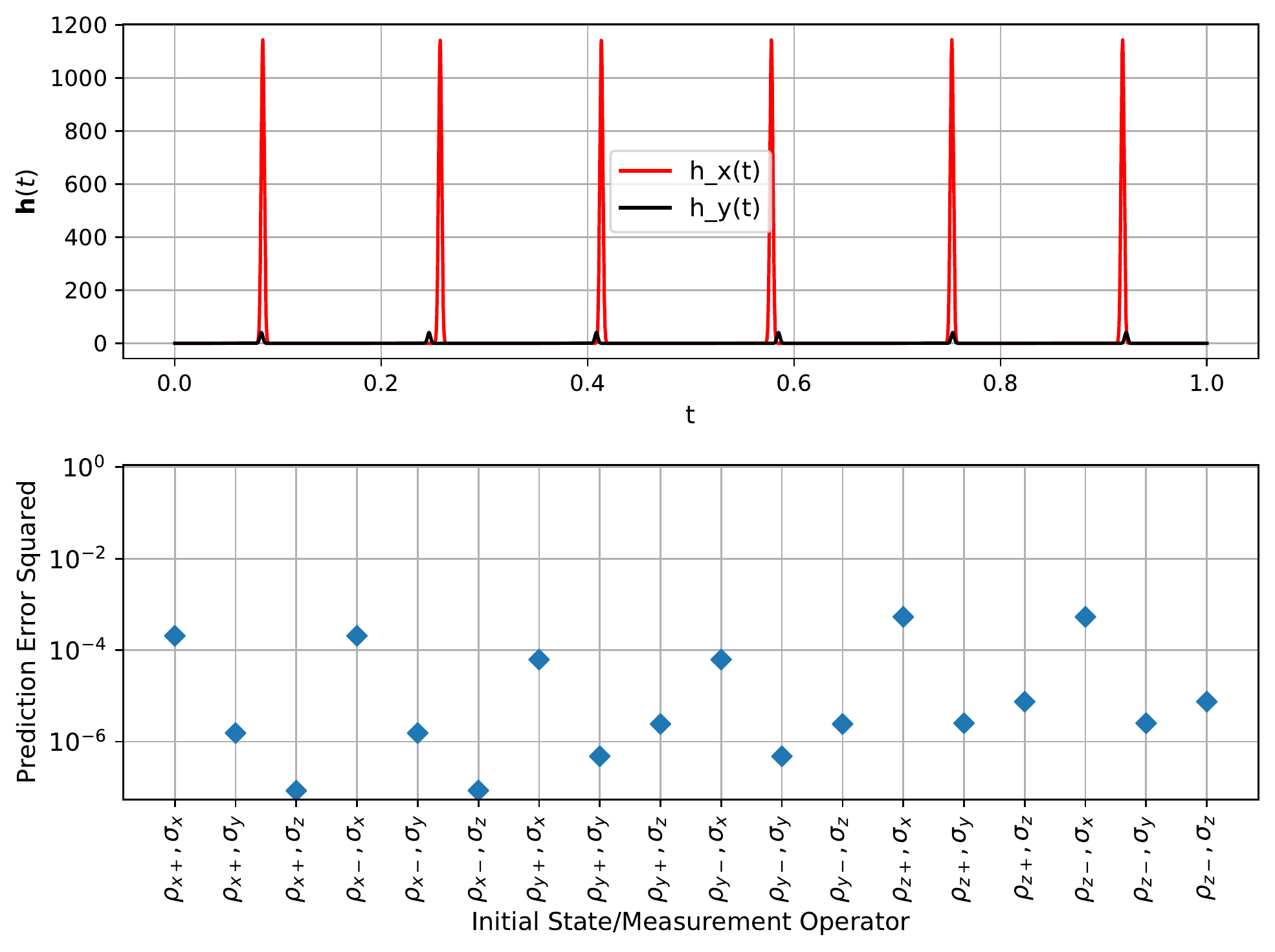}}\\
    \subfloat[Average Case]{\includegraphics[scale=0.5]{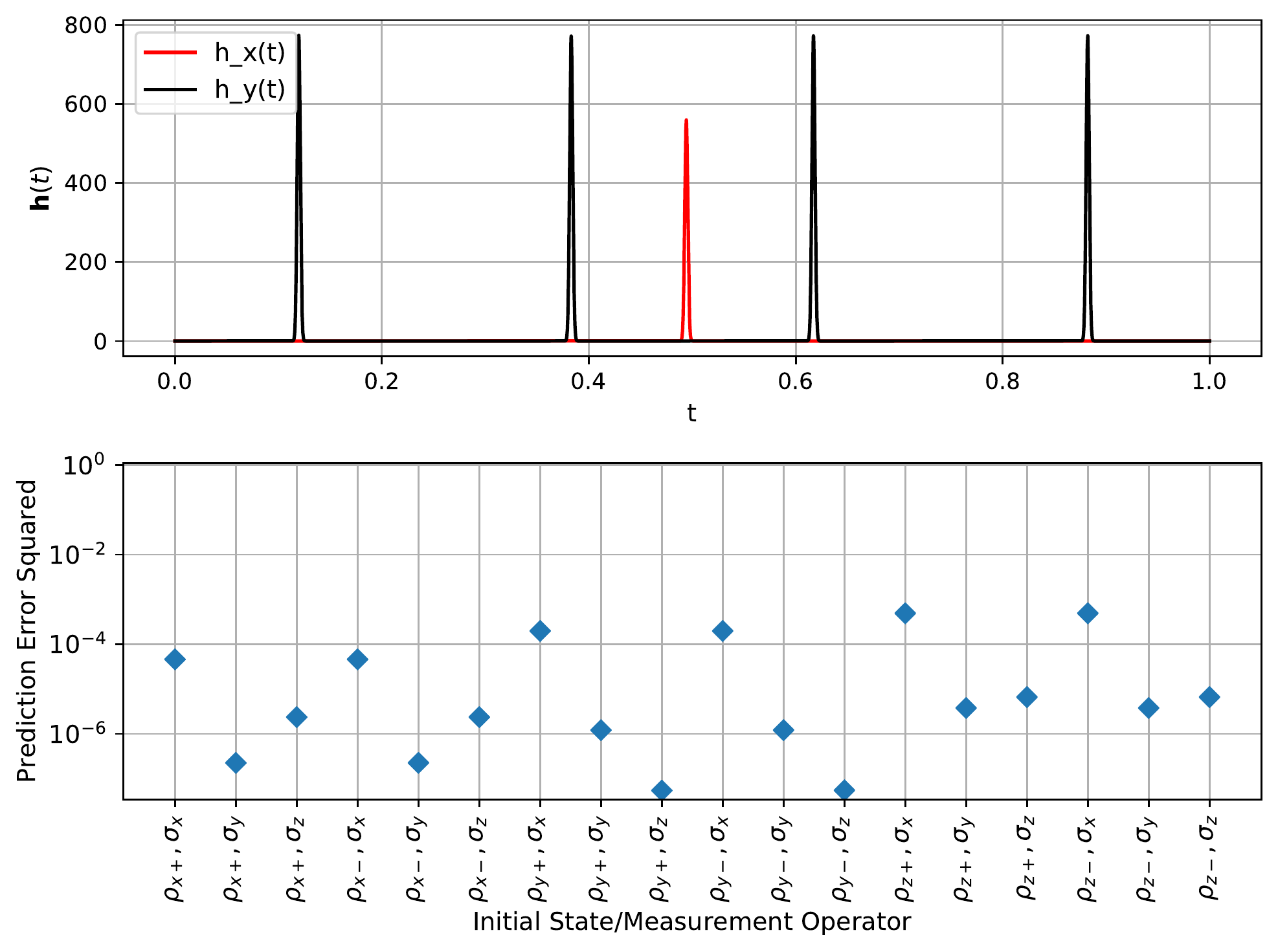}} \\
    \subfloat[Best Case]{\includegraphics[scale=0.5]{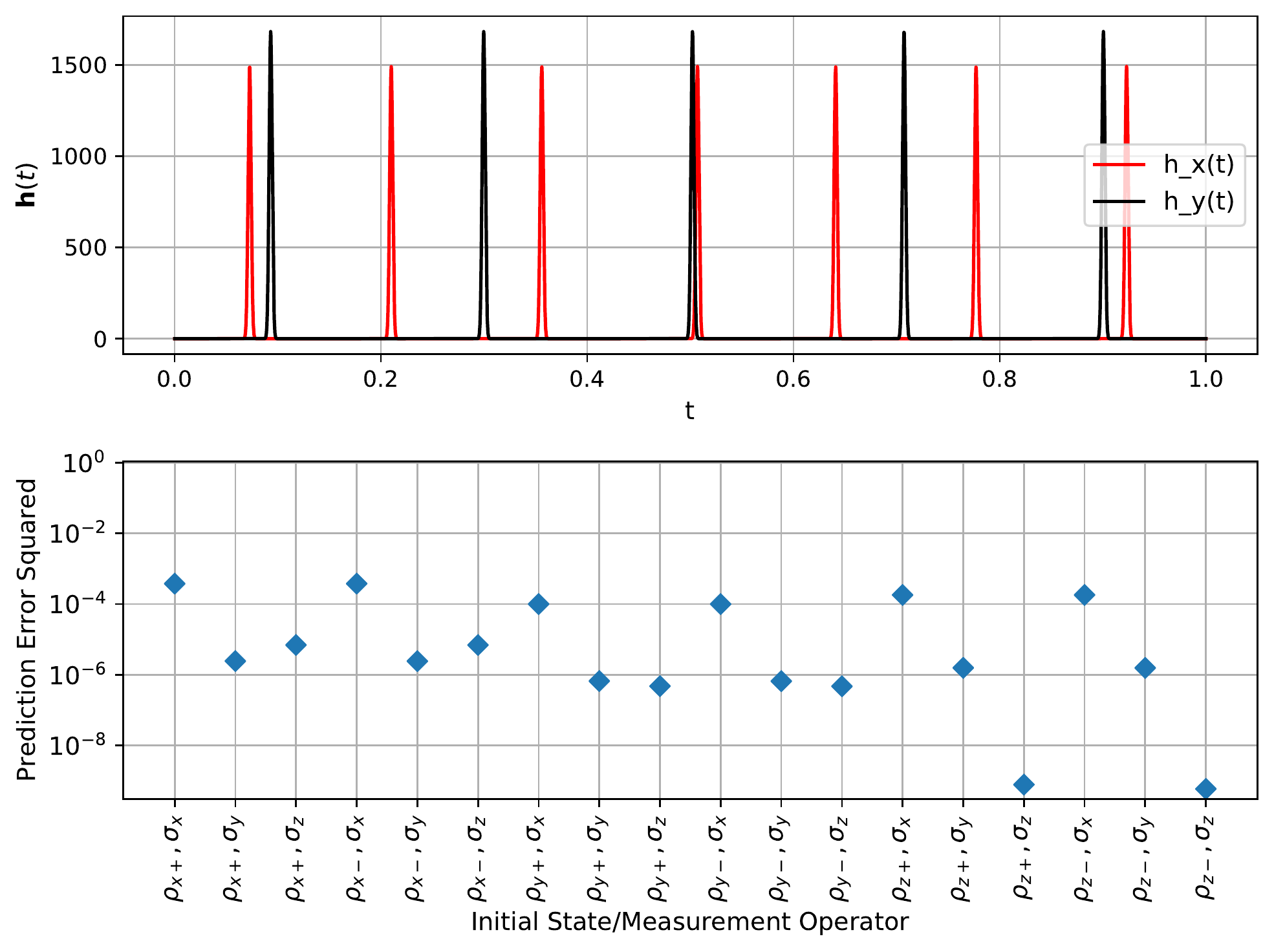}}
    \caption{The worst, average, and best case examples for the CPMG\_G\_XY\_7\_nl testing dataset.}
    \label{fig:ex_cat3a}
\end{figure}

\begin{figure}[h]
    \centering
    \subfloat[Worst case]{\includegraphics[scale=0.5]{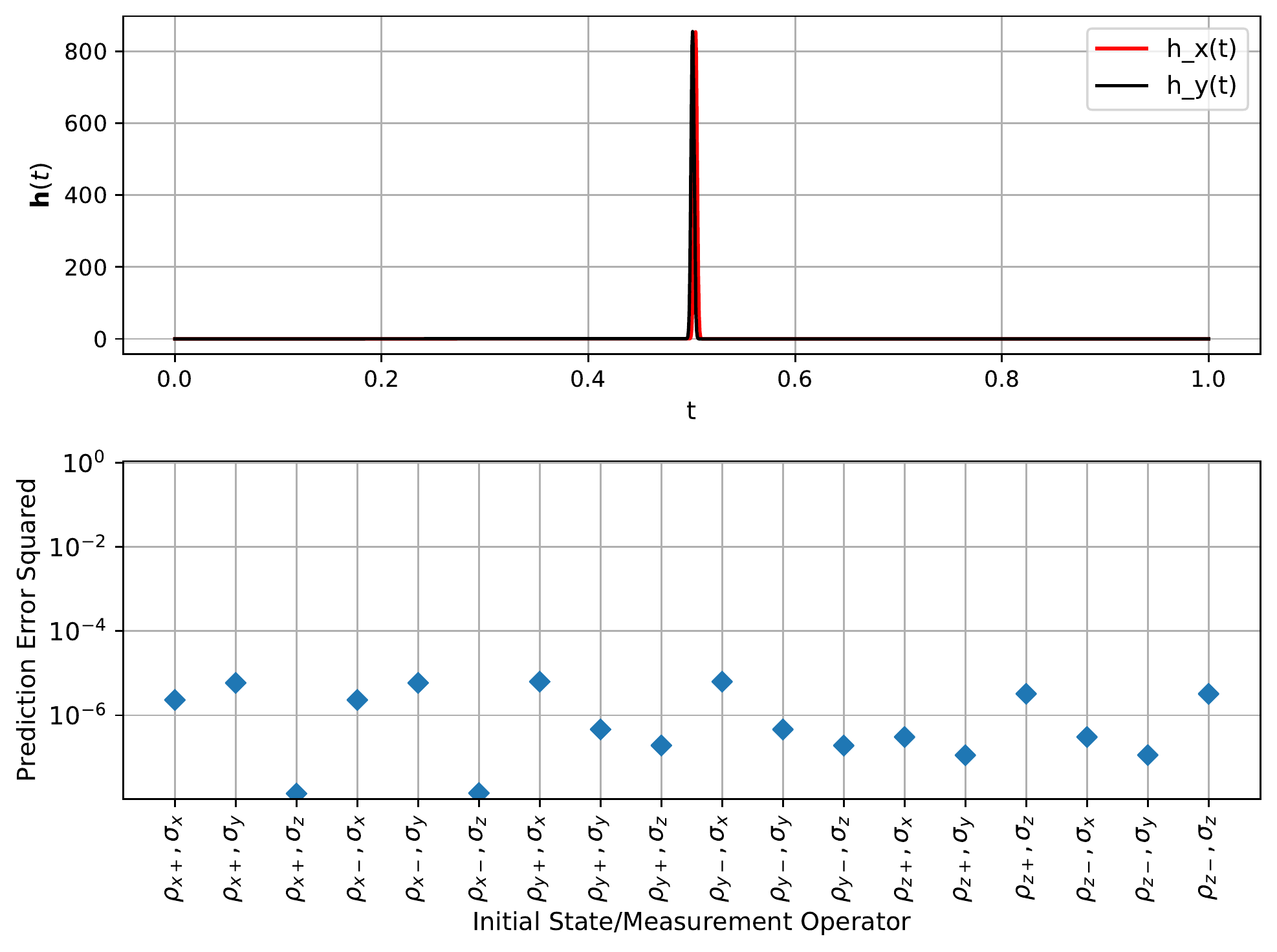}}\\
    \subfloat[Average Case]{\includegraphics[scale=0.5]{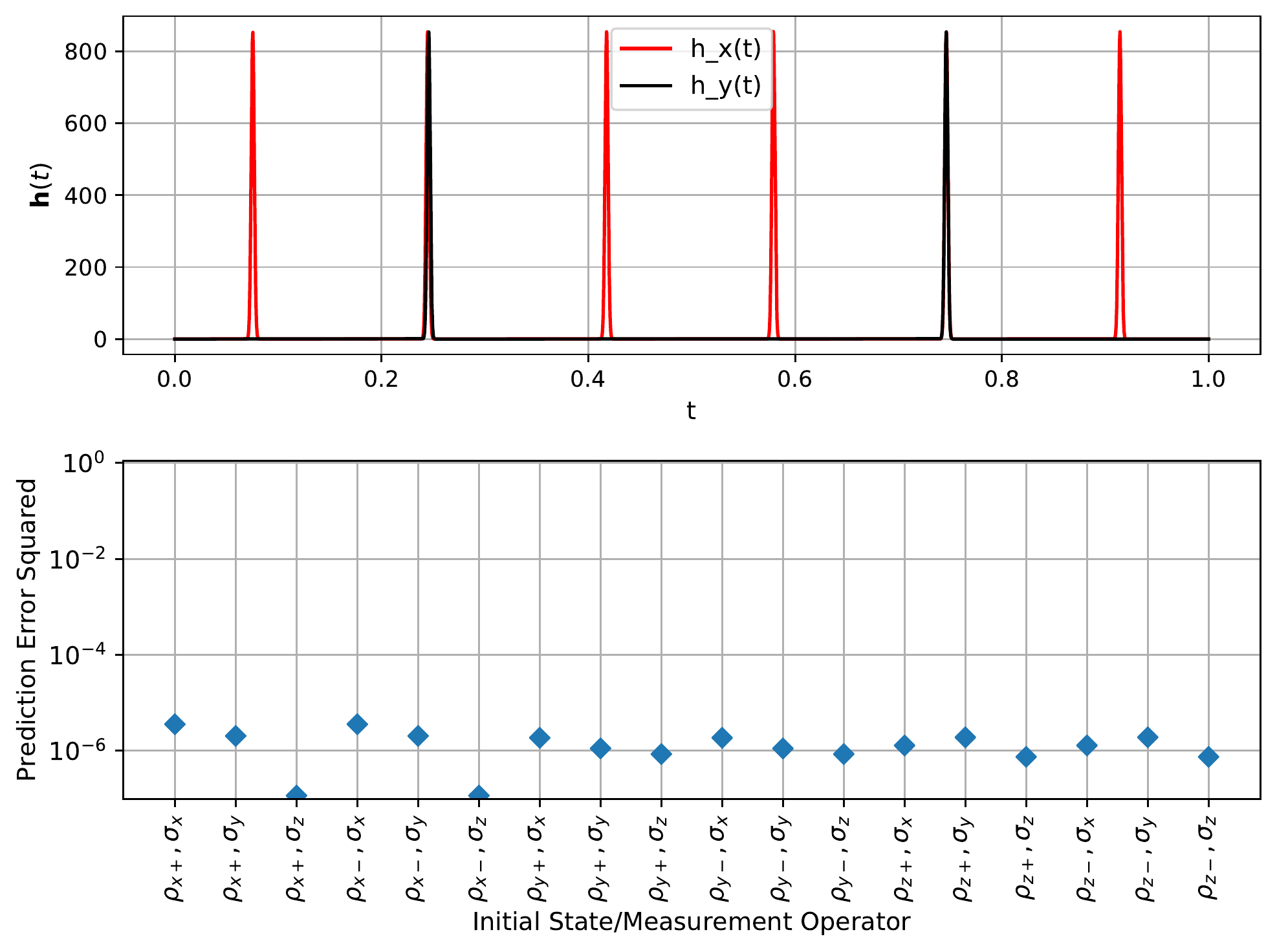}} \\
    \subfloat[Best Case]{\includegraphics[scale=0.5]{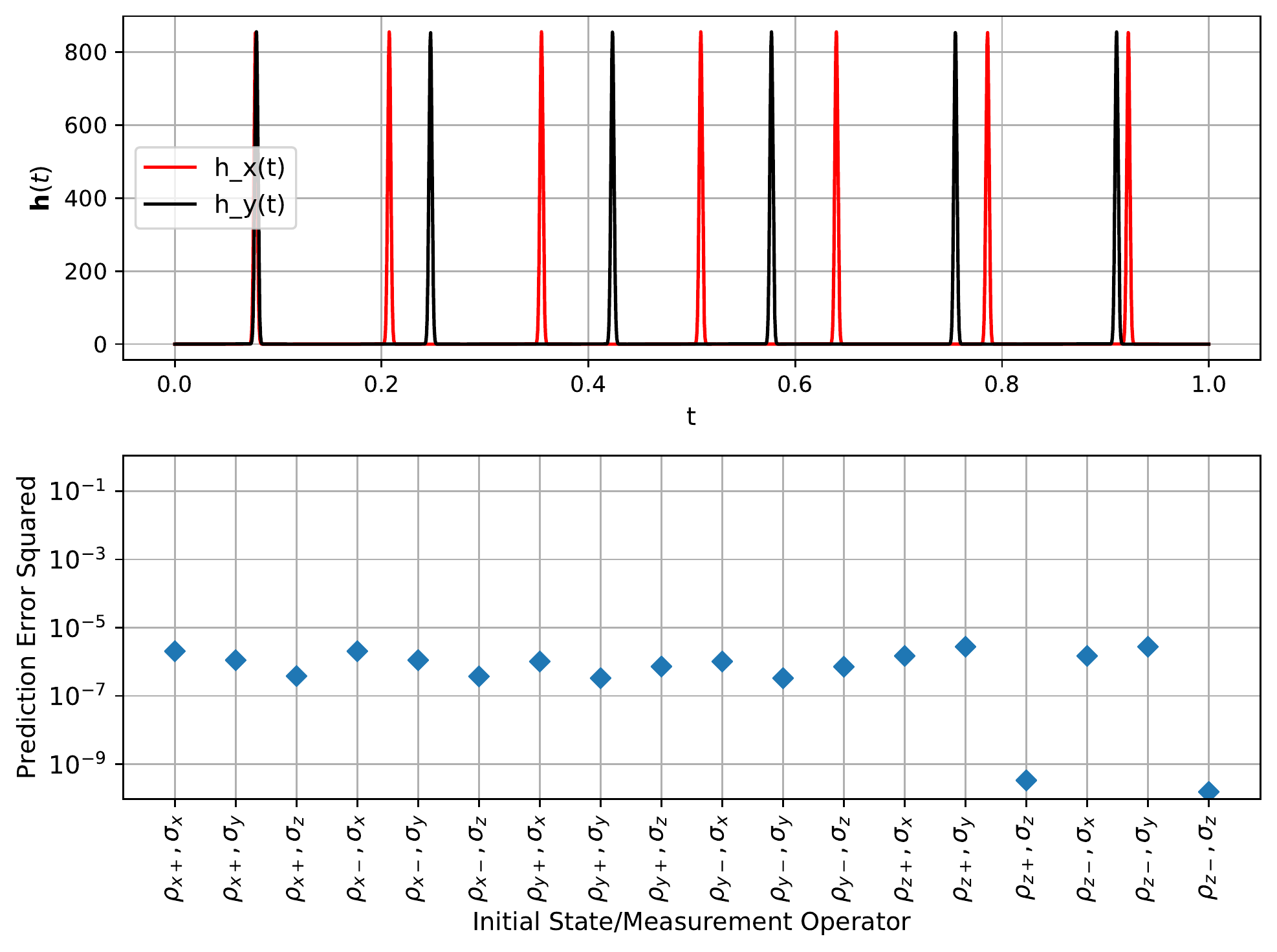}}
    \caption{The worst, average, and best case examples for the CPMG\_G\_XY\_pi\_7\_nl testing dataset.}
    \label{fig:ex_cat3b}
\end{figure}
\end{document}